\documentclass[aps,nofootinbib,noshowpacs,showkeys,preprintnumbers,amsmath,amssymb]{revtex4}
\def\bes{\begin{subequations}}
\def\ees{\end{subequations}}
\def\be{\begin{equation}}
\def\ee{\end{equation}}
\def\bea{\begin{eqnarray}}
\def\eea{\end{eqnarray}}
\def\ba{\begin{eqnarray}}
\def\ea{\end{eqnarray}}
\def\bear{\begin{array}}
\def\eear{\end{array}}
\def\p1sl{\displaystyle{\not}p_1}
\def\p2sl{\displaystyle{\not}p_2}

\newcommand{\K}{{\widetilde {\cal K}}}

\newcommand{\bG}{{\overline{\Gamma}}}

\newcommand{\Dst}{D^*}
\newcommand{\blam}{{\overline \lambda}}
\usepackage{mathtools}
\newcommand{\fsl}[1]{\ensuremath{\mathrlap{\!\not{\phantom{#1}}}#1}}
\usepackage{graphics}
\usepackage{graphicx}
\usepackage{dcolumn}
\usepackage{bm}
\usepackage{epsfig}
\usepackage{graphicx}
\usepackage{multirow}
\usepackage{dcolumn}
\usepackage{graphicx,epsfig}%

\begin{document}
\preprint{USM-TH-344}

\title{Rare decays of $B$ mesons via on-shell sterile neutrinos.\footnote{1606.04140v4: in v4, in Fig.4(b) the lower curve [for $({\bar B}^0 \to D^{*+} \tau^-{\bar N})$ decay] is now correct (in v3 an obsolete version of the curve was inadvertently included). A typo in Eq.~(C15a) is corrected. As published in Phys.\ Rev.\ D94, 053001 (2016) and Phys.\ Rev.\ D95, 039901(E) (2017).}}
\author{Gorazd Cveti\v{c}$^1$}
\email{gorazd.cvetic@usm.cl}
\author{C.~S.~Kim$^2$}
\email{cskim@yonsei.ac.kr}
\affiliation{$^1$
Department of Physics, Universidad T\'ecnica Federico Santa Mar\'ia, Valpara\'iso, Chile\\
$^3$Department of Physics and IPAP, Yonsei University, Seoul 120-749, Korea}

\date{\today}

\begin{abstract}
In view of the projected high number of produced $B$ mesons in
Belle II experiment  ($\sim 10^{10}$ per year),
in addition to the presently ongoing LHC-b,
we calculate the rate of
decay for the rare decays of $B$ mesons via a sterile on-shell neutrino
$N$, which subsequently may decay 
leptonically or semileptonically within the detector:
$B \to (D^{(*)}) \ell_1^{\pm} N$, then $N \to \ell_1^{\pm} \ell_2^{\mp} \nu$ or $N \to \ell^{\pm} \pi^\mp$.
Here, 
$\ell_1 \not= \ell_2$ in order to avoid serious QED background.
We account for the possible effects of the neutrino lifetime on the observability of the rare decays.
If no charmed mesons ($D^{(*)}$) are produced at the first vertex of the
sterile neutrino, a strong CKM-suppression becomes effective; this is not true
if we consider instead the decays of $B_c$ mesons which can be produced
copiously in LHC-b. The production of charmed mesons $D^{(*)}$ at the
first vertex offers an attractive possibility because it avoids strong
CKM-suppression. Such rare decays of $B$ mesons could be detected at
Belle II experiment, with $N$ neutrino either decaying within the
detector or manifesting itself as a massive missing momentum.
\end{abstract}

\maketitle


\section{Introduction}
\label{intr}

At present it is not known whether there exist sterile neutrinos. If they do, there exist additional neutrino mass eigenstates, and two of the central questions are: (1) How heavy are the new mass eigenstates $N$? (2) How strong (weak) are the corresponding heavy-light mixings $U_{\ell N}$, i.e., mixings of $N$ with the Standard Model (SM) flavor neutrinos $\nu_{\ell}$ ($\ell=e, \mu, \tau$)?
Further, it is not known whether the neutrinos are Majorana or Dirac fermions. Majorana fermions are at the same time their own antiparticles, which is not the case for Dirac fermions such as charged leptons or quarks. Majorana neutrinos allow both lepton number conserving (LNC) and lepton number violating (LNV) processes, while Dirac neutrinos allow only LNC processes to take place. The LNV processes are in general appreciable only if the Majorana neutrinos have appreciable masses.

The Majorana nature of (light) neutrinos can be established if neutrinoless double beta decays ($0\nu\beta\beta$) can be detected \cite{0NBB}. The existence of sterile (usually Majorana) neutrinos can be established by specific scattering processes \cite{scatt1,scatt2,scatt3,KimLHC} and via rare meson decays \cite{RMDs,HKS,Atre,CDKK,CDK,CKZ,symm,Quint,Mand}.

Evidence of the nonzero neutrino masses comes from neutrino oscillation, predicted by \cite{Pontecorvo} and later observed \cite{oscatm,oscsol,oscnuc}. The observation of oscillations can determine the mass differences of the light neutrinos. Nonetheless, if several sterile neutrinos exist, they may result in almost degenerate heavy mass states and such scenarios can lead to oscillation of heavy neutrinos \cite{Boya,CKZosc}.

Sterile neutrinos and the corresponding heavy neutrino particles appear naturally in several scenarios in which the small masses of the light neutrinos are explained. The very low masses $m_{\nu} \alt 1$ eV of the three light neutrinos can be explained by seesaw scenarios \cite{seesaw} where neutrinos are Majorana particles and the heavy neutrino mass eigenstates have very high masses $M \gg 1$ TeV. Other seesaw scenarios have lower masses of the heavy neutrinos, $M \sim 1$ TeV \cite{WWMMD} and
$M \sim 1$ GeV \cite{scatt2,nuMSM,HeAAS,KS,AMP,NSZ}, and their mixing with the SM flavors is in general less suppressed than in the original seesaw scenarios.

CP violation is also possible in the neutrino sector \cite{oscCP}. In the heavy neutrino sector, CP violation has been investigated in scattering processes
\cite{Pilaftsis} (resonant CP violation), coming from interference of the tree-level and one-loop neutrino propagator effects. Resonant CP violation was also investigated in the leptonic \cite{CKZ,symm} and semileptonic rare meson decays \cite{CKZ2,DCK,symm}, using a simplified (effectively tree-level) approach.
These effects are appreciable in scenarios where we have at least two heavy almost degenerate neutrinos (as is also required for oscillation, cf.~\cite{Boya,CKZosc}), with masses $\sim 1$ GeV. Such scenarios appear to be compatible with the neutrino minimal standard model ($\nu$MSM) \cite{nuMSM,Shapo} and some low-scale seesaw models \cite{lsseesaw}.

In this work we assume that there exists at least one sterile neutrino $N$, leading to a mass eigenstate with mass $M_N$ up to about $6$ GeV. In our previous work \cite{CDKK}
we investigated semihadronic rare decays of charged pseudoscalar mesons in such scenarios, with intermediate on-shell neutrino $N$ and final pseudoscalar meson, such as $B^+ \to e^+ N \to e^+ e^+ D^-$. CP violation in such type of decays (with two intermediate almost degenerated neutrinos $N_1$, $N_2$) were investigated in
Refs.~\cite{CKZ2,DCK}. Such processes are LNV and the neutrinos have to be Majorana. On the other hand, in Ref.~\cite{CDK} we investigated the
rare leptonic decays of charged pions, $\pi^+ \to e^+ N \to e^+ e^+ \mu^- \nu$, mediated by an on-shell $N$ (CP violation in Ref.~\cite{CKZ}, mediated by two
$N_j$'s). Such decays have LNV and LNC channels, and the intermediate
neutrino(s) $N$ can be either Majorana or Dirac. In Ref.~\cite{CDK} (cf.~also a review \cite{symm}) we showed that it is possible to distinguish
between the Majorana and Dirac nature of $N$ in such decays by measuring the differential decay width $d \Gamma/d E_{\mu}$ with respect of the muon energy $E_{\mu}$ (the latter being in the $N$-rest frame). Since the heavy-light neutrino
mixing coefficients are expected to be very suppressed, such differential decay
width is difficult to measure with sufficient statistics. Nonetheless, as argued
in Ref.~\cite{CDK}, this problem can be overcome if the decaying pions
are produced copiously (project X, $\sim 10^{29}$ pions per year,
\cite{ProjX}).
On the other hand, since pions are light, the produced on-shell neutrinos
$N$ are also light and have thus a long lifetime, thus most of them
escape through the detector before decaying. This effect accounts for
a significant suppression of the decay rate in the mentioned
rare decays  $\pi^+ \to e^+ N \to e^+ e^+ \mu^- \nu$.
Recently this issue of differentiating between Majorana and Dirac sterile neutrinos at the LHC has been revisited \cite{KimLHC} for the mass of sterile neutrino $m_N < m_W$.

On the other hand, if we consider the analogous rare leptonic decays of
heavier mesons, such as $B$ or $B_c$, the masses of the intermediate
on-shell neutrinos can be significantly larger, and the decay widths
of the rare decays are larger. Nonetheless, such mesons are not produced
copiously, except at LHC-b and in an upgrade of the Belle experiment,
Belle II \cite{BelleIIwww}. It is expected that Belle II
can produce $\sim 10^{10}$ $B$-pairs per year. Therefore, rare decays of $B$
mesons may give us a hint of the existence of heavier sterile neutrinos with
masses of up to $5$ GeV. Moreover, differential decay widths of such
decays may offer us a possibility of discerning the nature (Majorana or Dirac)
of such neutrinos. However, the rare leptonic decays of the type
$B \to \ell_1 N \to \ell_1 \ell_2 \ell_3 \nu$, where $\ell_j$ are
light charged leptons and $\nu$ is a light neutrino,
are strongly CKM-suppressed in comparison with the analogous decays of
$B_c$ mesons (because $|V_{u b}| \sim |V_{c b}|/10$). Belle II will produce
$B$ mesons but not $B_c$ mesons. Therefore, rare leptonic decays cannot
play an important role at Belle II, but rather at LHC-b where $B_c$ mesons are
produced copiously. The question that arises naturally, especially for Belle-II measurements, is whether we can have rare $B$-meson
decays which are not CKM-suppressed. The answer to this question is affirmative: Namely, semihadronic rare $B$-meson decays
$B \to D^{(*)} \ell_1 N$ and consecutively  $N \to \ell_2 \ell_3 \nu$ (or $N \to \ell_2 \pi$)
are not CKM-suppressed because they are proportional to $|V_{c b}|^2$.

In this work we calculate the branching ratios for the mentioned
rare decays $B_{(c)} \to (D^{(*)}) \ell_1 N$
where the produced on-shell neutrino $N$ may further decay
leptonically $N \to \ell_2 \ell_3 \nu$ or semileptonically
$N \to \ell \pi$. Some of
the formulas (those not involving $D^{(*)}$) have been known from our previous
works, while those with  $D^{(*)}$-meson are new and, to our knowledge, have
not been known in the literature (those with massless $N$ are known).
We take into account the effect of the decay probability of the intermediate
on-shell sterile neutrino $N$ within the detector, this probability can sometimes be significantly smaller than 1.

In Sec.~\ref{BtoN} we present formulas for the decay widths of
rare decays of $B_{(c)} \to (D^{(*)}) \ell_1 N$, those not involving the $D^{(*)}$ meson in Sec.~\ref{decnoD}, and in
Sec.~\ref{decD} those involving $D^{(*)}$ meson.
In Appendices \ref{app1}-\ref{appDst} we show more
detailed formulas relevant for these decays.
In Sec.~\ref{sec:Ndec}, formulas are presented for the subsequent decays of
the produced sterile neutrino, $N \to \ell_2 \ell_3 \nu$ and $N \to \ell_2 \pi$, for lepton number violating and lepton number conserving modes.\footnote{
For flavors of charged leptons, $\ell_1, \ell_2, \ell_3$, we can choose, e.g., $e^\pm e^\pm \mu^\mp$
having no opposite-sign same-flavor lepton pairs in the final state to avoid the serious SM radiative background $\gamma^*/Z^* \to e^+ e^-$ \cite{CDK}.}
In Sec.~\ref{sec:Br} we exhibit the branching ratios for these decays
as a function of the mass of $N$ neutrino, and the differential branching
ratios with respect to charged lepton energy when $N$ decays leptonically.
In Appendix \ref{appdiff} the relevant
formulas for the differential branching ratios are given.
In Sec.~\ref{sec:effBr} we account for the decay probability of neutrinos $N$ within the detector, and present the resulting
effective branching ratios for the mentioned rare
decays, as a function of the mass of $N$,
assuming that the probability for the neutrino decay within the
detector is significantly smaller than one.
In Sec.~\ref{sec:disc}, based on the results of the previous Sections,
we estimate values of the (effective) branching ratios of various
mentioned rare decays, for various
ranges of the mass $M_N$ and of the heavy-light mixing coefficients
$|U_{\ell N}|^2$ of the sterile neutrinos $N$. We
discuss the values of the mixing coefficients
$|U_{\ell N}|^2$ necessary for the detection of the mentioned rare decays
at Belle II and LHC-b,
and, implicitly, the upper bounds for these coefficients for the case that such decays are not detected. In Summary we
briefly recap the obtained results.


\section{Rare Decays of $B_{(c)}$ to On-shell Sterile Neutrino}
\label{BtoN}

\subsection{Decays $B_{(c)} \to \ell_1 N$}
\label{decnoD}

The decay width for the process $B_{(c)} \to \ell_1 N$, with the
subsequent decay of the on-shell $N$ neutrino to
$\ell_2 \ell_3 \nu$ or to $\ell_2 \pi$ (generically: to $XY$),
can be written in the
factorized form
\be
\Gamma(B_{(c)} \to \ell_1 N \to \ell_1 X Y)
=  \Gamma(B_{(c)} \to \ell_1 N) \frac{\Gamma(N \to XY)}{\Gamma_N} \ ,
\label{fact}
\ee
where $\Gamma_N$ is the total decay width of the sterile neutrino $N$,
and $\ell_1$ is a charged lepton ($\ell_1 = e, \mu, \tau$).
The first factor on the right-hand side is well known
\be
\Gamma(B_{(c)}^{\pm} \to \ell_1^{\pm} N) =  |U_{\ell_1 N}|^2
{\overline \Gamma(B_{(c)}^{\pm} \to \ell_1^{\pm} N)} \  ,
\label{GBlN}
\ee
where the canonical width ${\overline \Gamma}$ (i.e., without the
heavy-light mixing factor $|U_{\ell_1 N}|^2$) is
\be
{\overline \Gamma(B_{(c)}^{\pm} \to \ell_1^{\pm} N)} =
\frac{G_F^2 f_{B_{(c)}}^2}{8 \pi} |V_{Q_u Q_d}|^2 M_{B_{(c)}}^3 \lambda^{1/2}(1,y_N,y_1)
\left[ (1 - y_N) y_N + y_1 (1 + 2 y_N - y_1) \right] \ ,
\label{bGBlN}
\ee
where $G_F = 1.1664 \times 10^{-5} \ {\rm GeV}^{-2}$ is the Fermi coupling constant, $f_{B_{(c)}}$ is the decay constant of the meson $B^{\pm}$ (or $B_c^{\pm}$),
$V_{Q_u Q_d}$ is the corresponding CKM matrix element ($V_{u b}$ for $B$,
$V_{cb}$ for $B_c$), and we use the notations
\be
y_N = \frac{M_N^2}{M_{B_{(c)}}^2} \ , \qquad y_1 = \frac{M_{\ell_1}^2}{M_{B_{(c)}}^2} \ ,
\label{yNyell}
\ee
and the function $\lambda^{1/2}$ is given in Eq.~(\ref{sqlam}) in Appendix
\ref{app1}. The coefficient $U_{\ell N}$ is the heavy-light mixing coefficient
of the (extended) PMNS matrix, i.e., the light flavor neutrino state
$\nu_{\ell}$ (with flavor $\ell = e, \mu, \tau$) is
\be
\nu_{\ell} = \sum_{k=1}^3 U_{\ell \nu_k} \nu_k + U_{\ell N} N \ .
\label{mixN}
\ee
For simplicity, we assume that there is only one sterile (heavy) neutrino $N$, in addition to the three light neutrinos $\nu_k$.

\begin{figure}[htb] 
\begin{minipage}[b]{.49\linewidth}
\centering\includegraphics[width=85mm]{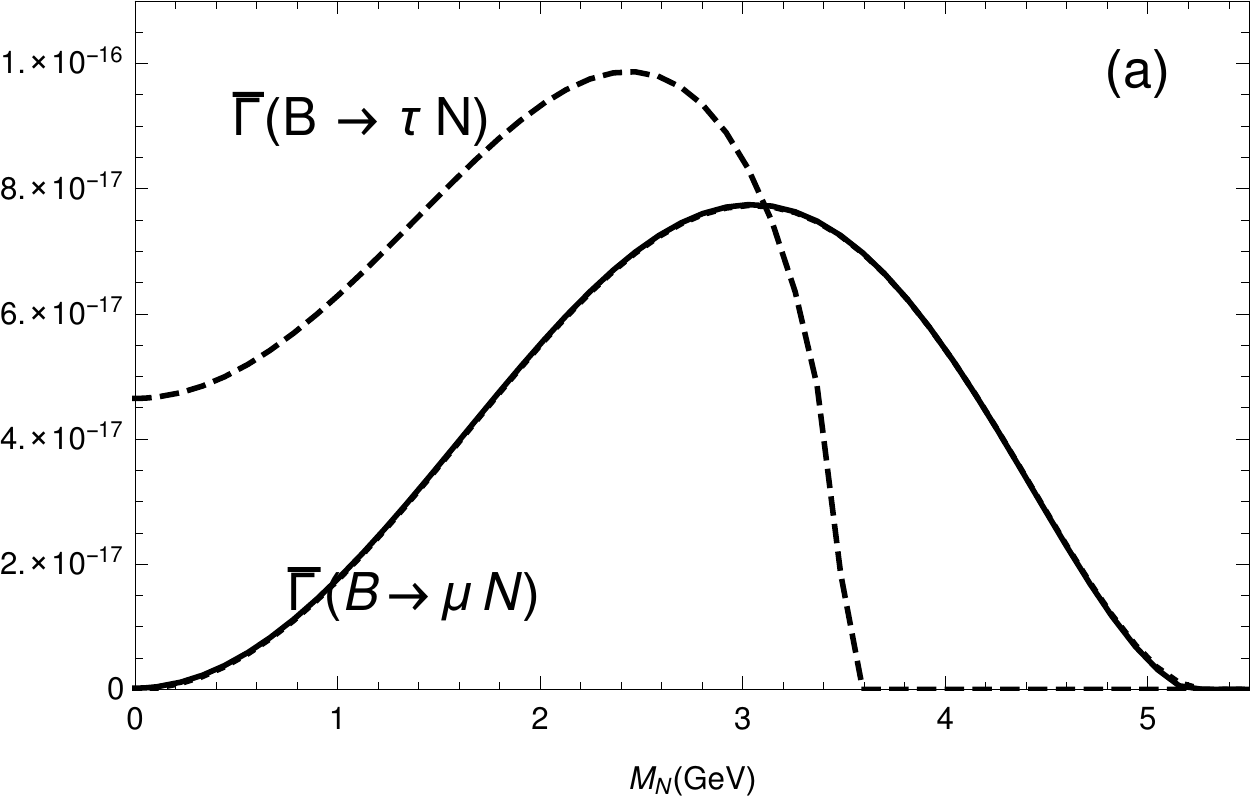}
\end{minipage}
\begin{minipage}[b]{.49\linewidth}
\centering\includegraphics[width=85mm]{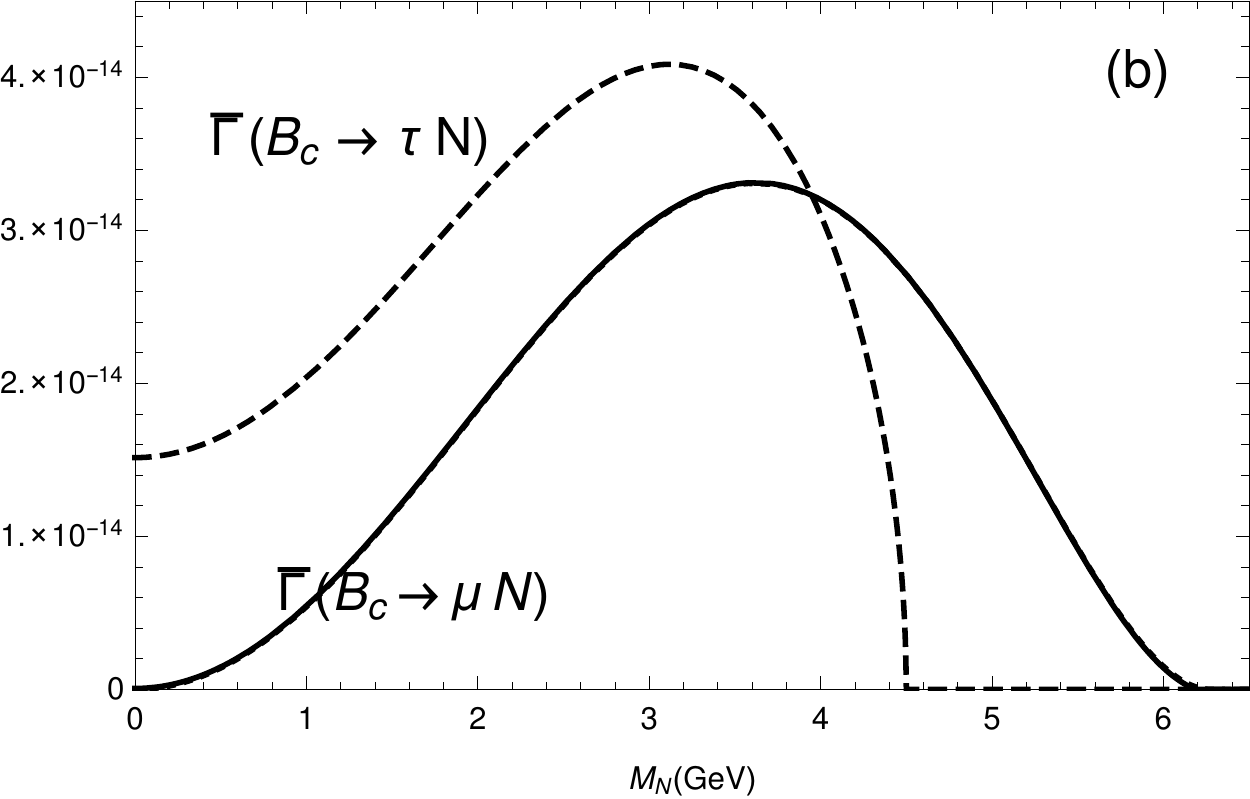}
\end{minipage}
\caption{The canonical decay width Eq.~(\ref{bGBlN}), in units of GeV, as a function of the mass of heavy neutrino $N$, for (a) $B^{\pm} \to \ell^{\pm} N$, (b) $B_c^{\pm} \to \ell^{\pm} N$, where $\ell = \mu$ (solid),  $\ell=\tau$ (dashed), and $\ell=e$ (dotted). The dotted line is practically indistinguishable from the solid one.}
\label{FigbGBmuN}
 \end{figure}

In Figs.~\ref{FigbGBmuN}(a), (b) we present the canonical decay width (\ref{bGBlN}), for the decays $B^{\pm} \to \ell^{\pm} N$ and $B_c^{\pm} \to \ell^{\pm} N$ ($\ell=\mu, \tau$) as a function of the mass of $N$. We used the values $f_B=0.196$ GeV and $f_{B_c}=0.322$ GeV (the central values of Ref.~\cite{CKWN}), $|V_{ub}|=4.13 \times 10^{-3}$ \cite{PDG2014} and $|V_{cb}|=4.012 \times 10^{-2}$ \cite{Belle1}.
We see clearly that the decays of $B$ are strongly suppressed, due to the small CKM matrix element.

\subsection{Decays $B \to D^{(*)} \ell_1 N$}
\label{decD}

As already mentioned in the Introduction, the type of rare decays
of the $B^{\pm}$ meson described in the previous Section, i.e.,
the meson produced at the dedicated Belle(II)
experiment, will have strong CKM-suppression ($V_{ub} \approx 0.004$),
as seen from Eq.~(\ref{bGBlN}).
The corresponding rare decays of $B_c$ have much weaker CKM-suppression
($V_{cb} \approx 0.04$), but, unfortunately, they are not produced
at Belle(II). They are copiously produced at LHC, though.

In order to use the potential of the new Belle II experiment for
the detection of rare $B$-meson decays, we need to consider a somewhat
more complicated variant of such decays, a variant in which the
CKM-suppression ($\propto |V_{ub}|^2$) does not take place. This
suppression is avoided if, at the first stage of the decay,
$B$ meson decays into a charmed $D^{(*)}$-meson and an off-shell $W$ which
decays into $\ell_1$ and $N$, i.e., we consider the decays
$B \to D^{(*)} \ell_1 N \to  D^{(*)} \ell_1 XY$, where at the
second stage the intermediate on-shell neutrino $N$ decays
into $X Y$ as in the previous Section, i.e., either leptonically as
$N \to \ell_2 \ell_3 \nu$, or semileptonically as $N \to \ell \pi$.

This means that, according to the expression (\ref{fact}), we have to
calculate the first factor $\Gamma(B \to  D^{(*)} \ell_1 N)$ which
appears in the decay widths
\bes
\label{BDlNXY}
\bea
\Gamma(B \to D^{(*)} \ell_1 N \to \ell_1 \ell_2 \ell_3 \nu)
&=&  \Gamma(B \to D^{(*)} \ell_1 N) \frac{\Gamma(N \to \ell_2 \ell_3 \nu)}{\Gamma_N} \ ,
\label{BDlNllnu}
\\
\Gamma(B \to D^{(*)} \ell_1 N \to \ell_1 \ell_2 \pi)
&=&  \Gamma(B \to D^{(*)} \ell_1 N) \frac{\Gamma(N \to \ell_2 \pi)}{\Gamma_N} \ ,
\label{BDlNlPi}
\eea
\ees
and where the second factor $\Gamma(N \to XY)$ has already been given in Eqs.~(\ref{GNllnu})-(\ref{bGNllnu}) when $XY = \ell_2 \ell_3 \nu$, and in Eqs.~(\ref{GNlPi})-(\ref{bGNlPi}) when $XY = \ell_2 \pi$.

The expressions necessary for evaluation of the
first factor $\Gamma(B \to D^{(*)} \ell_1 N)$ in Eqs.~(\ref{BDlNXY})
are obtained in Appendix \ref{appD} for the case of
$B \to D \ell_1 N$, and in Appendix \ref{appDst} for $B \to D^{*} \ell_1 N$.
The latter decay is theoretically more complicated because $D^{*}$ is
a vector while $D$ is a pseudoscalar meson. We note that in the
literature, these decays are known for the case of zero masses of
the neutrino $N$ and of the charged lepton $\ell_1$. On the other hand, here
we obtained formulas for the more general case of massive $N$ and $\ell_1$.

\subsubsection{$B \to D \ell_1 N$}
\label{subs:BDell1N}

The process $B(p_B) \to D(p_D) \ell_1(p_1) N(p_N)$ is depicted schematically in Fig.~\ref{FigBDW} in Appendix \ref{appD}. The expression for the corresponding differential decay width $(d/d q^2)\Gamma(B \to D \ell_1 N)$ is given in Eqs.~(\ref{dGBDNl})-(\ref{dbGBDNl}), in terms of the form factors $F_1(q^2)$ and $F_0(q^2)$, Eqs.~(\ref{FFBD}), where $q^2$ is square of the momentum of the virtual $W$ (i.e., of the $\ell_1$-$N$ pair, cf.~Fig.~\ref{FigBDW}). The resulting decay width $\Gamma(B \to D \ell_1 N)$ is obtained upon (numerical) integration of the differential decay width over the kinematically allowed values of $q^2$
\bes
\label{GBDNl}
\bea
\lefteqn{
\Gamma(B \to D \ell_1 N) = |U_{\ell_1 N}|^2 \; {\overline \Gamma}(B \to D \ell_1 N) \ ,
}
\label{GBDNla}
\\
{\overline \Gamma}(B \to D \ell_1 N) & = &
\frac{1}{384 \pi^3} G_F^2 |V_{c b}|^2 \frac{1}{M_B}
\int_{(M_N+M_1)^2}^{(M_B-M_D)^2}  d q^2 \;
\frac{1}{(q^2)^2}
\lambda^{1/2} \left(1, \frac{q^2}{M_B^2}, \frac{M_D^2}{M_B^2} \right)
\lambda^{1/2} \left(1, \frac{M_1^2}{q^2}, \frac{M_N^2}{q^2} \right)
\nonumber\\
&& \times
    {\bigg \{} F_1(Q^2)^2 \left[ 2 (q^2)^2 - q^2 M_N^2  + M_1^2 (2 M_N^2- q^2) - M_N^4 - M_1^4 \right] \left[ (q^2 - M_D^2)^2 - 2 M_B^2 (q^2 + M_D^2) + M_B^4 \right]
    \nonumber\\
    && + F_0(q^2)^2 3 (M_B^2 - M_D^2)^2 \left[ q^2 M_N^2 + M_1^2 (2 M_N^2 + q^2) - M_N^4 - M_1^4 \right]
        {\bigg \}} \ .
\label{GBDNlb}
\eea
\ees
The form factor $F_1(q^2)$ is well known \cite{CLN}. It can be expressed in terms of the variable $w$
\bes
\label{wz}
\bea
w & = & \frac{(M_B^2 + M_D^2 - q^2)}{2 M_B M_D} \ ,
\label{w}
\\
z(w) & = & \frac{\sqrt{w+1} - \sqrt{2}}{\sqrt{w+1} + \sqrt{2}} \ ,
\label{z}
\eea
\ees
in the following approximate form \cite{CLN}:
\be
F_1(q^2) = F_1(w=1) \left( 1 - 8 \rho^2 z(w) + (51 \rho^2 - 10) z(w)^2 - (252 \rho^2 - 84) z(w)^3 \right) \ ,
\label{CLNF1}
\ee
where the free parameters $\rho^2$ and $F_1(w=1)$ have been recently determined with high precision by the Belle Collaboration, Ref.~\cite{Belle1}
\bes
\label{rho2F1max}
\bea
\rho^2 &= & 1.09 \pm 0.05 \ ,
\label{rho2}
\\
|V_{cb}| F_1(w=1) &=& (48.14 \pm 1.56) \times 10^{-3} \ .
\label{F1max}
\eea
\ees
The value (\ref{F1max}) was deduced from their value of $\eta_{EW} {\cal G}(1) |V_{cb}| = \eta_{EW} F_1(w=1) \sqrt{4 r}/(1+r) = (42.29 \pm 1.37) \times 10^{-3}$, where $r=M_D/M_B$ and $\eta_{EW} =1.0066 \approx 1$ \cite{Sirlin}. In our numerical evaluations, we will use the central values $\rho^2 =  1.09$ and $|V_{cb}| F_1(w=1)  = 48.14 \times 10^{-3}$.

When the masses of $\ell_1$ and $N$ are simultaneously zero, only $F_1(q^2)$ form factor contributes, and consequently the form of $F_0(q^2)$ is not well known in the literature. In our massive case, however, $F_0(q^2)$ contributes significantly as well. Nonetheless, we can get a reasonably good approximation for $F_0(q^2)$ by using the (truncated) expansion for $F_0$ in powers of $(w-1)$, Ref.~\cite{CaNeu}
\bes
\label{F0}
\bea
F_0(q^2) & = & \frac{(M_B+M_D)}{2 \sqrt{M_B M_D}}
\left[ 1 - \frac{q^2}{(M_B+M_D)^2} \right] f_0(w(q^2)) \ ,
\label{F0a}
\\
f_0(w) & \approx & f_0(w=1) \left[ 1 + {\rho}_0^2 (w - 1) + (0.72 \rho_0^2 - 0.09) (w - 1)^2 \right] \ ,
\label{F0b}
\eea
\ees
where the value $f_0(w=1) \approx 1.02$ \cite{NeuPRps,CaNeu} is obtained from
the heavy quark limit.
The variable $w$ was defined in Eq.~(\ref{w}). The endpoint value $(w=1)$
corresponds to the maximal value of $q^2$, $q^2=(M_B-M_D)^2$. The remaining free parameter $\rho_0$ in the expression (\ref{F0b}) can then be fixed by the condition of absence of spurious poles at $q^2=0$: $F_0(0)=F_1(0)$ ($\approx 0.690$). This results in the value $\rho_0^2 \approx 1.102$ and $(0.72 \rho_0^2 - 0.09) \approx 0.704$.

\begin{figure}[htb]
\centering\includegraphics[width=90mm]{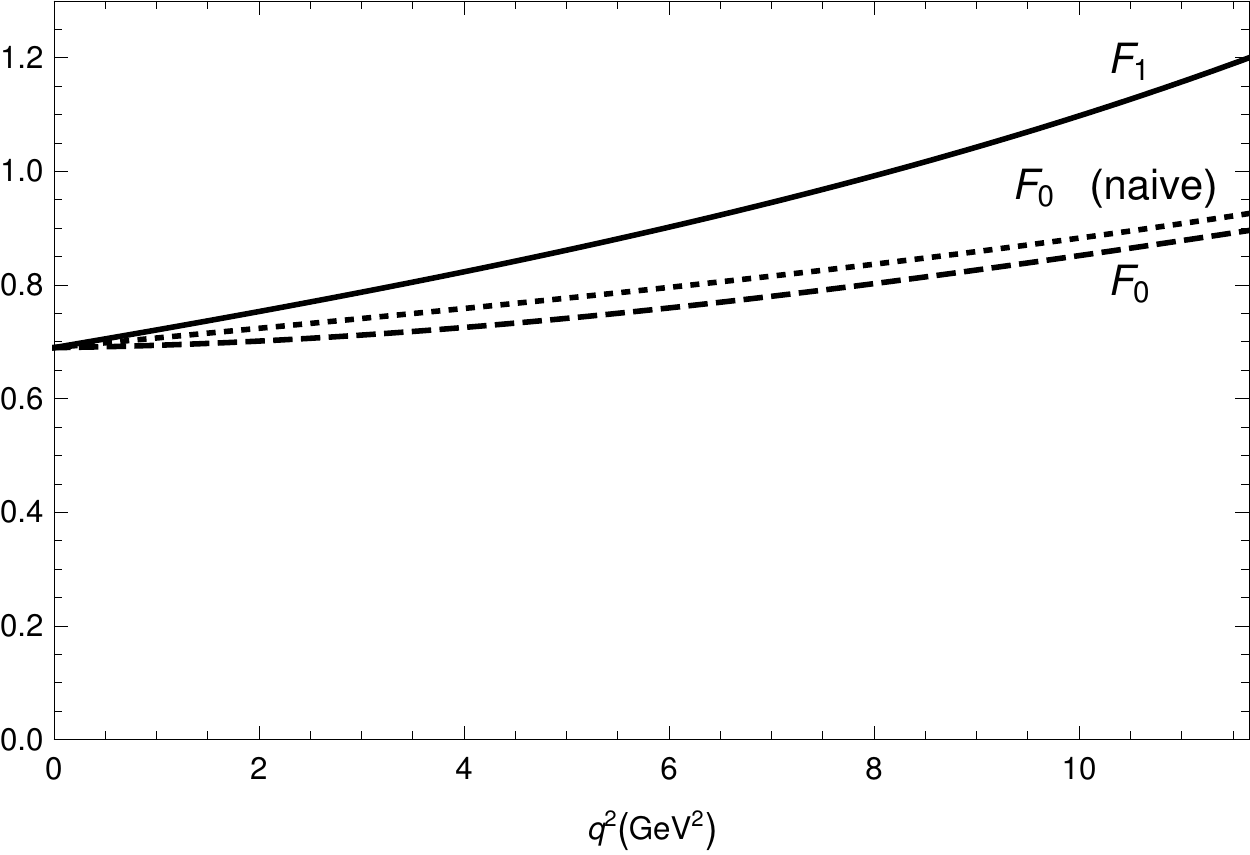}
\caption{The form factors $F_1(q^2)$ and $F_0(q^2)$ in the range $0 \leq q^2 \leq (M_B-M_D)^2$.}
\label{FigF1F0}
\end{figure}

We present the resulting Form factors $F_1(q^2)$ and $F_0(q^2)$ for
$0 \leq q^2 \leq (M_B-M_D)^2$ in Fig.~\ref{FigF1F0}.
For comparison, we also included the more naive form factor $F_0(q^2)$ obtained from the heavy-quark limit from $F_1(q^2)$
\be
F_0(q^2)^{\rm (naive)} = \left[ 1 - \frac{q^2}{(M_B+M_D)^2} \right] F_1(q^2) \ ,
\label{F0naive}
\ee
where for $F_1(q^2)$ the (optimal) form described above is used.

\subsubsection{$B \to D^* \ell_1 N$}
\label{subs:BDstell1N}

As mentioned, the decay width for the process
$B(p_B) \to D^*(p_D) \ell_1(p_1) N(p_N)$ has a more complicated expression,
because, due to the vector character of $D^*$, more form factors appear,
cf.~Eqs.~(\ref{FFBDst})-(\ref{A3}) in Appendix \ref{appDst} - altogether,
four independent form factors feature now: $V(q^2)$ and $A_j(q^2)$ ($j=1,2,0$),
while $A_3$ is a combination of $A_1$ and $A_2$.
Three of the four independent form factors ($V$, $A_1$ and $A_2$) are well known, and have been determined recently to high precision by the Belle Collaboration \cite{Belle2} using the parametrization of Ref.~\cite{CLN}
\bes
\label{A1VA2}
\bea
A_1(q^2) & = & \frac{1}{2} R_* (w+1) F_*(1) \left[ 1 - 8 \rho_*^2 z(w) + (53 \rho_*^2 - 15) z(w)^2 - (231 \rho_*^2 - 91) z(w)^3 \right] \ ,
\label{A1}
\\
V(q^2) & = & A_1(q^2)  \frac{2}{R_*^2 (w+1)} \left[ R_1(1) - 0.12 (w-1) + 0.05 (w-1)^2 \right] \ ,
\label{V}
\\
A_2(q^2) & = & A_1(q^2) \frac{2}{R_*^2 (w+1)} \left[ R_2(1) + 0.11 (w-1) - 0.06 (w-1)^2 \right] \ .
\label{A2}
\eea
\ees
Here, $R_* = 2 \sqrt{ M_B M_{\Dst}}/(M_B+M_{\Dst})$, the variables $w$ and $z(w)$ are given by Eqs.~(\ref{wz}), and
the values of the free parameters determined in Ref.~\cite{Belle2} are
\bes
\label{paramsDst}
\bea
\rho_*^2 & = & 1.214(\pm 0.035) \ , \qquad 10^3 F_*(1) |V_{cb}| = 34.6(\pm 1.0) \ ,
\label{rhostFst}
\\
R_1(1) & = &1.401(\pm 0.038) \ , \qquad R_2(1) = 0.864(\pm 0.025) \ .
\label{R1R2}
\eea
\ees
We will use the central values of these parameters.

When the masses of final fermions ($\ell_1$ and $N$) are simultaneously zero, only the above three of the four independent form factors contribute ($V$, $A_1$, $A_2$). However, in our (massive) case, the form factor $A_0$ also contributes. It is not well known. In order to get a reasonable approximation for the form factor $A_0(q^2)$, we can employ the heavy-quark-limit relations
\be
A_1(q^2)  \approx A_2(q^2) \left[ 1 - \frac{q^2}{(M_B+M_{\Dst})^2} \right]
\label{A1A2hql}
\ee
in the general expression (\ref{A3}) for $A_3(q^2)$, resulting in the
approximate relation between $A_2$ and $A_3$
\be
A_2(q^2) \approx A_3(q^2)/\left[1 - \frac{q^2}{2 M_{\Dst} (M_B+M_{\Dst})} \right] \ ,
\label{A2appr}
\ee
Using, in addition, the heavy-quark-limit relation $A_0(q^2) \approx A_2(q^2)$,
we obtain the following approximation for the (otherwise unknown) form factor $A_0(q^2)$:
\be
A_0(q^2) \approx A_3(q^2)/\left[1 - \frac{q^2}{2 M_{\Dst} (M_B+M_{\Dst})} \right]
= \frac{(M_B+M_{\Dst})^2}{\left( 2 M_{\Dst} (M_B+M_{\Dst}) - q^2 \right)}
\left( 1 - \frac{(M_B-M_{\Dst})}{(M_B+M_{\Dst})} \frac{A_2(q^2)}{A_1(q^2)} \right) A_1(q^2)
\ ,
\label{A0appr}
\ee
This relation also fulfills the obligatory relation $A_0(0)=A_3(0)$ which reflects the condition of the absence of pole at $q^2=0$ in the hadronic matrix element (\ref{FFBDst}). Therefore, $A_0(q^2)$ can now be evaluated, too.

\begin{figure}[htb]
\centering\includegraphics[width=90mm]{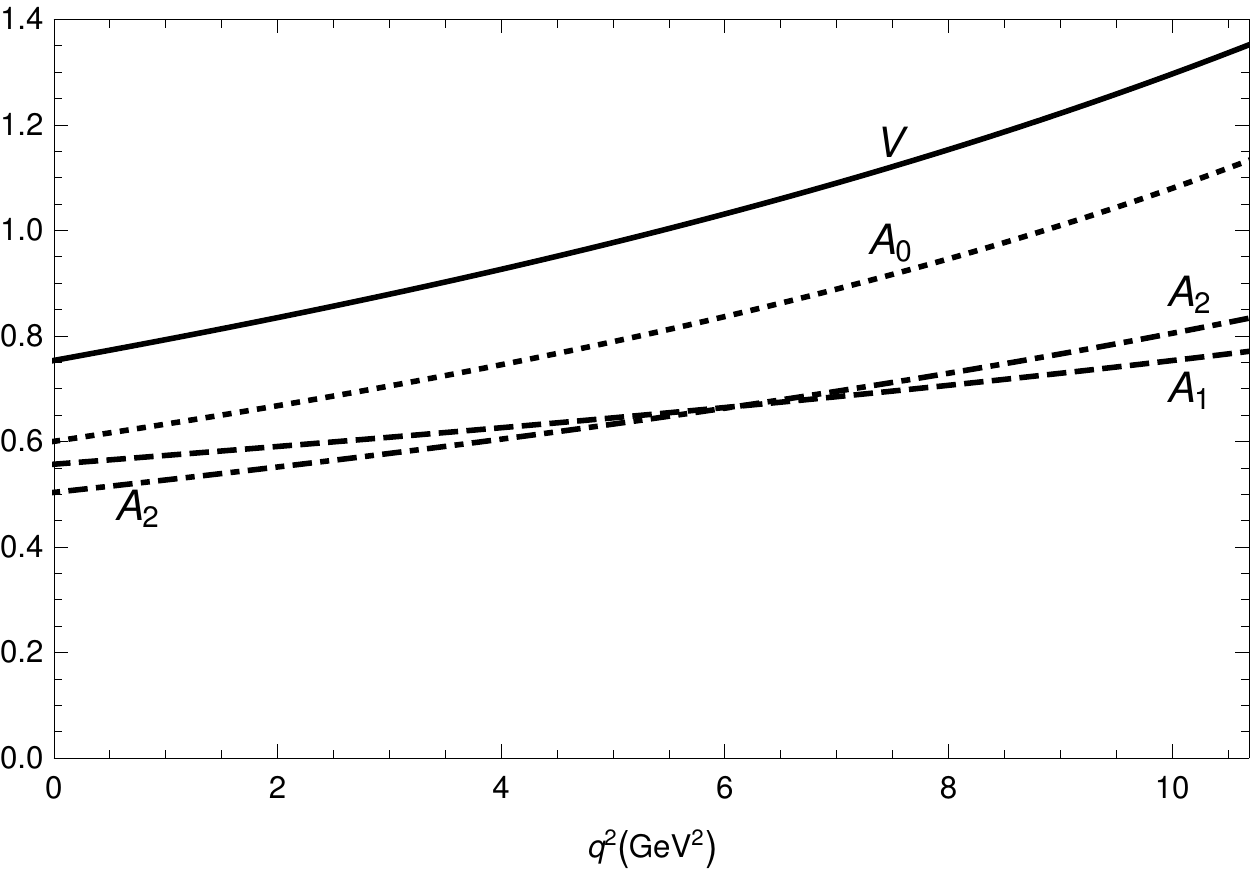}
\caption{The form factors $V(q^2)$ and $A_j(q^2)$ ($j=1,2,0$)
in the range $0 \leq q^2 \leq (M_B-M_{\Dst})^2$.}
\label{FigVAs}
\end{figure}

The decay width for the considered process,
 $\Gamma(B \to D^* \ell_1 N)$, is obtained numerically by integration of the differential decay width over the kinematically allowed values of $q^2$, where we use the obtained differential decay width (\ref{dGdq2b}), and the expression (\ref{A0appr}) of $A_0$ in terms of $A_1$. We then get
\be
\label{GBDstNldef}
\Gamma(B \to D^* \ell_1 N) = |U_{\ell_1 N}|^2 \; {\overline \Gamma}(B \to D^* \ell_1 N) \ ,
\ee
where the canonical decay width (i.e., without the heavy-light neutrino mixing) is
\bea
\lefteqn{
{\overline \Gamma}(B \to D^* \ell_1 N) =
\frac{1}{64 \pi^3} \frac{G_F^2 |V_{cb}|^2}{M_B^2}
\int_{(M_N+M_1)^2}^{(M_B-M_{\Dst})^2}  d q^2 \;
\blam^{1/2} |{\vec q}| q^2 {\Bigg \{}
 \left( 1 - \frac{(M_N^2+M_1^2)}{q^2} - \frac{1}{3} \blam \right)
 {\bigg [} 2 (M_B+M_{\rm D})^2 A_1(q^2)^2
 }
 \nonumber\\
 &&
 + \frac{8 M_B^2 |{\vec q}|^2}{(M_B+M_{\Dst})^2} V(q^2)^2 +  \frac{M_B^4}{4 M_{\Dst}^2 q^2} \left( (M_B+M_{\Dst})
   \left( 1 - \frac{(q^2+M_{\Dst}^2)}{M_B^2} \right) A_1(q^2) -  \frac{4 |{\vec q}|^2}{(M_B+M_{\Dst})} A_2(q^2) \right)^2 {\bigg ]}
 \nonumber\\
 &&
 + \left[ - \left(\frac{M_N^2-M_1^2}{q^2} \right)^2 + \frac{(M_N^2+M_1^2)}{q^2} \right] \frac{M_B^2 |{\vec q}|^2}{q^2}
 \left[ \frac{2 (M_B+M_{\Dst})^2}{\left(2 M_{\Dst} (M_B+M_{\Dst}) - q^2 \right)} \right]^2 \left[ 1 - \frac{(M_B - M_{\Dst}) A_2(q^2)}{(M_B + M_{\Dst}) A_1(q^2)} \right]^2 A_1(q^2)^2 {\Bigg \}},
 \nonumber\\
 \label{bGBDstNl}
 \eea
where $\blam$ and $|{\vec q}|$, as a function of $q^2$, are given in Appendix \ref{appDst} in Eqs.~(\ref{blam}) and (\ref{magq}), respectively.
In Fig.~\ref{FigVAs} we present the form factors $V$ and $A_j$ ($j=1,2,0$) as a function of $q^2$ between $0 \leq q^2 \leq (M_B-M_{\Dst})^2$, where we took the masses of the neutral $B$ and charged $D^*$ ($M_B=5.280$ GeV and $M_{\Dst}=2.010$ GeV; $M_B-M_{\Dst}=3.269$ GeV).

\begin{figure}[htb] 
\begin{minipage}[b]{.49\linewidth}
\centering\includegraphics[width=85mm]{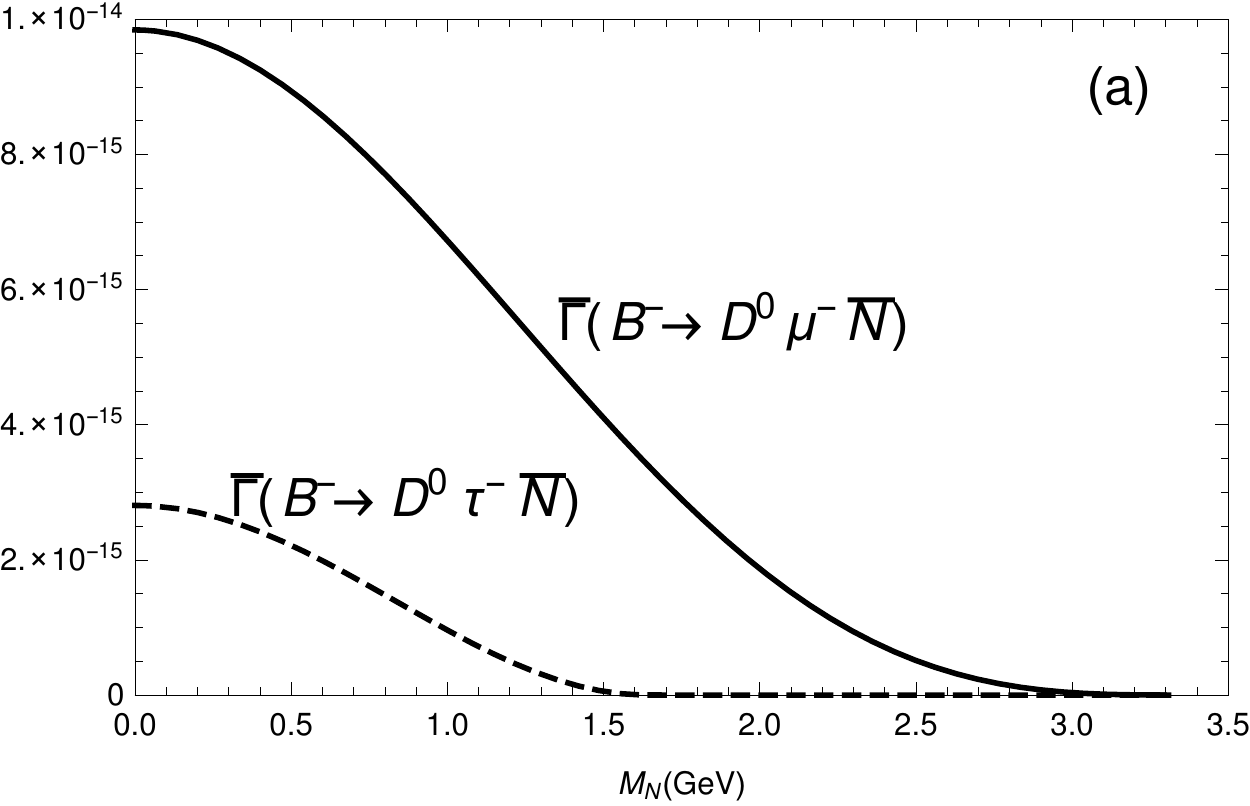}
\end{minipage}
\begin{minipage}[b]{.49\linewidth}
\centering\includegraphics[width=85mm]{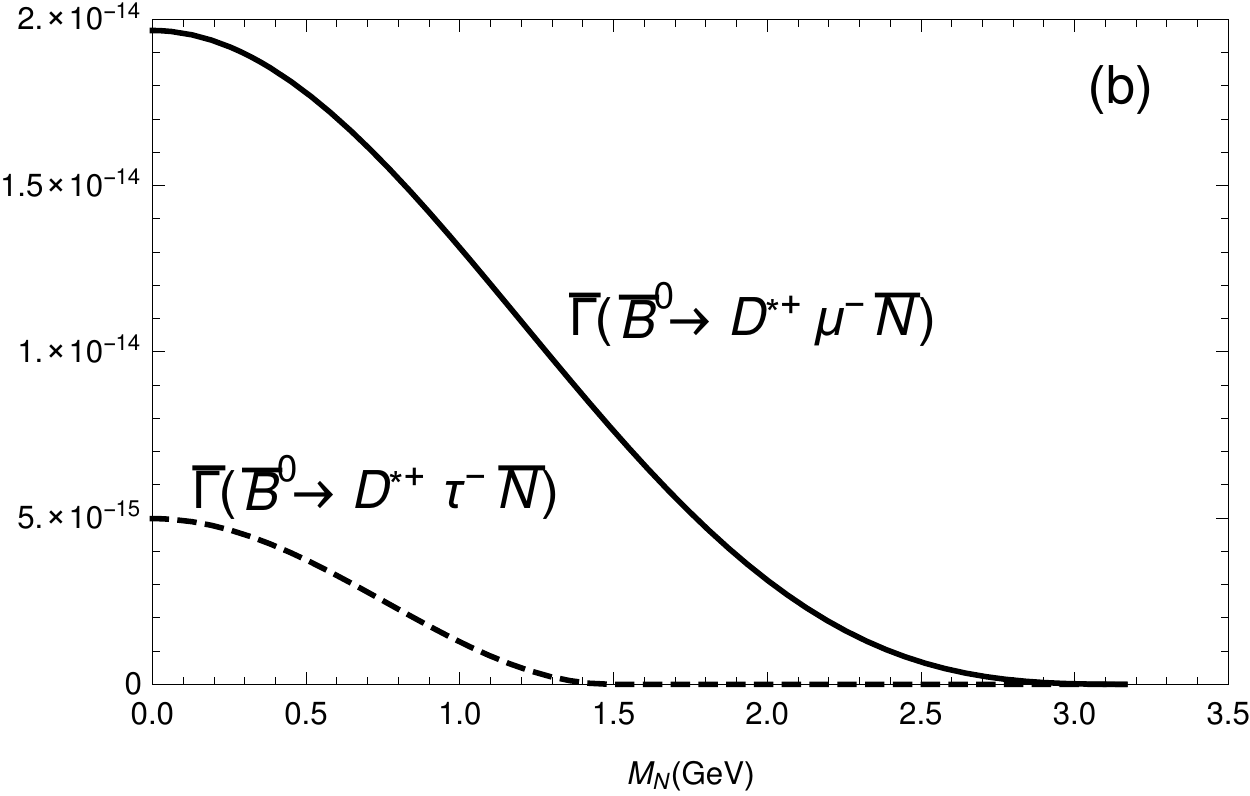}
\end{minipage}
\caption*{The canonical decay widths of $B$ meson into (a) $D \ell N$ [Eq.~(\ref{GBDNl})], (b) $D^* \ell N$ [Eq.~(\ref{bGBDstNl})], as a function of the mass of $N$ neutrino. The specific cases of $\ell=\mu, \tau$ are presented. The decay widths are in units of GeV.}
\label{FigbGBDlN}
 \end{figure}

We present in Figs.~\ref{FigbGBDlN}(a),(b) the main results of this Section, i.e., the canonical decay widths Eqs.~(\ref{GBDNl}) and (\ref{bGBDstNl}). The used values of the masses of $D^{(*)}$ are from Ref.~\cite{PDG2014}: $1.8648$ GeV ($D^0$); $2.0103$ GeV ($D^{* \pm}$).

\section{Subsequent Decays of $N$ to Lepton Number Violating and Conserving Modes}
\label{sec:Ndec}

\subsection{Leptonic decays  $N \to \ell_1^\pm \ell_2^\mp \nu$}
\label{lepdecN}

The sterile on-shell neutrino $N$, produced by $B_{(c)} \to \ell_1^{\pm} N$, subsequently decays into various channels. First we are interested in the leptonic decays of the lepton-number-conserving (LNC) type $N \to \ell_2^{\mp} \ell_3^{\pm} \nu_{\ell_3}$ and of the lepton-number-violating
(LNV) type $N \to \ell_3^{\pm} \ell_2^{\mp} \nu_{\ell_2}$, cf.~Ref.~\cite{symm} (cf.~also Refs.~\cite{CDKK,CKZ})
\bes
\bea
\Gamma^{\rm (LNC)}(N \to \ell_2^{\mp} \ell_3^{\pm} \nu_{\ell_3}) & = & |U_{\ell_2 N}|^2 {\overline \Gamma}(N \to \ell_2 \ell_3 \nu) \ ,
\label{GNllnu.LC}
\\
 \Gamma^{\rm (LNV)}(N \to \ell_3^{\pm} \ell_2^{\mp} \nu_{\ell_2}) & = & |U_{\ell_3 N}|^2 {\overline \Gamma}(N \to \ell_2 \ell_3 \nu) \ ,
\label{GNllnu.LV}
\eea
\label{GNllnu}
\ees
where the charged leptons ($\ell_2, \ell_3$) are in general $e, \mu, \tau$,
and the expression for the canonical (i.e., without the heavy-light mixing factor)
decay width  ${\overline \Gamma}(N \to \ell_2 \ell_3 \nu)$ is given by
\be
 {\overline \Gamma}(N \to \ell_2 \ell_3 \nu) =  \frac{G_F^2 M_N^2}{192 \pi^3}
{\cal F}(x_2,x_3) \ ,
\label{bGNllnu}
\ee
where the dimensionless notations are used
\be
x_j = \frac{M_{\ell_j}^2}{M_N^2} \ ,
\label{xj}
\ee
and the function ${\cal F}(x_2,x_3)$ for nonzero lepton masses was calculated in Ref.~\cite{CKZ} and is given here in Appendix \ref{app1} (see also Ref.~\cite{symm}). It is symmetric under the exchange $x_2 \leftrightarrow x_3$ ($\ell_2 \leftrightarrow \ell_3$).

In order to obtain the width for the rare decays $B_{(c)} \to (D^{(*)})\ell_1 N \to (D^{(*)})\ell_1 \ell_2 \ell_3 \nu$ (with $\nu$ a light, practically massless neutrino),
we combine the results (\ref{GBlN})- (\ref{bGBlN}) and (\ref{GNllnu})-(\ref{bGNllnu}) in the expression (\ref{fact})
\be
\Gamma \left( B_{(c)}^{\pm} \to (D^{(*)})\ell_1^{\pm} N \to (D^{(*)})\ell_1^{\pm} \ell_2^{\mp} \ell_3^{\pm} \nu \right)^{\rm (X)}
= |U_{\ell_1 N}|^2 |U_{\ell_X N}|^2 \frac{1}{\Gamma_N} {\overline \Gamma} \left( B_{(c)}^{\pm} \to (D^{(*)})\ell_1^{\pm} N \right) {\overline \Gamma}(N \to \ell_2^{\mp} \ell_3^{\pm} \nu) \ ,
\label{factnoDnoPi}
\ee
where X stands for X=LNC or X=LNV process, and $U_{\ell_X N} = U_{\ell_2 N}$
for X=LNC and $U_{\ell_X N} = U_{\ell_3 N}$ for X=LNV.

We will see later that the total decay width of the $N$ neutrino, $\Gamma_N$, cancels out in the effective decay rates where the decay probability of the intermediate on-shell $N$ is accounted for.

\begin{figure}[htb] 
\begin{minipage}[b]{.49\linewidth}
\centering\includegraphics[width=65mm]{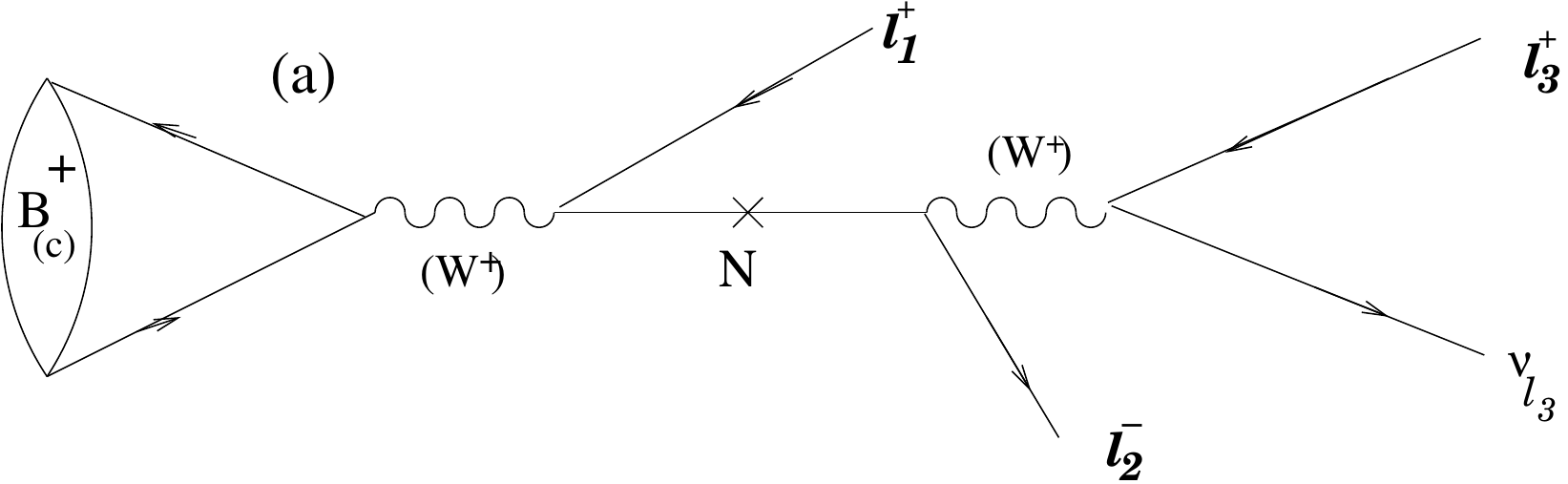}
\end{minipage}
\begin{minipage}[b]{.49\linewidth}
\centering\includegraphics[width=65mm]{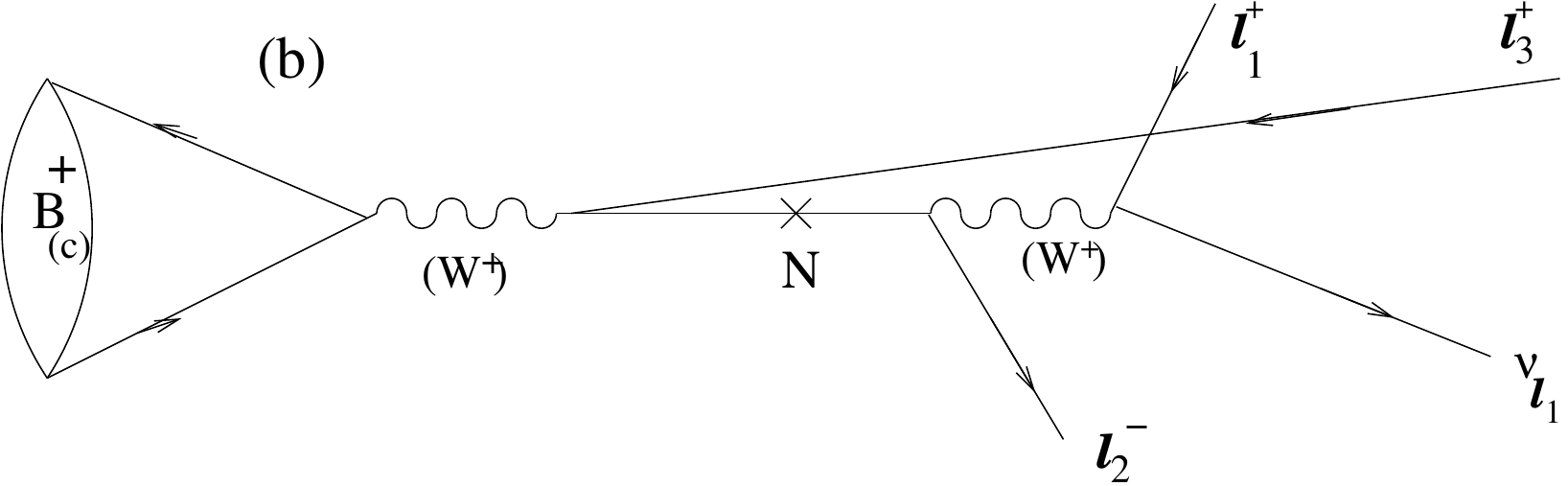}
\end{minipage}
\vspace{-0.4cm}
\caption{The lepton number conserving (LNC) process: (a) the direct (D) channel; (b) the crossed (C) channel; the crossed channel appears only if $\ell_1=\ell_3$.}
\label{FigLNC}
 \end{figure}

\begin{figure}[htb] 
\begin{minipage}[b]{.49\linewidth}
\centering\includegraphics[width=65mm]{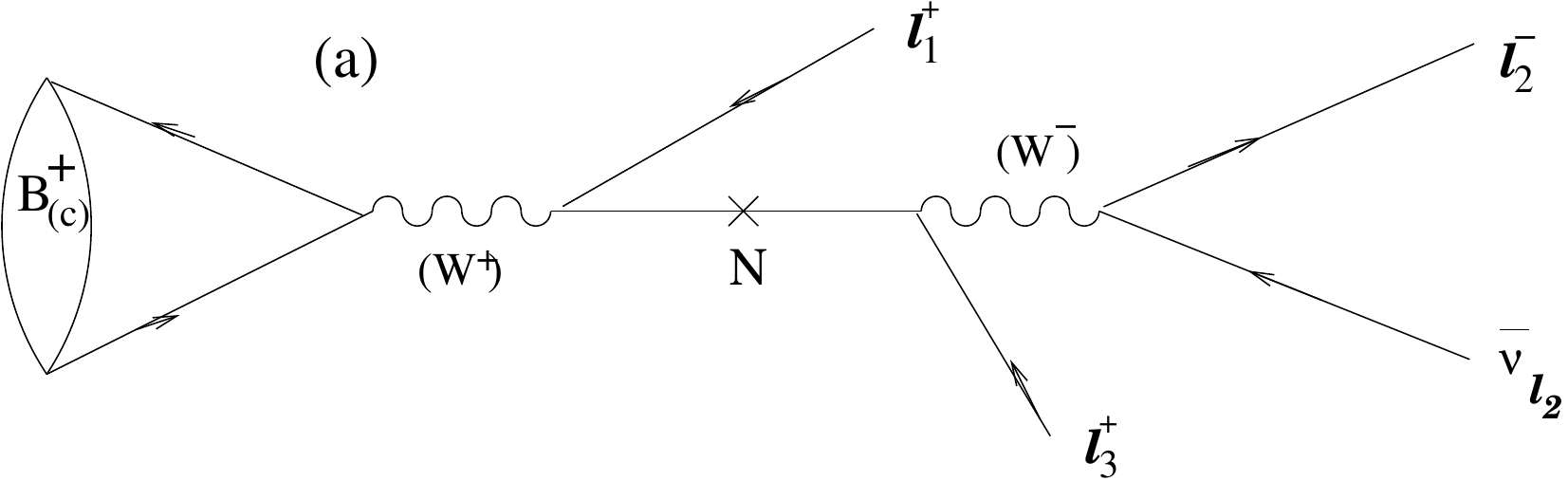}
\end{minipage}
\begin{minipage}[b]{.49\linewidth}
\centering\includegraphics[width=65mm]{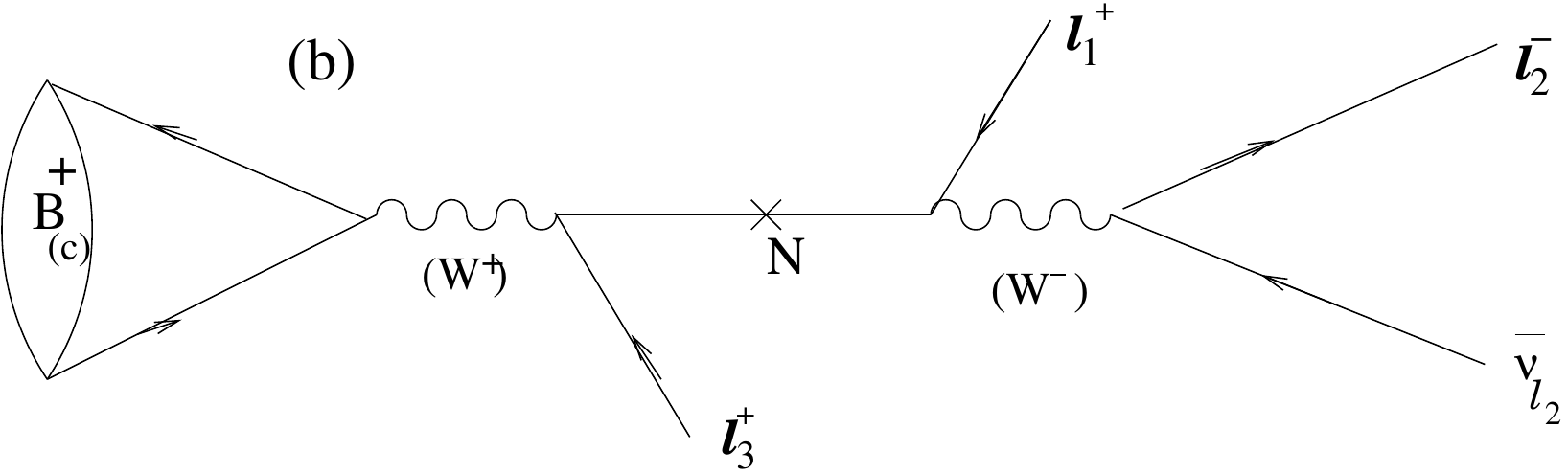}
\end{minipage}
\vspace{-0.4cm}
\caption{The lepton number violating (LNV) process: (a) the direct (D) channel; (b) the crossed (C) channel; the crossed channel appears only if $\ell_1=\ell_3$.}
\label{FigLNV}
 \end{figure}

The LNC and LNV processes contained in the expression (\ref{factnoDnoPi}), are
depicted in Figs.~\ref{FigLNC} and \ref{FigLNV}.
If $\ell_1=\ell_3$, then we have a statistical factor $1/2!$ in front of the expression for the decay width. However, in such a case two different channels (direct and crossed) contribute, with equal strength, giving a factor $2$ which cancels the aforementioned factor. Therefore, the formula (\ref{factnoDnoPi}) is valid for both cases, when $\ell_1=\ell_3$ and $\ell_1 \not= \ell_3$.

If the intermediate $N$ is Dirac (Dir.), then only the LNC processes contribute, and thus we must apply in the formula (\ref{factnoDnoPi}) for the total factor simply
the X=LNC contribution; if $N$ is Majorana (Maj.), both LNC and LNV processes contribute
\bes
\label{noDnoPi}
\bea
  \Gamma \left( B_{(c)}^{\pm} \to (D^{(*)})\ell_1^{\pm} N \to (D^{(*)})\ell_1^{\pm} \ell_2^{\mp} \ell_3^{\pm} \nu \right)^{\rm (Dir.)}
&=&
  |U_{\ell_1 N}|^2 |U_{\ell_2 N}|^2 \frac{1}{\Gamma_N} {\overline \Gamma} \left( B_{(c)}^{\pm} \to (D^{(*)}) \ell_1^{\pm} N \right) {\overline \Gamma}(N \to \ell_2^{\mp} \ell_3^{\pm} \nu_{\ell_3}) \ ,
  \nonumber\\
\label{noDnoPiDir}
\\
  \Gamma \left( B_{(c)}^{\pm} \to (D^{(*)})\ell_1^{\pm} N \to (D^{(*)})\ell_1^{\pm} \ell_2^{\mp} \ell_3^{\pm} \nu \right)^{\rm (Maj.)} &=&
  |U_{\ell_1 N}|^2 (|U_{\ell_2 N}|^2 + |U_{\ell_3 N}|^2)
  \nonumber\\
  && \times
\frac{1}{\Gamma_N} {\overline \Gamma} \left( B_{(c)}^{\pm} \to (D^{(*)})\ell_1^{\pm} N \right) {\overline \Gamma}(N \to \ell_2^{\mp} \ell_3^{\pm} \nu) \ .
\label{noDnoPiMaj}
\eea
\ees

\subsection{Semileptonic decays  $N \to \ell^\pm \pi^\mp$}
\label{semilepdecN}

Other possibilities for the decay of $N$ (produced by $B_{(c)}^{\pm} \to \ell_1^{\pm} N$) are semileptonic, and among them
kinematically the most favorable is the decay into a (charged) pion:
$N \to \ell_2^{\mp} \pi^{\pm}$ (LNC case of $B$-decay);
$N \to \ell_3^{\pm} \pi^{\mp}$ (LNV case of $B$-decay). In this case,
the second factor in the expression (\ref{fact}) is now different
\be
\Gamma(N \to \ell^{\pm} \pi^{\mp}) = |U_{\ell N}|^2 {\overline \Gamma}(N \to \ell^{\pm} \pi^{\mp}) \ ,
\label{GNlPi}
\ee
where the canonical expression ${\overline \Gamma}$ is (e.g., cf.~Refs.~\cite{CDKK,CKZ2,symm,CKZosc})
\be
{\overline \Gamma}(N \to \ell^{\pm} \pi^{\mp}) =
\frac{1}{16 \pi} G_F^2 f_{\pi}^2 M_N^3 \lambda^{1/2}(1, x_{\pi}, x_{\ell})
\left[ 1 - x_{\pi} - 2 x_{\ell} - x_{\ell}  (x_{\pi}-x_{\ell}) \right] \ ,
\label{bGNlPi}
\ee
where we use the notations analogous to Eq.~(\ref{xj})
\be
x_{\pi} = \frac{M_{\pi}^2}{M_N^2} \ , \qquad x_{\ell}=\frac{M_{\ell}^2}{M_N^2} \ ,
\label{xPixell}
\ee
and $f_{\pi}$ ($\approx 0.1304$ GeV) is the decay constant of pion. It turns
out that the corresponding rare decay $B_{(c)} \to \ell_1 N \to \ell_1 \ell_2 \pi$,
as in the previous Subsection,
can be either LNC
($B_{(c)}^{\pm} \to (D^{(*)})\ell_1^{\pm} N \to (D^{(*)})\ell_1^{\pm} \ell_2^{\mp} \pi^{\pm}$)
or LNV
 ($B_{(c)}^{\pm} \to (D^{(*)})\ell_1^{\pm} N \to (D^{(*)})\ell_1^{\pm} \ell_2^{\pm} \pi^{\mp}$).
If $N$ is Dirac, it is only LNC; if $N$ is Majorana, it is the sum of
LNC and LNV.
Again, as in the previous Subsection, if $\ell_2 = \ell_1$ (i.e., equal flavor and
equal charge), the statistical factor $1/2!$ appears in front of the integration, but is then cancelled by factor $2$ appearing from the two equal contributions of the direct and crossed channel. Therefore, the following formula holds for both cases of $\ell_2 = \ell_1$ and $\ell_2 \not= \ell_1$, partly analogous to Eqs.~(\ref{noDnoPi})
\bea
\Gamma \left( B_{(c)}^{\pm} \to (D^{(*)})\ell_1^{\pm} N \to (D^{(*)})\ell_1^{\pm} \ell_2^{\mp} \pi^{\pm} \right)^{\rm (LNC)}
&=&
\Gamma \left( B_{(c)}^{\pm} \to (D^{(*)})\ell_1^{\pm} N \to (D^{(*)})\ell_1^{\pm} \ell_2^{\pm} \pi^{\mp} \right)^{\rm (LNV)}
\nonumber\\
& = &
 |U_{\ell_1 N}|^2 |U_{\ell_2 N}|^2
\frac{1}{\Gamma_N} {\overline \Gamma} \left( B_{(c)}^{\pm} \to (D^{(*)})\ell_1^{\pm} N \right) {\overline \Gamma}(N \to \ell_2^{\mp} \pi^{\pm}).
\label{noDPi}
\eea

The two factors ${\overline {\Gamma}}$ on the right-hand side here are
given in Eqs.~(\ref{bGBlN}) and (\ref{bGNlPi}), they are independent of
the charges involved, and thus the LNC and LNV decays here are equal.
If $N$ is Dirac, only LNC process contributes; if $N$ is Majorana,
both LNC and LNV processes contribute, amounting in doubling the value
of the width.

\begin{figure}[htb] 
\begin{minipage}[b]{.49\linewidth}
\centering\includegraphics[width=85mm]{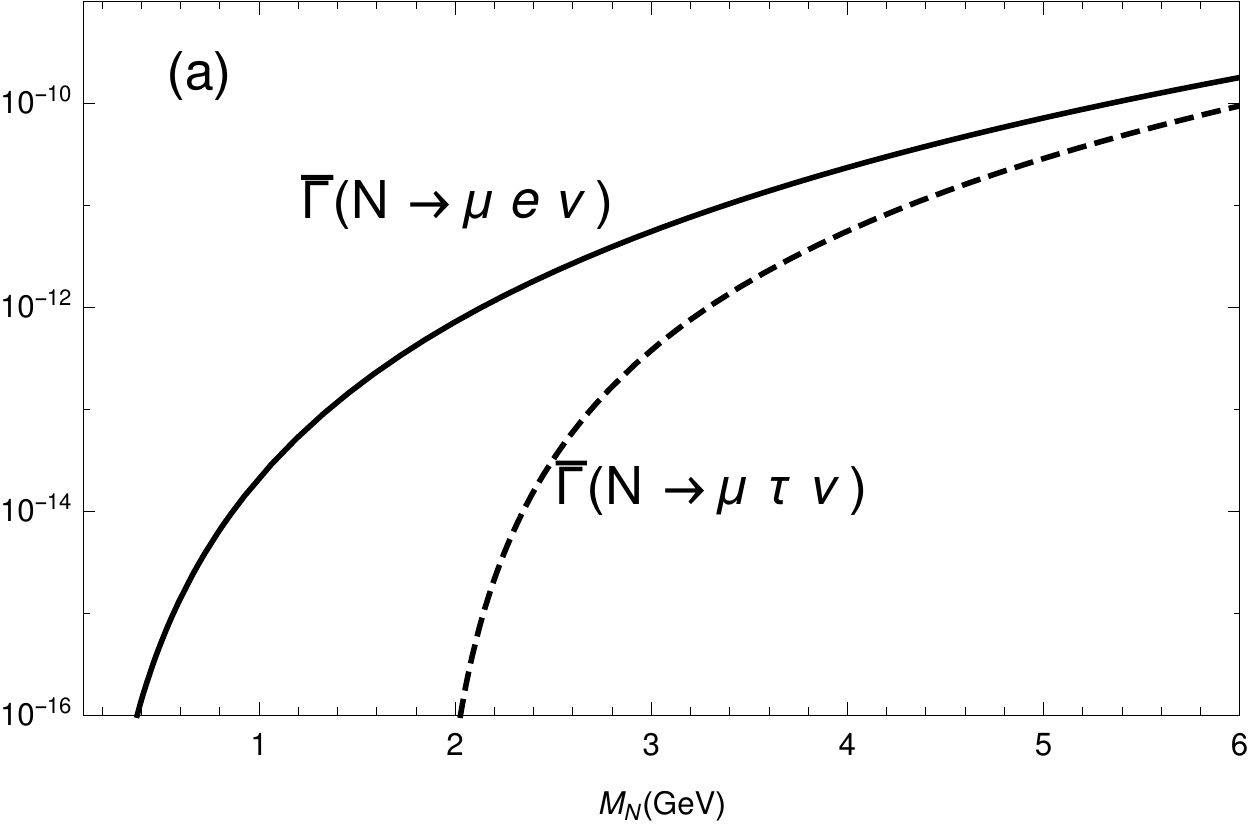}
\end{minipage}
\begin{minipage}[b]{.49\linewidth}
\centering\includegraphics[width=85mm]{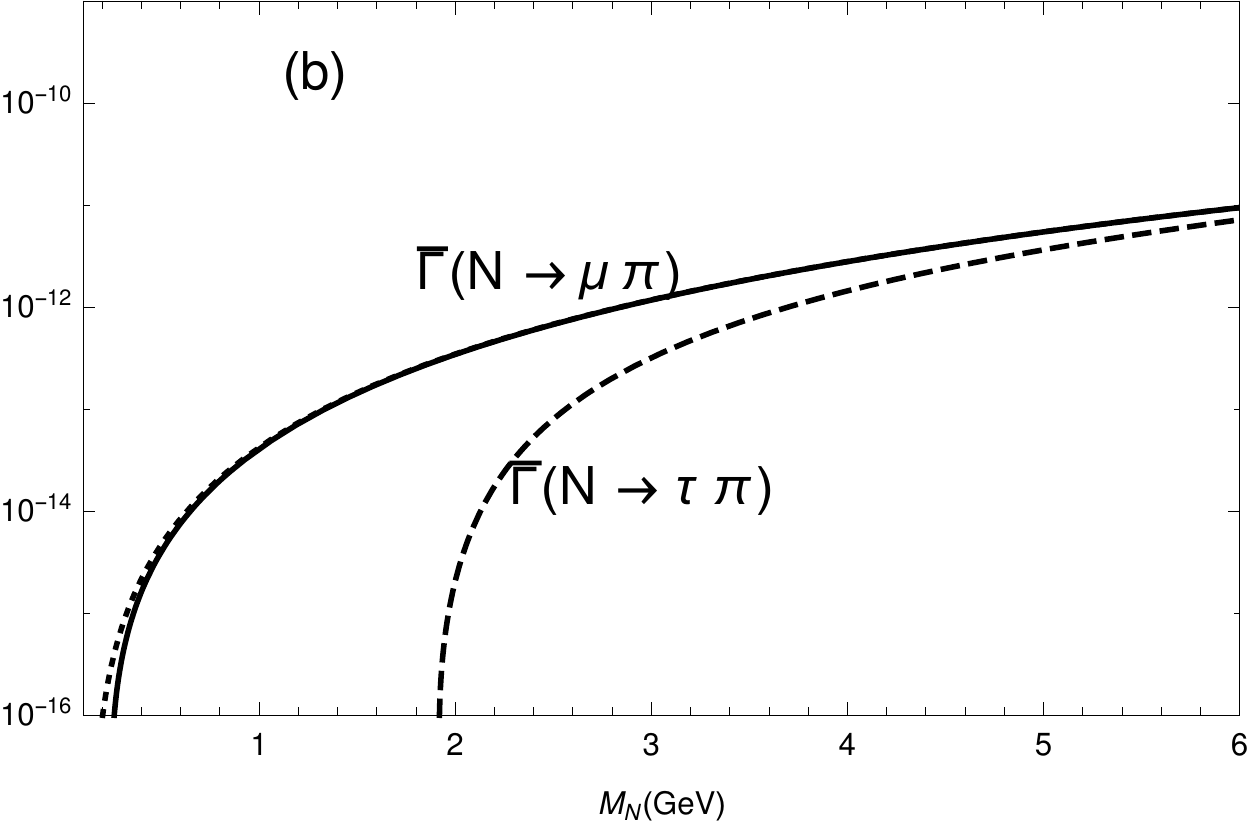}
\end{minipage}
\vspace{-0.2cm}
\caption{The canonical decay widths, in units of GeV, as a function of mass of $N$, for (a) leptonic decays $N \to \mu \ell \nu$ where $\ell=e$ (solid) and $\ell=\tau$ (dashed); (b) semileptonic  decays $N \to \ell \pi$ where $\ell=e$ (solid) and $\ell=\tau$ (dashed). In (b) we also included the curve for $N \to e \pi$ (dotted), and it is close to the solid line $N \to \mu \pi$.}
\label{FigbGNXY}
\end{figure}

In Fig.~\ref{FigbGNXY} we present the canonical decay widths of heavy neutrino $N$, ${\overline \Gamma}(N \to \ell_2 \ell_3 \nu)$ (for $\ell_2=\mu$ and $\ell_3=e$ or $\tau$) and ${\overline \Gamma}(N \to \ell \pi)$ (for $\ell=\mu$ or $\tau$), as a function of mass $M_N$, cf.~Eqs.~(\ref{bGNllnu}) and (\ref{bGNlPi}). In all these decay widths, we assume that the charged leptons have specific electric charges (e.g.., $\mu^+ e^- \nu$, $\mu^+ \pi^-$).

\section{Branching ratios}
\label{sec:Br}

We recall that the decay width of the considered processes is written in the factorized form (\ref{fact}), and the theoretical branching ratio is obtained by dividing it by the decay width of $B_{(c)}$
\be
 {\rm Br}\left( B_{(c)} \to (D^{(*)}) \ell_1 N \to (D^{(*)}) \ell_1 X Y \right)
  =  \Gamma \left( B_{(c)} \to (D^{(*)}) \ell_1 N \right) \frac{\Gamma(N \to XY)}{\Gamma_N \Gamma_{{B_{(c)}}}} \ .
\label{Br}
\ee
The factor $\Gamma(N \to XY)$ is given for leptonic decay of $N$ ($XY= \ell_2 \ell_3 \nu$) in Eqs.~(\ref{GNllnu})-(\ref{bGNllnu}), and for semileptonic decay of $N$ ($XY=\ell_2 \pi$) in Eqs.~(\ref{GNlPi})-(\ref{bGNlPi}), cf.~Figs.~\ref{FigbGNXY}. The first factor $\Gamma \left( U_{(c)} \to (D^{(*)}) \ell_1 N \right)$ is given for the process without $D^{(*)}$ in Eqs.~(\ref{GBlN})-(\ref{bGBlN}) (cf.~Figs.~\ref{FigbGBmuN}), for process with $D$ and $D^*$ in Eqs.~(\ref{GBDNl}) and (\ref{GBDstNldef})-(\ref{bGBDstNl}), respectively (cf.~Figs.~\ref{FigbGBDlN}).

\subsection{Total decay width of $N$}
\label{subs:GN}

\begin{figure}[htb] 
\begin{minipage}[b]{.49\linewidth}
\includegraphics[width=85mm]{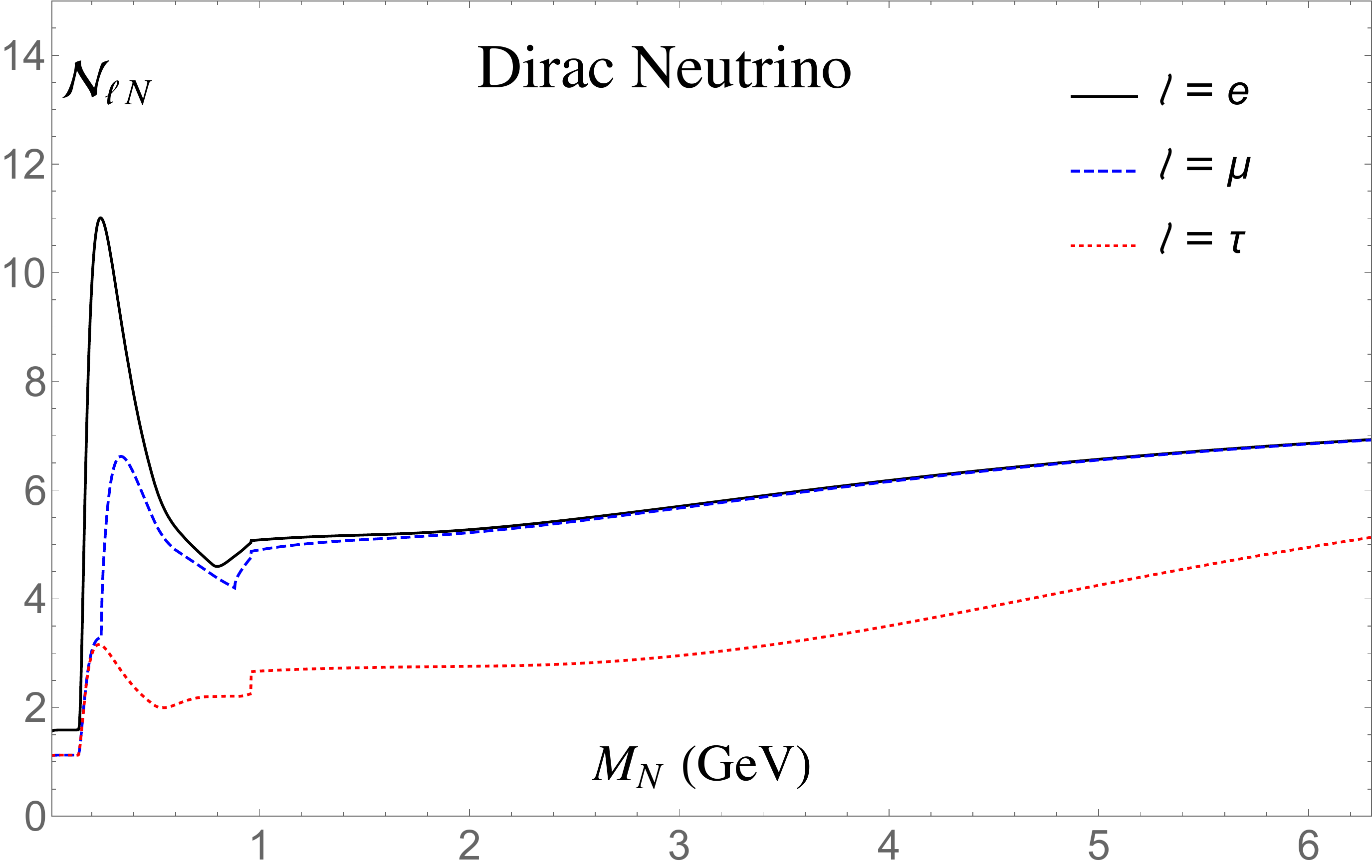}
\end{minipage}
\begin{minipage}[b]{.49\linewidth}
\includegraphics[width=85mm]{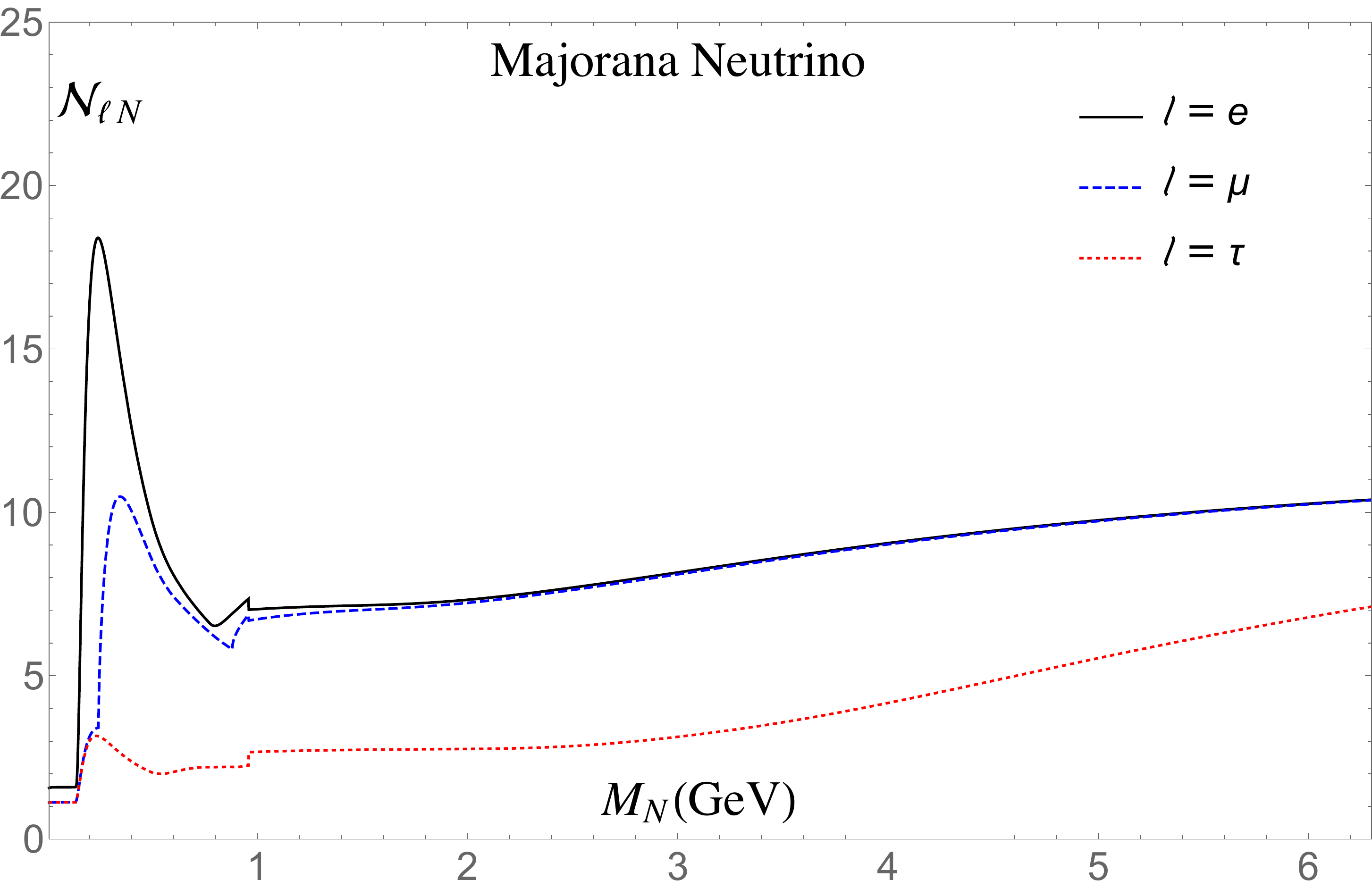}
\end{minipage} \vspace{12pt}
\caption{The coefficients
${\cal N}_{\ell N}$ ($\ell = e, \mu, \tau$)
appearing in Equations~(\ref{GNwidth})--(\ref{calK}),
as a function of the mass of the sterile neutrino $N$.
The left-hand figure is for Dirac
neutrino, and the right-hand figure for Majorana neutrino.}
\label{FigcNellN}
\end{figure}

The only factor that remains to be evaluated to obtain the branching ratios (\ref{Br}) is the decay width of the sterile neutrino $N$. This decay width can be written in the following form
\begin{equation}
\Gamma_{N} = \K \; {\overline \Gamma}_N(M_{N}) \ ,
\label{GNwidth}
\end{equation}
where the corresponding canonical decay width is
\begin{equation}
 {\overline \Gamma}_N(M_{N}) \equiv \frac{G_F^2 M_{N}^5}{96 \pi^3} \ ,
\label{barGN}
\end{equation}
and the factor $\K$ contains all the dependence on
the heavy-light mixing factors
\begin{equation}
\K(M_{N}) \equiv \K = {\cal N}_{e N} \; |U_{e N}|^2 + {\cal N}_{\mu N} \; |U_{\mu N}|^2 + {\cal N}_{\tau N} \; |U_{\tau N}|^2  \
\label{calK}
\end{equation}
Here, the coefficients ${\cal N}_{\ell N}$ turn out to be numbers $\sim 1$-$10$ which depend on the mass $M_N$ and on the character of $N$ neutrino (Dirac or Majorana). We refer to Ref.~\cite{CKZ2} for details of the calculation of ${\cal N}_{\ell N}$, based on expressions of Ref.~\cite{HKS} (see also Refs.~\cite{GKS,Atre}). In Figs.~\ref{FigcNellN} we present the resulting coefficients - the figures were taken from Ref.~\cite{CKZ2} for Majorana $N$, and \cite{symm} for Dirac $N$.
We can see from that for $1 \ {\rm GeV} \leq M_N \leq 6 \ {\rm GeV}$, which is the relevant mass range for the rare $B_{(c)}$-decays considered here, we have approximately
\bes
\label{calKest}
\bea
{\K}^{\rm (Dir.)} & \approx &   6  (|U_{e N_j}|^2 + |U_{\mu N_j}|^2) + 3  |U_{\tau N_j}|^2 \ ,
\label{calKestDir}
\\
{\K}^{\rm (Maj.)} & \approx &   8  (|U_{e N_j}|^2 + |U_{\mu N_j}|^2) + 3  |U_{\tau N_j}|^2 \ .
\label{calKestMaj}
\eea
\ees
We note that this factor is for Majorana neutrino not simply twice the factor for Dirac neutrino.

\subsection{Canonical branching ratios}
\label{subs:bBr}

If we factor out all the heavy-light mixing factors in the branching ratios (\ref{Br}), we end up with the canonical branching ratio  (i.e., without any heavy-light mixing dependence)  ${\overline {\rm Br}}$
\be
   {\overline {\rm Br}} \left( B_{(c)} \to (D^{(*)}) \ell_1 N \to  (D^{(*)}) \ell_1 X Y \right) \equiv  \frac{1}{\Gamma_{B_{(c)}} {\overline \Gamma}_N(M_{N})} {\overline \Gamma} \left( B_{(c)} \to (D^{(*)}) \ell_1 N \right) {\overline \Gamma}(N \to XY)  \ .
\label{barBr}
\ee
In the case of leptonic decays of $N$ ($XY  = \ell_2 \ell_3 \nu$), the branching ratio (\ref{Br}) can then be written [cf.~Eqs.~(\ref{noDnoPi})]
\bes
\label{cBr}
\bea
 {\rm Br}^{\rm (Dir.)} \left( B_{(c)} \to (D^{(*)}) \ell_1^{\pm} N \to (D^{(*)}) \ell_1^{\pm} \ell_2^{\mp} \ell_3^{\pm} \nu_{\ell_3} \right) & = &
 \frac{1}{{\K}} |U_{\ell_1 N}|^2 |U_{\ell_2 N}|^2  {\overline {\rm Br}} \left( B_{(c)} \to (D^{(*)}) \ell_1^{\pm} \ell_2^{\mp} \ell_3^{\pm} \nu_{\ell_3} \right),
 \label{cBrDir}
\\
 {\rm Br}^{\rm (Maj.)} \left( B_{(c)} \to (D^{(*)}) \ell_1^{\pm} N \to (D^{(*)}) \ell_1^{\pm} \ell_2^{\mp} \ell_3^{\pm} \nu \right) & = &
 \frac{1}{{\K}} |U_{\ell_1 N}|^2 (|U_{\ell_2 N}|^2+|U_{\ell_3 N}|^2)
      {\overline {\rm Br}} \left( B_{(c)} \to (D^{(*)}) \ell_1^{\pm} \ell_2^{\mp} \ell_3^{\pm} \nu \right).
      \nonumber\\
 \label{cBrMaj}
 \eea
 \ees
The structure of the heavy-light mixing coefficients is different in the cases when $N$ is Majorana and when it is Dirac. This is so because in the case of Majorana $N$ we have contributions of both LNC and LNV processes (Figs.~\ref{FigLNC} and \ref{FigLNV}), and in the case of Dirac $N$ we have only LNC contributions (Fig.~\ref{FigLNC}).

If, however, $N$ decays semileptonically ($XY = \ell_2 \pi$), then the relation between ${\rm Br}$ and ${\overline {\rm Br}}$ is simpler
 \bea
\label{cBrPi}
  {\rm Br} \left( B_{(c)} \to (D^{(*)}) \ell_1 N \to (D^{(*)}) \ell_1 \ell_2 \pi \right) & = &
  \frac{1}{{\K}}  |U_{\ell_1 N}|^2 |U_{\ell_2 N}|^2  {\overline {\rm Br}}  \left(B_{(c)} \to (D^{(*)}) \ell_1 \ell_2 \pi \right) \ .
 \eea
We note that the heavy-light mixing factors in the expressions (\ref{cBr})-(\ref{cBrPi}) are not $\sim |U_{\ell N}|^4$, but $\sim |U_{\ell N}|^2$, because ${\K} \sim |U_{\ell N}|^2$ ($\ell=e, \mu$ or $\tau$). The enhancement effect $1/\Gamma_N \propto 1/{\K}$ ($\propto 1/|U_{\ell N}|^2$) has its origin in the on-shellness of the intermediate $N$ neutrino.

If we assume that all or most of the on-shell neutrinos $N$ decay within the detector (see the next Section when this is not so), then the branching ratios (\ref{cBr})-(\ref{cBrPi}) are those directly measured in the experiment. In the considered rare decays, we have to exclude those decays where among the produced particles are $e^+ e^-$ or $\mu^+ \mu^-$ pairs, since such pairs  represent appreciable QED background. If $\ell_1=\ell_2$ in Eqs.~(\ref{cBr}), such background would appear. However, if $\ell_1 = \ell_3$ ($\not = \ell_1$), no such background appears.
\begin{figure}[htb] 
\begin{minipage}[b]{.49\linewidth}
\centering\includegraphics[width=85mm]{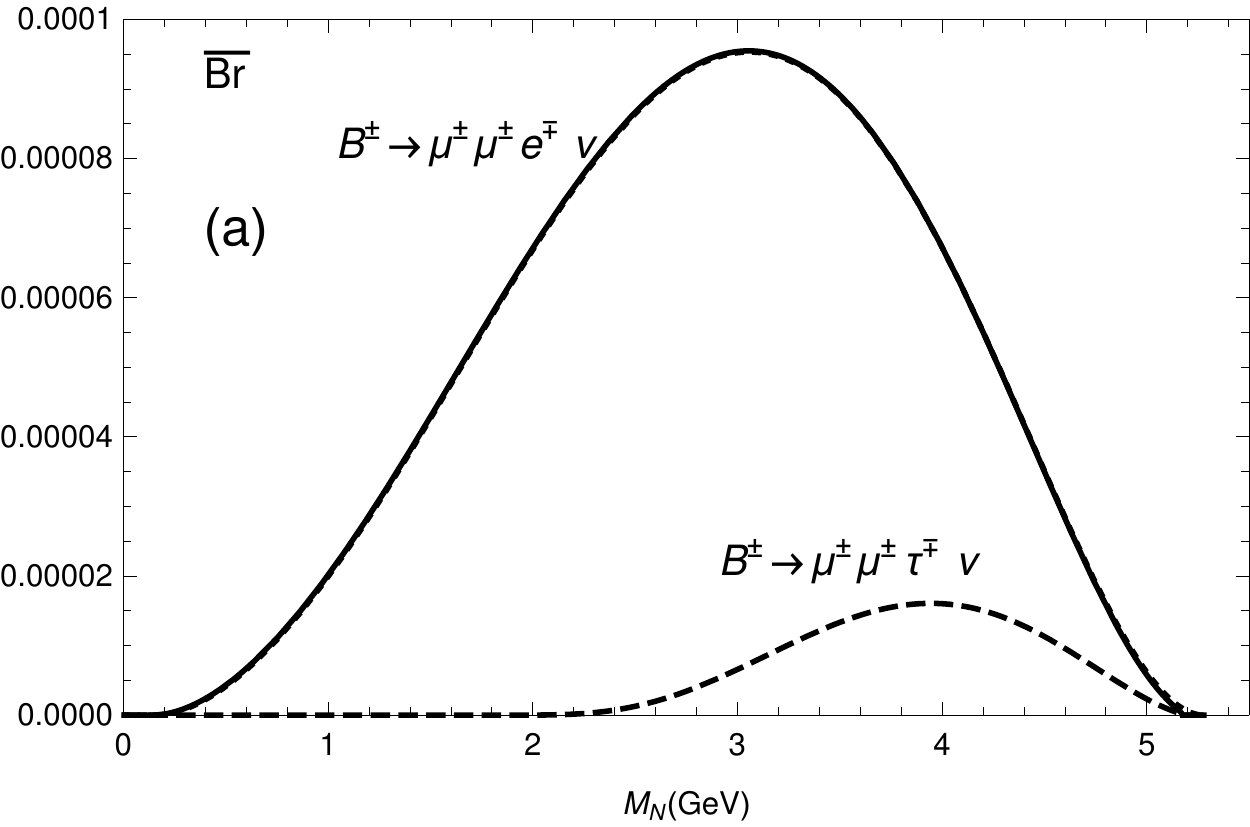}
\end{minipage}
\begin{minipage}[b]{.49\linewidth}
\centering\includegraphics[width=85mm]{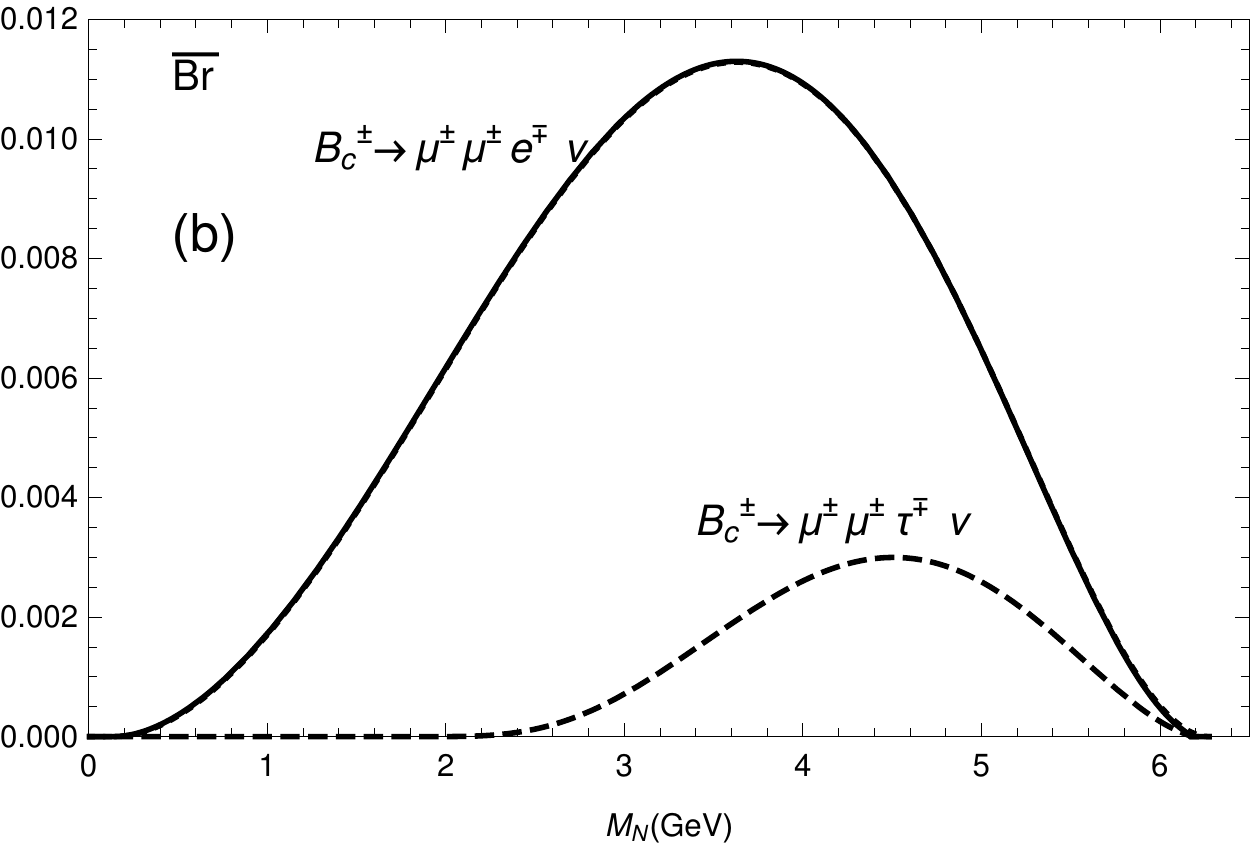}
\end{minipage}
\vspace{-0.2cm}
\caption{The canonical branching ratio ${\overline {\rm Br}}$, as a function of mass of the on-shell neutrino $N$, for the leptonic decays (a) $B^{\pm} \to \mu^{\pm} \mu^{\pm} \ell^{\mp} \nu$, (b) $B_c^{\pm} \to \mu^{\pm} \mu^{\pm} \ell^{\mp} \nu$, where $\ell = e$ (solid) and $\ell=\tau$ (dashed). Included is also the curve for $B_{(c)}^{\pm} \to e^{\pm} e^{\pm} \mu^{\mp} \nu$ (dotted), which is almost indistinguishable from the solid curve.}
\label{FigbBrB}
 \end{figure}
\begin{figure}[htb] 
\begin{minipage}[b]{.49\linewidth}
\centering\includegraphics[width=85mm]{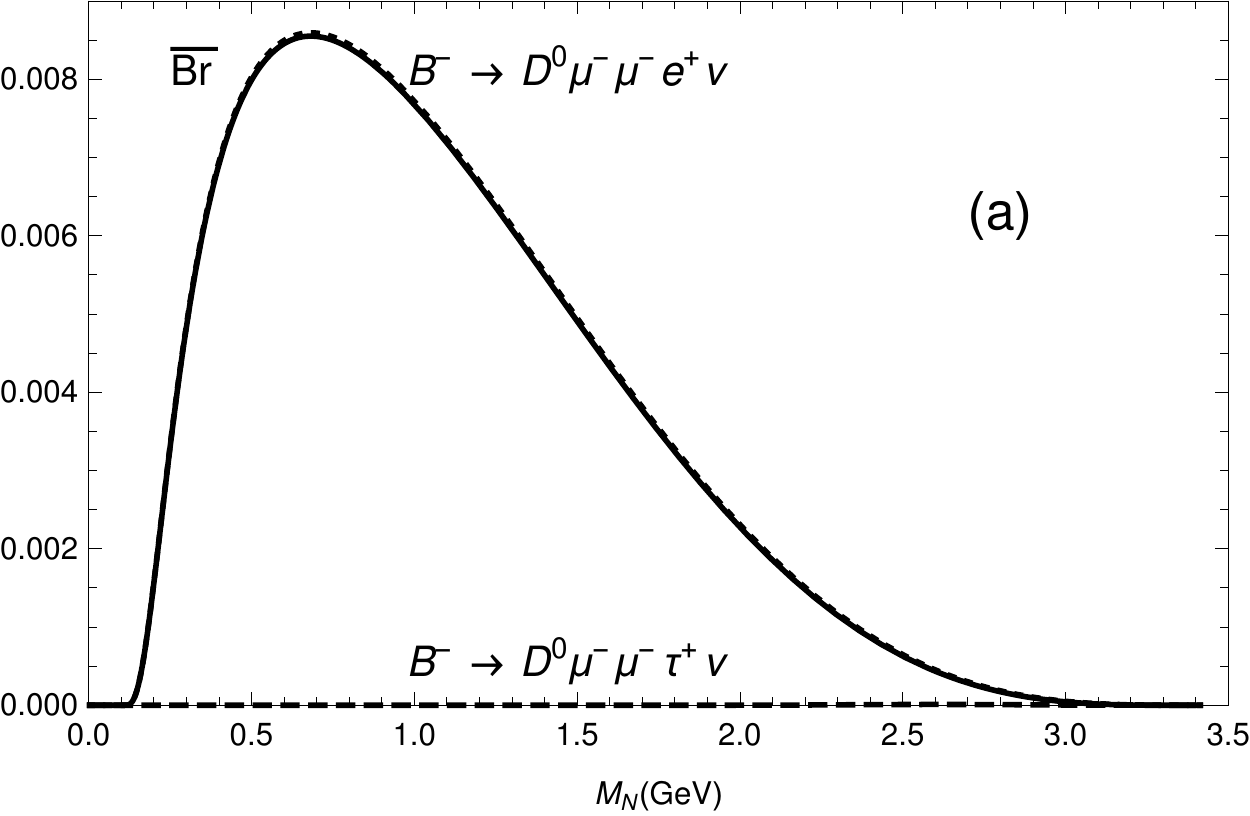}
\end{minipage}
\begin{minipage}[b]{.49\linewidth}
\centering\includegraphics[width=85mm]{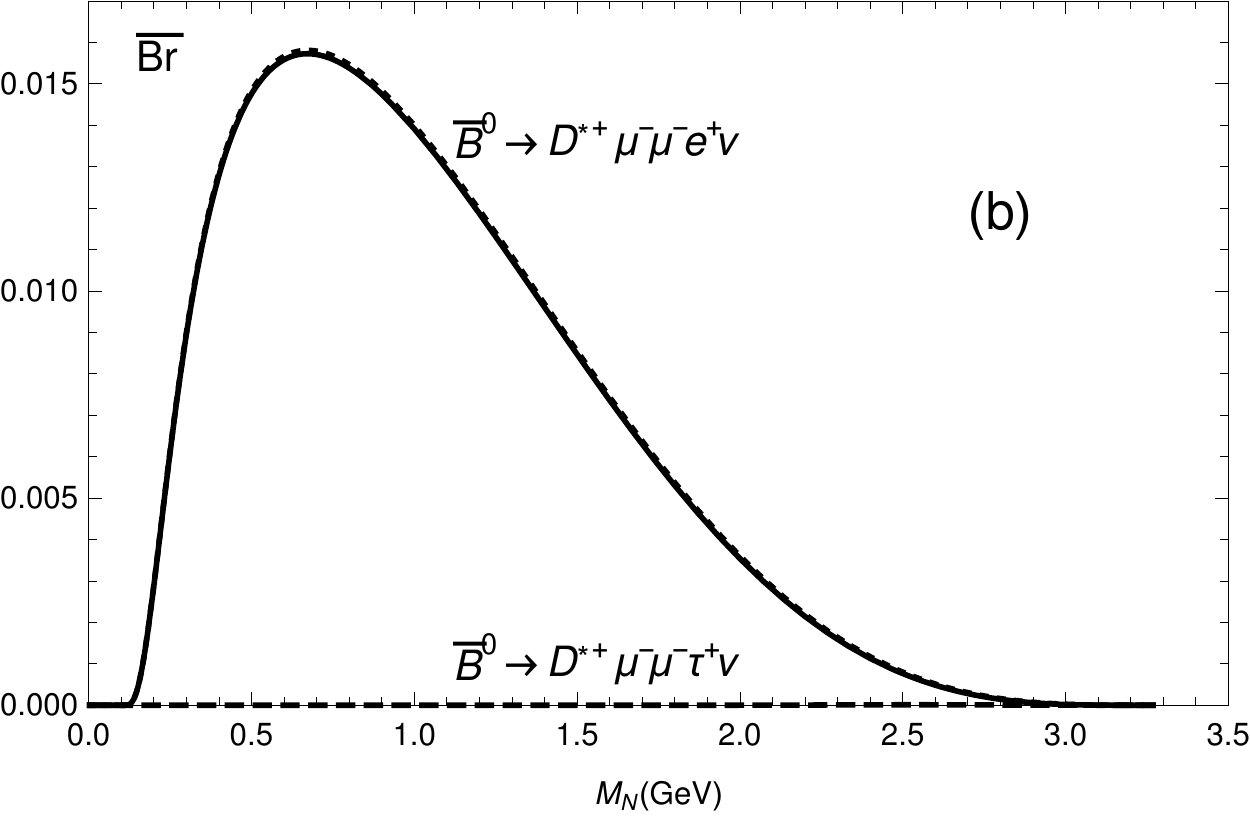}
\end{minipage}
\vspace{-0.2cm}
\caption{The canonical branching ratio, as a function of mass of the on-shell neutrino $N$, for the decays (a) $B^{-} \to D^0 \mu^{-} \mu^{-} \ell^+ \nu$, (b) ${\bar B}^0 \to D^{*+} \mu^{-} \mu^{-} \ell^{+} \nu$, where $\ell = e$ (solid) and $\ell=\tau$ (dashed). The case of $\ell=\tau$ is kinematically strongly suppressed, due to the analogous suppression in Fig.~\ref{FigbGNXY}(a), and it is practically invisible in the Figure.  Included is also the curve for $B \to (D^{(*)}) e^{-} e^{-} \mu^{+} \nu$ (dotted), which is almost indistinguishable from the solid curve.}
\label{FigbBrBD}
 \end{figure}

In Fig.~\ref{FigbBrB}(a) we present the branching ratios for the decays $B^{\pm} \to \mu^{\pm} \mu^{\pm} \ell^{\mp} \nu$ (i.e., $\ell_1 = \ell_3 = \mu$) for $\ell = e, \tau$, and in Fig.~\ref{FigbBrB}(b) the analogous decays of $B_c^{\pm}$. In Figs.~\ref{FigbBrBD}(a) and (b) the analogous decays of $B$ mesons are presented when there is a $D$ or $D^*$ meson among the final particles, thus avoiding the mentioned CKM suppression. We see that in Figs.~\ref{FigbBrB} and especially in \ref{FigbBrBD} the case $\ell = \tau$ is suppressed. This is due to the analogous (kinematical) suppression of ${\overline \Gamma}(N \to \mu \tau \nu)$ in comparison with ${\overline \Gamma}(N \to \nu e \nu)$, especially at low $M_N$, cf.~Eq.~(\ref{bGNllnu}) and Fig.~\ref{FigbGNXY}(a).

\begin{figure}[htb] 
\begin{minipage}[b]{.49\linewidth}
\centering\includegraphics[width=85mm]{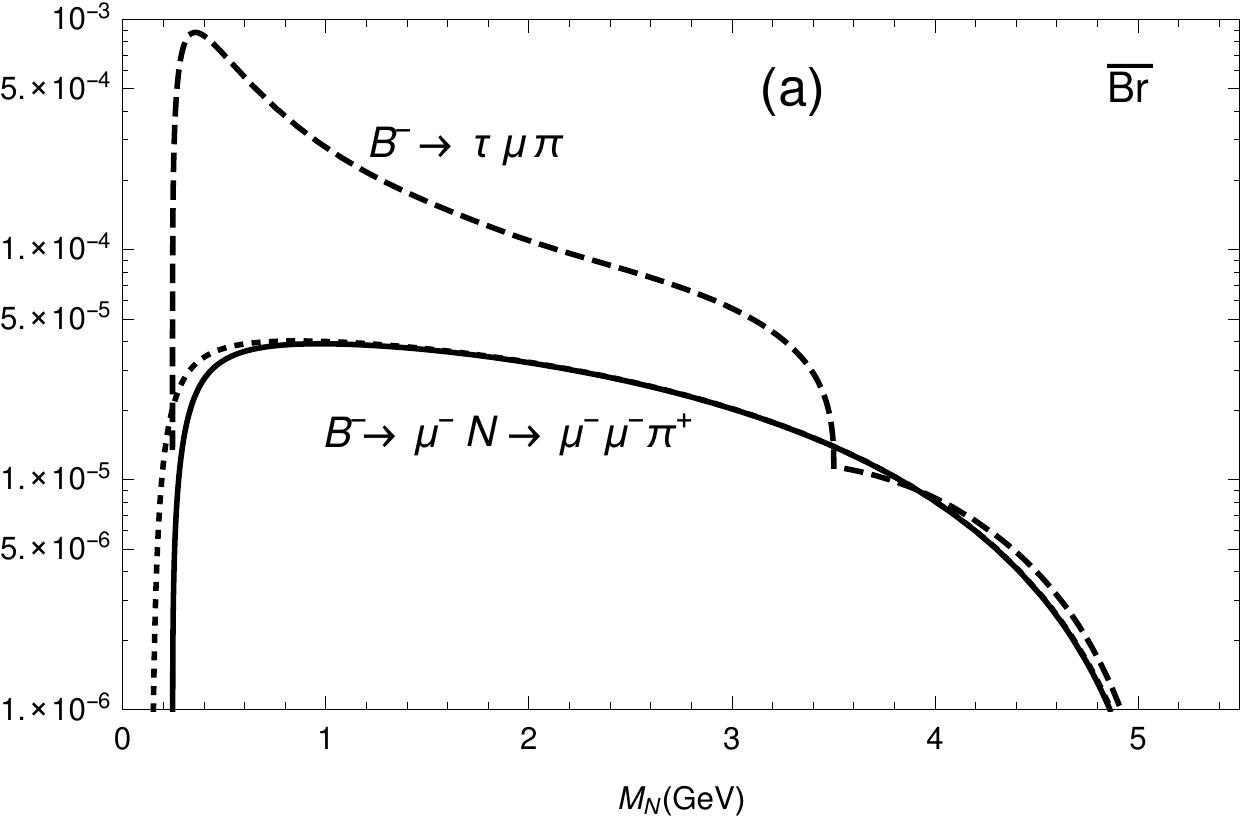}
\end{minipage}
\begin{minipage}[b]{.49\linewidth}
\centering\includegraphics[width=85mm]{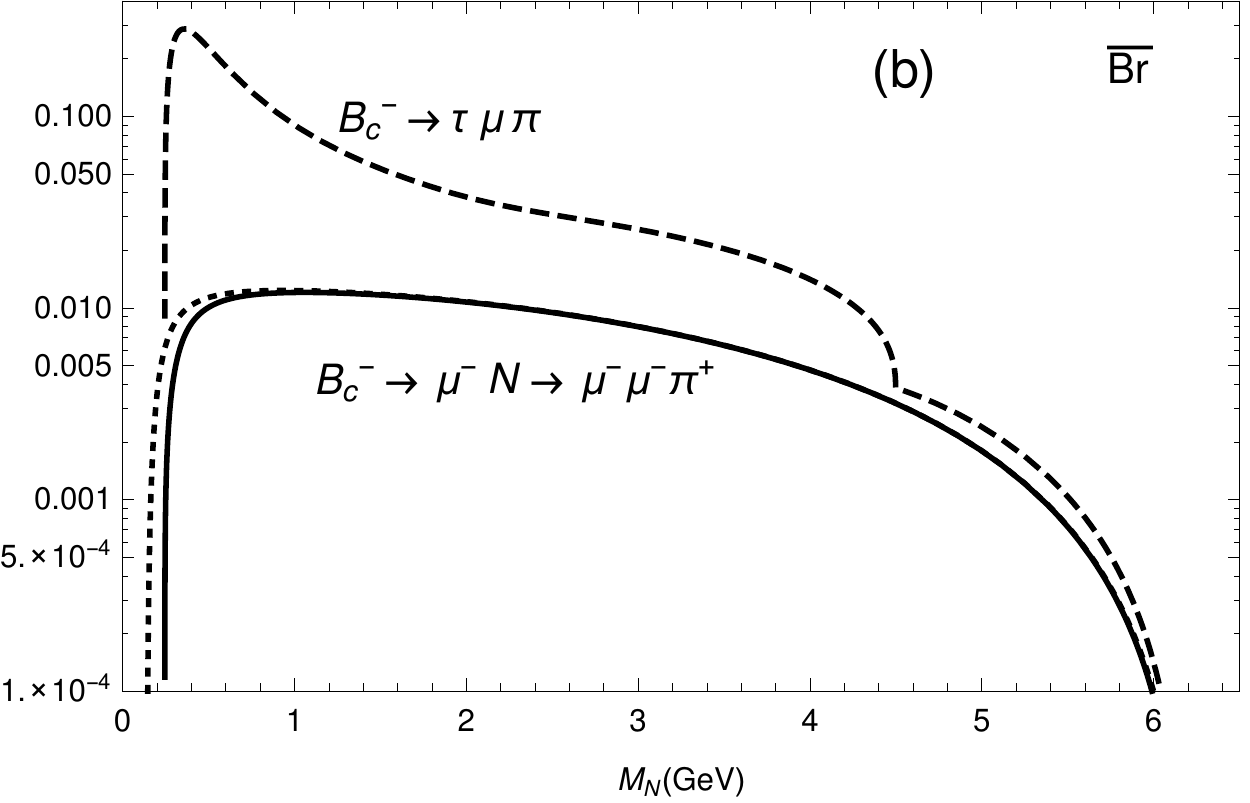}
\end{minipage}
\vspace{-0.2cm}
\caption{The canonical branching ratio, as a function of mass of the on-shell neutrino $N$, for the (LNV) decays (a) $B^- \to \mu^- N \to \mu^- \mu^- \pi^+$ (solid) and $B^- \to e^- N \to e^- e^- \pi^+$ (dotted); (b) $B_c^- \to \mu^- N \to \mu^- \mu^- \pi^+$ (solid) and  $B_c^- \to e^- N \to e^- e^- \pi^+$ (dotted). The dotted curves are close to the solid ones. Included in (a), as a dashed line, is the canonical branching ratio for the decays $B^- \to \tau^- N \to \tau^- \mu^{\mp} \pi^{\pm}$ and $B^- \to \mu^- N \to \mu^- \tau^{\mp} \pi^{\pm}$ (sum of all four decays), and in (b) the analogous quantity for $B_c^-$. Logarithmic scale is used for better visibility.}
\label{FigbBrBPi}
 \end{figure}

\begin{figure}[htb] 
\begin{minipage}[b]{.49\linewidth}
\centering\includegraphics[width=85mm]{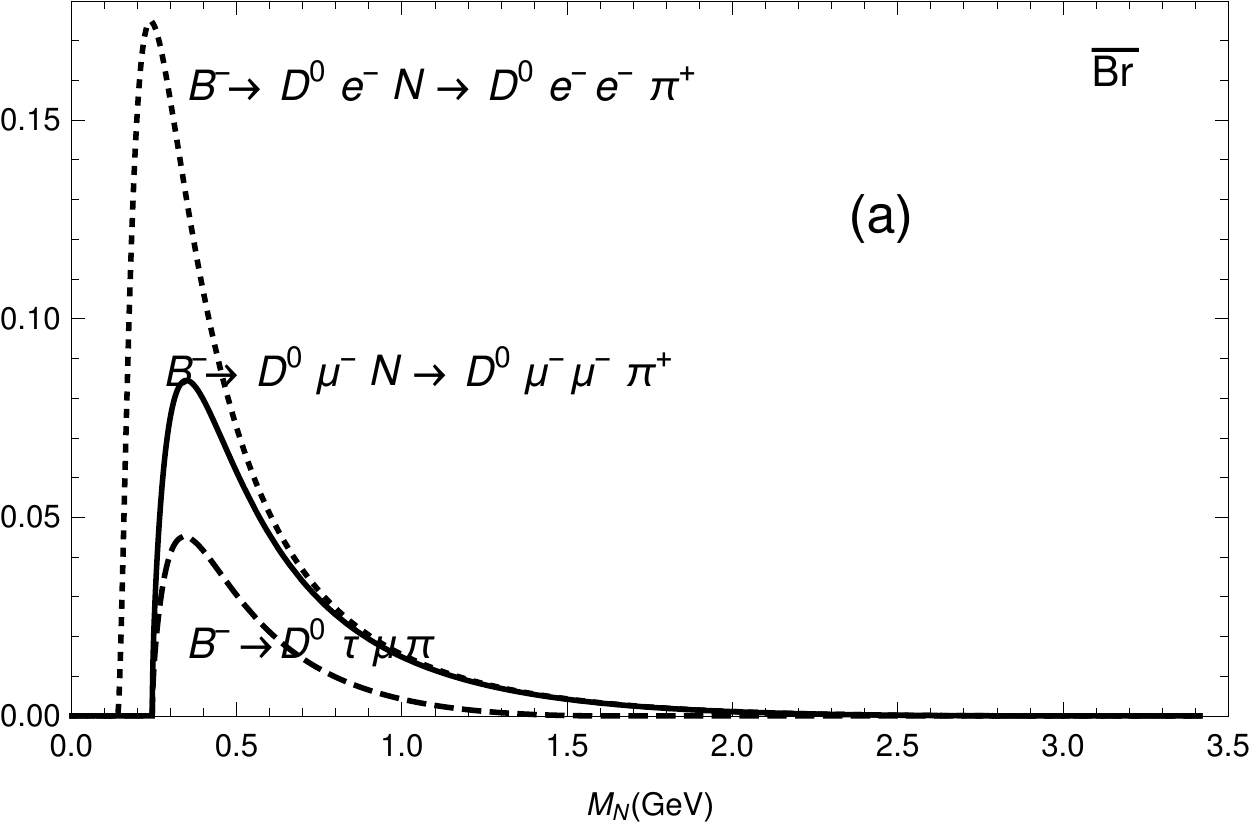}
\end{minipage}
\begin{minipage}[b]{.49\linewidth}
\centering\includegraphics[width=85mm]{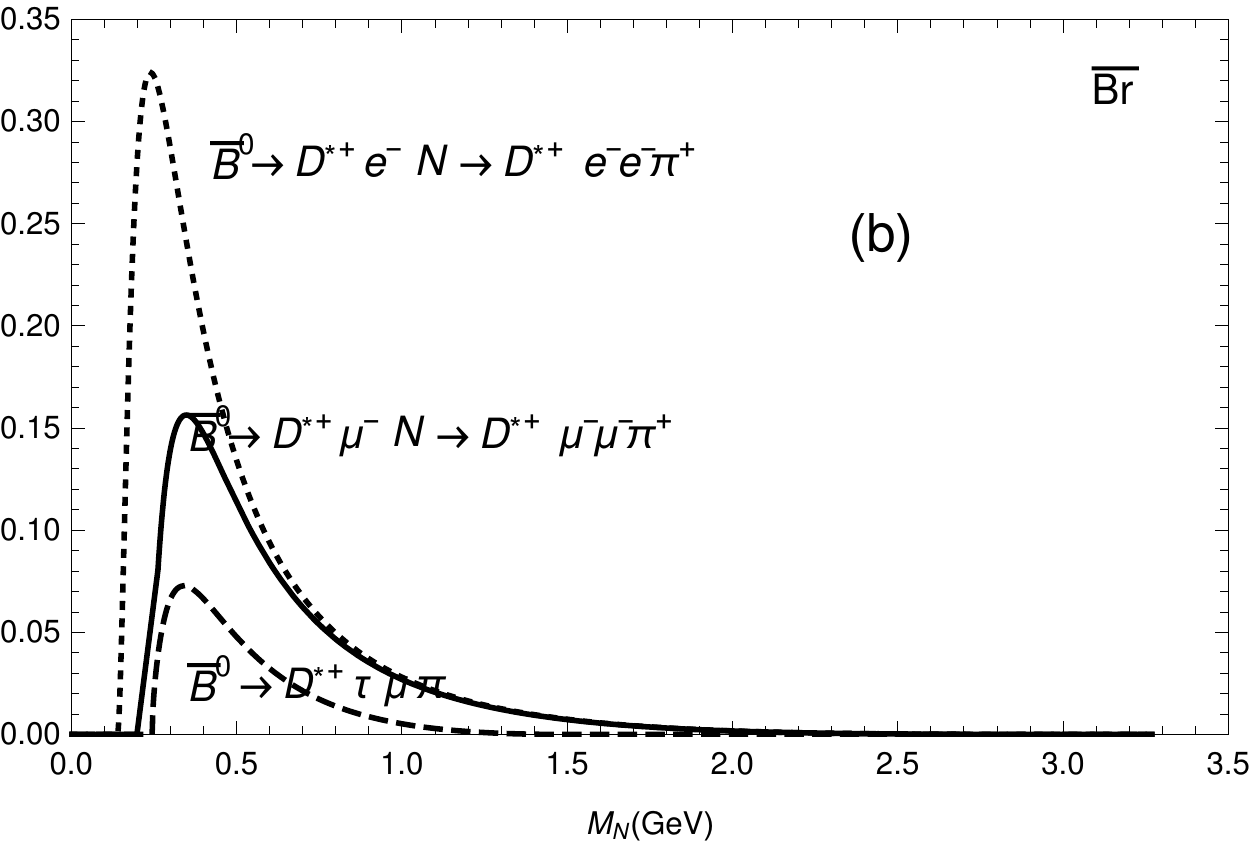}
\end{minipage}
\vspace{-0.2cm}
\caption{The canonical branching ratio, as a function of mass of the on-shell neutrino $N$, for the (LNV) decays (a) $B^-  \to D^0 \mu^- N \to D^0 \mu ^-\mu^- \pi^+$ (solid) and $B^-  \to D^0 e^- N \to D^0 e^- e^- \pi^+$ (dotted); (b) ${\bar B}^0 \to D^{*+} \mu^- N  \to D^{*+} \mu^- \mu^- \pi^+$ (solid) and  ${\bar B}^0 \to D^{*+} e^- N  \to D^{*+} e^- e^- \pi^+$ (dotted). Included in (a), as a dashed line, is the canonical branching ratio for the decays $B^- \to D^0 \tau^- N \to D^0 \tau^- \mu^{\mp} \pi^{\pm}$ and $B^- \to D^0 \mu^- N \to D^0 \mu^- \tau^{\mp} \pi^{\pm}$ (sum of all four decays), and in (b) the analogous quantity with $B^0$ and $D^{*+}$.}
\label{FigbBrBDPi}
 \end{figure}

In Figs.~\ref{FigbBrBPi} and \ref{FigbBrBDPi} the analogous branching ratios are shown, but now for the cases when the intermediate $N$ decays semileptonically ($N \to \ell_2 \pi$).
In Figs.~\ref{FigbBrBPi}, we notice that at low $M_N < 1$ GeV ($M_N \ll M_{\tau}$) a relatively strong enhancement occurs for some processes involving $\tau$ lepton, namely $B_{(c)}^- \to \tau^- N \to \tau^- \mu^{\mp} \pi^{\pm}$. This is so because $\bG(B_{(c)}^- \to \tau^- N)$ is enhanced there, cf.~Eq.~(\ref{bGBlN}) and Figs.~\ref{FigbGBmuN}. Further, we note that the LNV process $B_{(c)} \to ( D^{(*)} ) \mu^- N \to  ( D^{(*)} ) \mu^- \mu^- \pi^+$ gives the same rates as the corresponding LNC process $B_{(c)} \to ( D^{(*)} ) \mu^- N \to  ( D^{(*)} ) \mu^- \mu^+ \pi^-$. If $N$ is Majorana neutrinos both processes contribute, and if $N$ is Dirac only the latter process contributes. However, this LNC process has large QED background due to the produced $\mu^- \mu^+$ pair. Furthermore, if $N$ is Dirac neutrino, only two (LNC) channels of the four (LNC+LNV) channels for $B_{(c)} \to ( D^{(*)}) \tau \mu \pi$ presented in Figs.~\ref{FigbBrBPi} and \ref{FigbBrBDPi} take place, i.e., the dashed curves must be reduced by factor 2 if $N$ is Dirac Neutrino.

\subsection{Differential decay branching ratios for leptonic decays of $N$}
\label{subs:diff}

\begin{figure}[htb]
\centering\includegraphics[width=120mm]{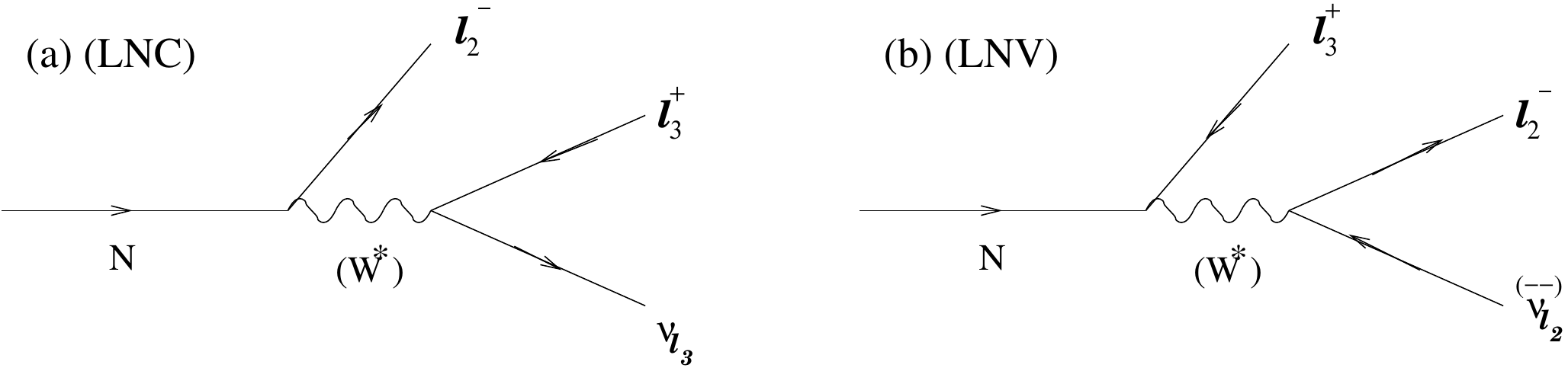}
\caption{(a) LNC decay $N \to \ell_2^- \ell^+ \nu_{\ell_3}$; (b) LNV decay $N \to \ell_3^+ \ell_2^- \nu_{\ell_2}$.}
\label{FigNLNCV}
\end{figure}

Measurement of the branching ratios of those of the considered rare decays in which $N$ decays leptonically ($N \to \ell_2 \ell_3 \nu$) does not give us a direct indication of whether the neutrino $N$ is Majorana or Dirac. This is so because the final light (practically massless) neutrino $\nu$ is not detected. We recall that, for example, the decays $B \to (D^{(*)}) \mu^{+} N \to (D^{(*)}) \mu^{+} \mu^{+} e^{-} \nu$ can be LNC ($\nu=\nu_{\mu}$) or LNV ($\nu={\bar \nu}_e$), and LNV processes are possible only if $N$ is Majorana. However, if in this process we can measure the differential decay width $d \Gamma/d E_e$ with respect to the energy $E_e$ of electron (energy in the rest frame of $N$), then $d \Gamma/d E_e$ is different in the case when $N$ is Majorana or when it is Dirac neutrino. This has been shown, for the decays of the light mesons $\pi \to e^+ e^+ \mu^- \nu$, in Refs.~\cite{CDK,symm}.
For the process $B_{(c)} \to  (D^{(*)}) \mu^{\pm} N \to (D^{(*)}) \mu^{\pm} \mu^{\pm} e^{\mp} \nu$ we have
\bes
\label{diff1}
\bea
\lefteqn{
\frac{d \Gamma \left( B_{(c)} \to \left( D^{(*)} \right) \mu^{\pm} \mu^{\pm} e^{\mp} \nu \right)}{d E_e}  =  \Gamma \left( B_{(c)} \to \left( D^{(*)} \right) \mu^{\pm} N \right) \frac{1}{\Gamma_N} \frac{\Gamma(N \to \mu^{\pm} e^{\mp} \nu)}{d E_e}
}
\label{diff1a}
\\
  & = & \frac{|U_{\mu N}|^2}{{\K}} {\overline \Gamma} \left( B_{(c)} \to \left( D^{(*)} \right) \mu^{\pm} N \right) \frac{1}{{\overline \Gamma}_N}
\left\{ |U_{\mu N}|^2 \frac{ d {\overline \Gamma}^{\rm (LNV)}(N \to \mu^{\pm} e^{\mp} \nu_e)}{d E_e} +|U_{e N}|^2 \frac{ d {\overline \Gamma}^{\rm (LNC)}(N \to e^{\mp} \mu^{\pm} \nu_{\mu})}{d E_e} \right\}.
\nonumber\\
  \label{diff1b}
  \eea
  \ees
  In Eq.~(\ref{diff1b}), the LNV term does not appear if $N$ is Dirac. The differential decay width is thus proportional to the following canonical differential branching ratios:
   \bea
  \frac{ d {\overline {\rm Br}}_N(N \to \mu e \nu ; \alpha )}{d E_e} & \equiv &
  \alpha \frac{1}{{\overline \Gamma}_N} \frac{ d {\overline \Gamma}^{\rm (LNV)}(N \to \mu^{\pm} e^{\mp} \nu_e)}{d E_e} + (1 - \alpha) \frac{1}{{\overline \Gamma}_N} \frac{ d {\overline \Gamma}^{\rm (LNC)}(N \to e^{\mp} \mu^{\pm} \nu_{\mu})}{d E_e}
  \label{dBrbar}
  \eea
  where
  \bes
  \label{alpha}
  \bea
  \alpha^{\rm (Maj.)} & = & \frac{ |U_{\mu N}|^2 }{(|U_{\mu N}|^2  +|U_{e N}|^2 )} \ ,
  \label{alphaMaj}
  \\
  \alpha^{\rm (Dir.)} & = & 0 \ .
  \label{alphaDir}
  \eea
  \ees
  The LNC and LNV hadronic decays of $N$ are shown in Figs.~\ref{FigNLNCV},
for the general cases $N \to \ell_2^- \ell_3^+ \nu$. In the considered specific case of Eqs.~(\ref{diff1}), we have $\ell_2 = e$ and $\ell_3 = \mu$.
  The explicit expressions for $d \bG/d E_e$ for LNC and LNV decay of $N$ are given in Appendix \ref{appdiff}.
  The differential decay width (\ref{diff1}) is then rewritten in terms of the above canonical differential branching ratio as
  \bes
     \label{diff2}
  \bea
 \frac{d \Gamma^{\rm (Dir.)}}{d E_e} \left( B_{(c)} \to \left( D^{(*)} \right) \mu^{\pm} \mu^{\pm} e^{\mp} \nu_{\mu} \right)
 & = &
 \frac{|U_{\mu N}|^2 |U_{e N}|^2 }{{\K}}
      {\overline \Gamma} \left( B_{(c)} \to \left( D^{(*)} \right) \mu^{\pm} N \right)
      \frac{ d {\overline {\rm Br}}_N(N \to \mu e \nu ; \alpha=0)}{d E_e},
      \label{diff2Dir}
\\
      \frac{d \Gamma^{\rm (Maj.)}}{d E_e} \left( B_{(c)} \to \left( D^{(*)} \right) \mu^{\pm} \mu^{\pm} e^{\mp} \nu \right) & = &
 \frac{|U_{\mu N}|^2}{{\K}} (|U_{\mu N}|^2  +|U_{e N}|^2 )
      {\overline \Gamma} \left( B_{(c)} \to \left( D^{(*)} \right) \mu^{\pm} N \right)
      \frac{ d {\overline {\rm Br}}_N(N \to \mu e \nu ; \alpha^{\rm (Maj.)}) }{d E_e} .
      \nonumber\\
      \label{diff2Maj}
      \eea
 \ees
\begin{figure}[htb]
\begin{minipage}[b]{.49\linewidth}
\centering\includegraphics[width=\linewidth]{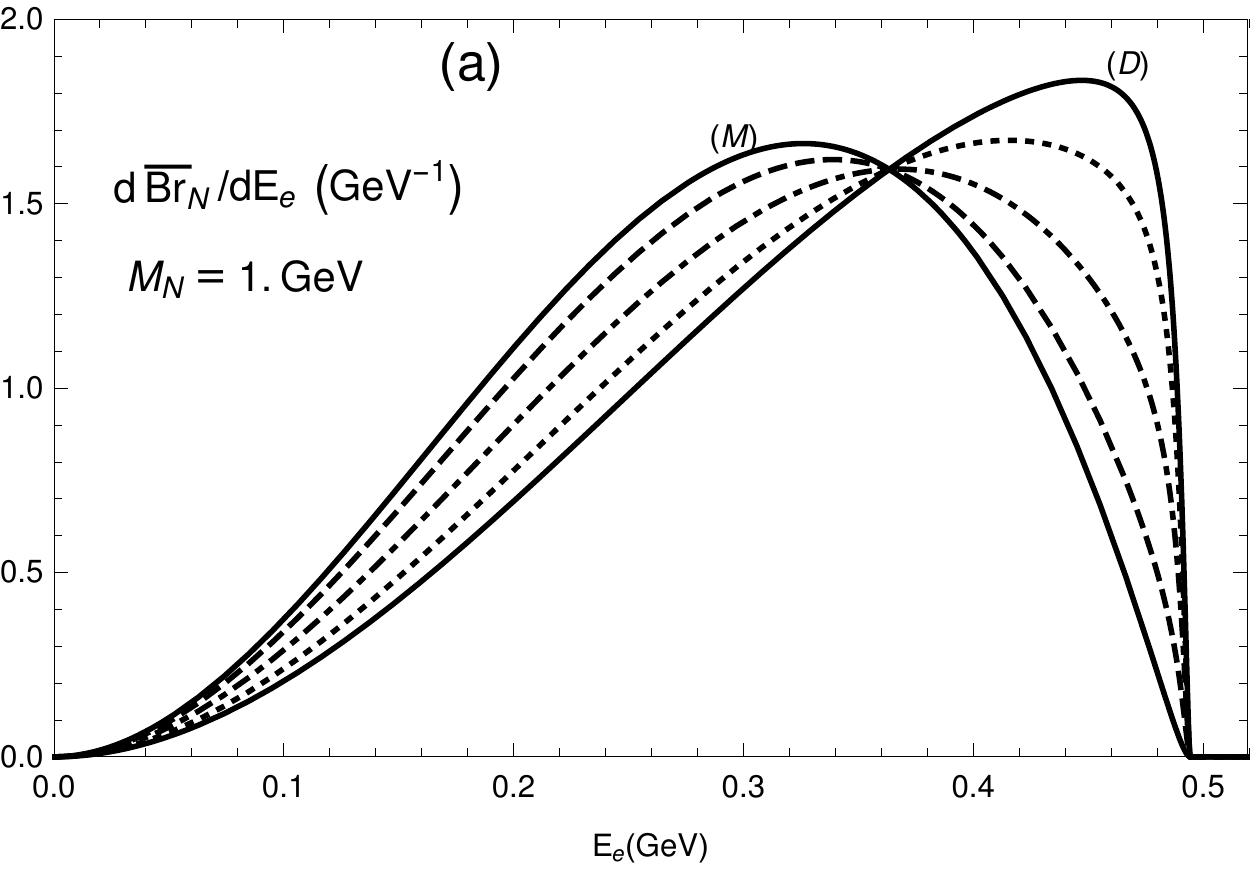}
\end{minipage}
\begin{minipage}[b]{.49\linewidth}
\centering\includegraphics[width=\linewidth]{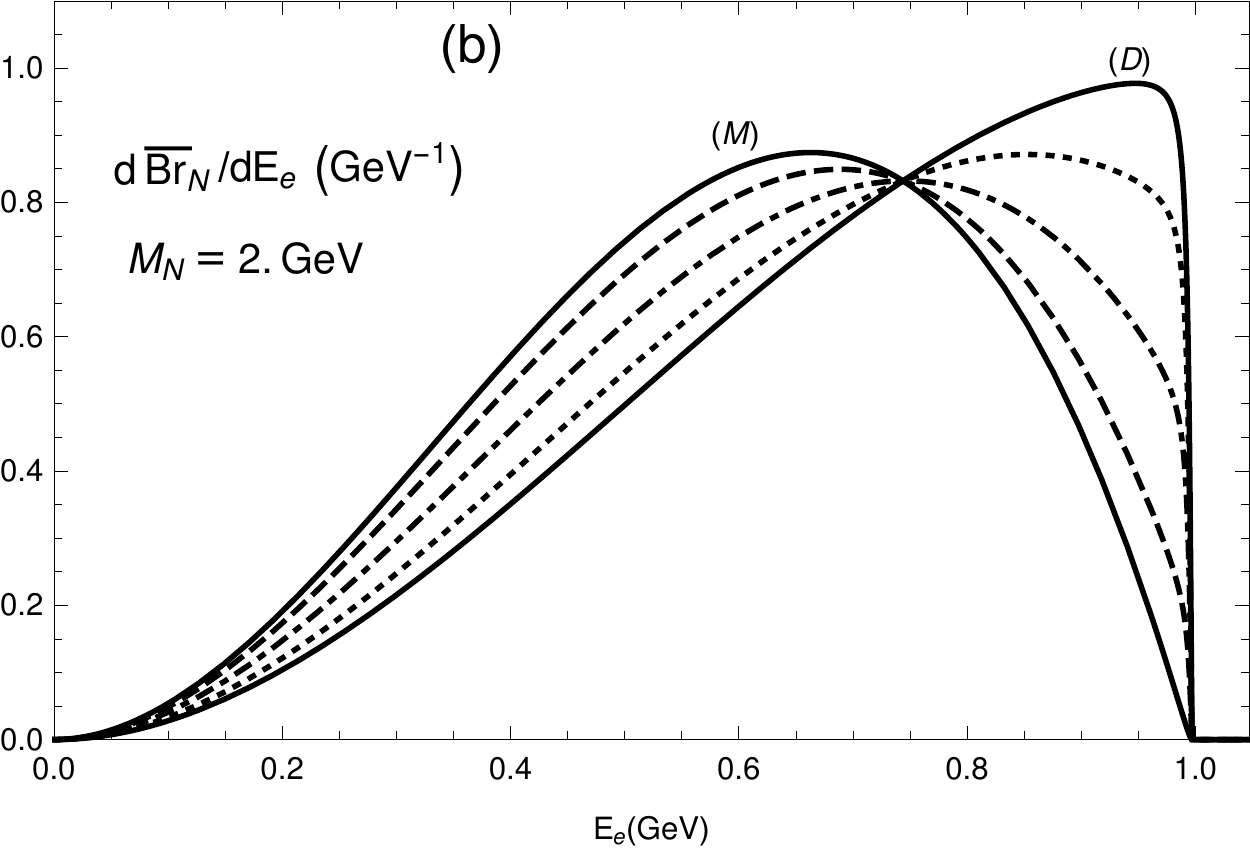}
\end{minipage} \vspace{6pt}
\begin{minipage}[b]{.49\linewidth}
\centering\includegraphics[width=\linewidth]{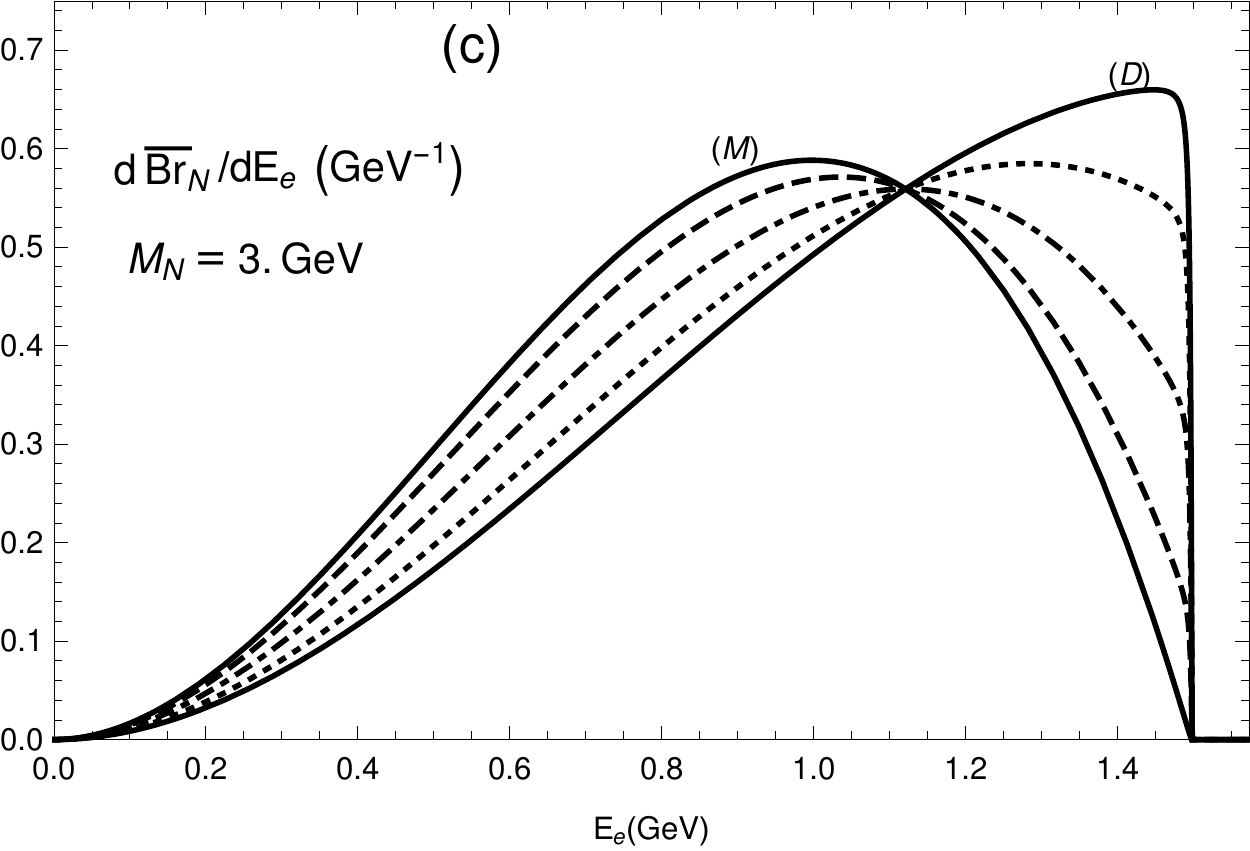}
\end{minipage}
\begin{minipage}[b]{.49\linewidth}
\centering\includegraphics[width=\linewidth]{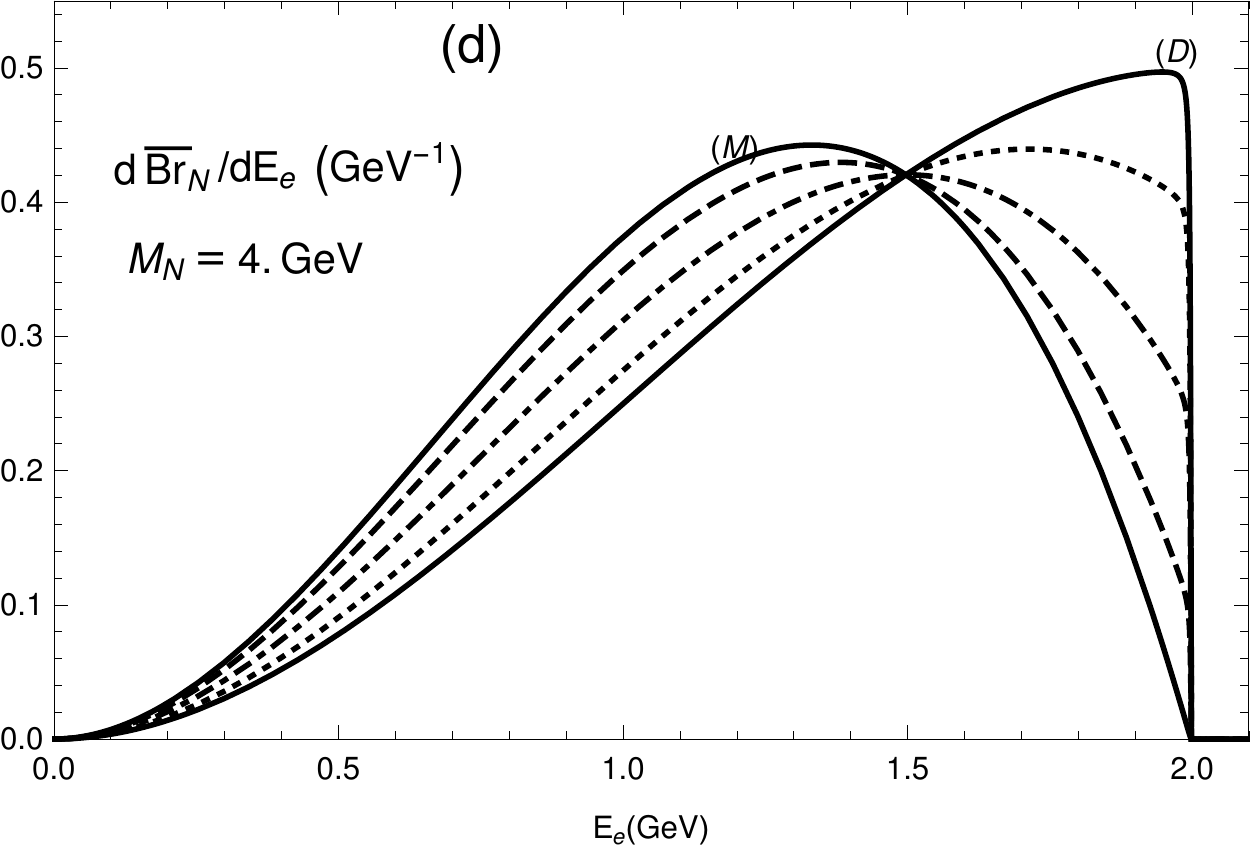}
\end{minipage}
\caption{The canonical differential branching ratio
  $d {\overline {\rm Br}}_{N}(\alpha)/d E_{e}$, Eq.~(\ref{dBrbar}),
  as a function of the electron energy in the neutrino $N$ rest frame,
  relevant for the decays
  $B_{(c)} \to   (D^{(*)}) \mu^{\pm} N \to (D^{(*)}) \mu^{\pm} \mu^{\pm} e^{\mp} \nu$, for various on-shell neutrino
  masses: (a) $M_N=1$ GeV, (b) $M_N=2$ GeV, (c) $M_N=3$ GeV, (d) $M_N=4$ GeV.
  In each figure there are five curves, corresponding to different values of the admixture parameter $\alpha$ [Eqs.~(\ref{dBrbar})-(\ref{alpha})]:
  $\alpha_M=1.0$ is the solid (M) curve; $0.8$ (dotted); $0.5$ (dot-dashed); $0.2$ (dashed); the Dirac case, $\alpha_M=0$ is the solid line labelled (D).}
\label{FigdBrNdE2}
\end{figure}
\begin{figure}[htb]
\begin{minipage}[b]{.49\linewidth}
\centering\includegraphics[width=\linewidth]{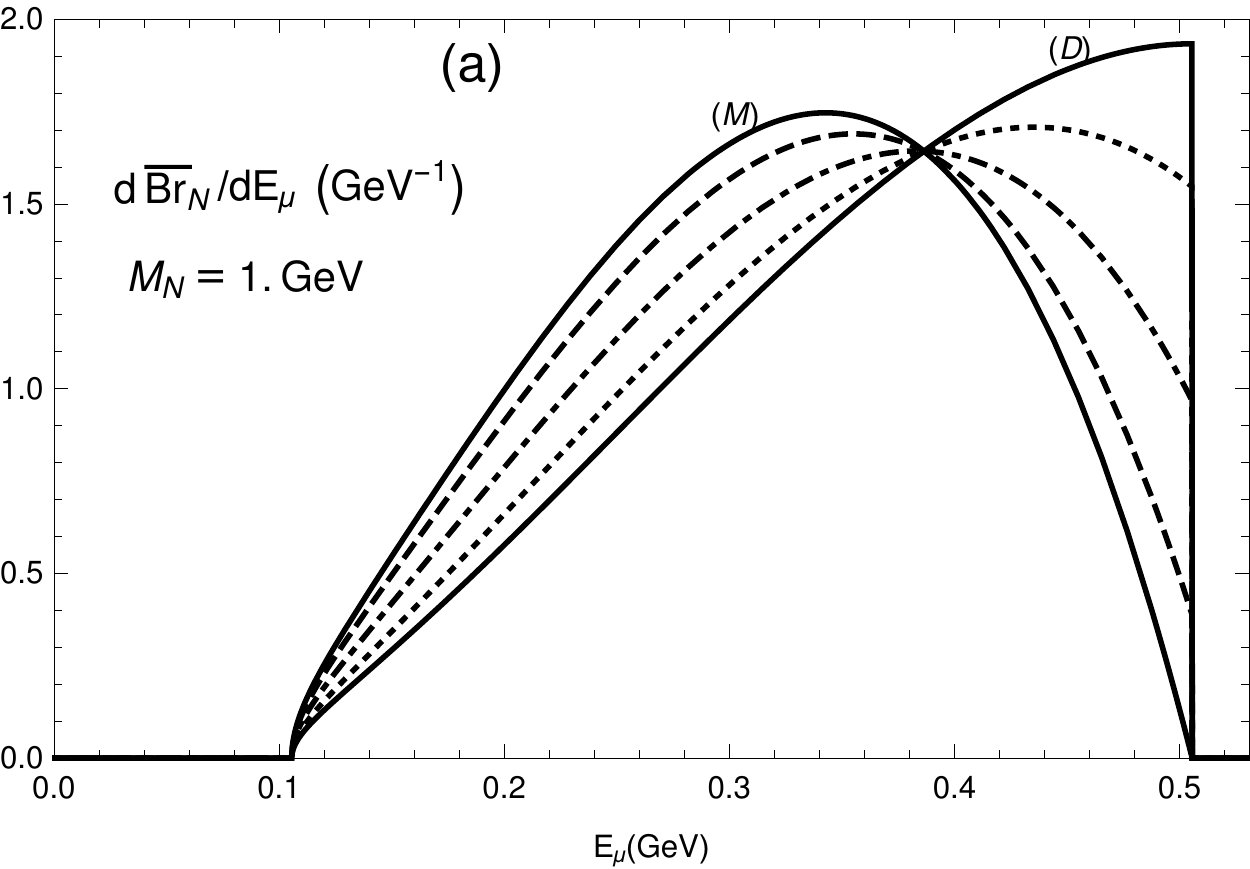}
\end{minipage}
\begin{minipage}[b]{.49\linewidth}
\centering\includegraphics[width=\linewidth]{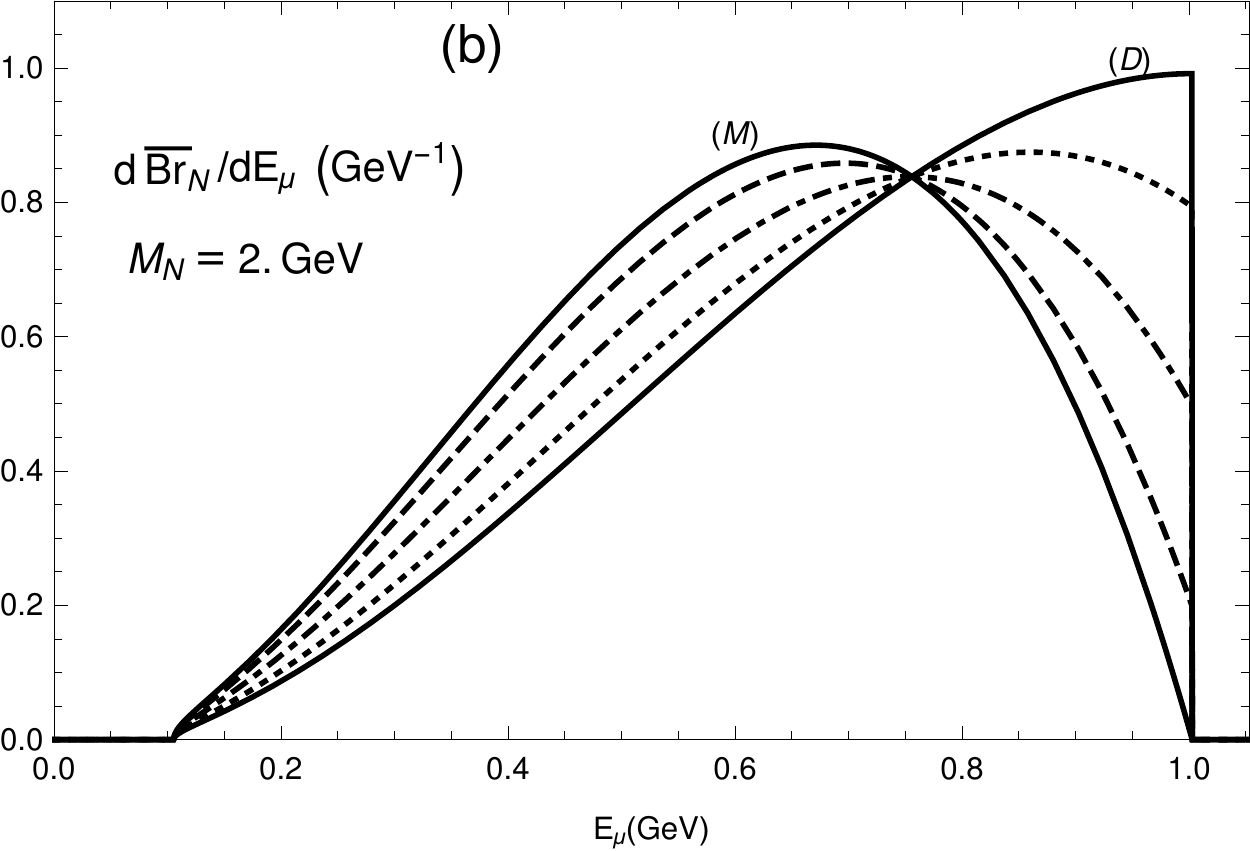}
\end{minipage} \vspace{6pt}
\begin{minipage}[b]{.49\linewidth}
\centering\includegraphics[width=\linewidth]{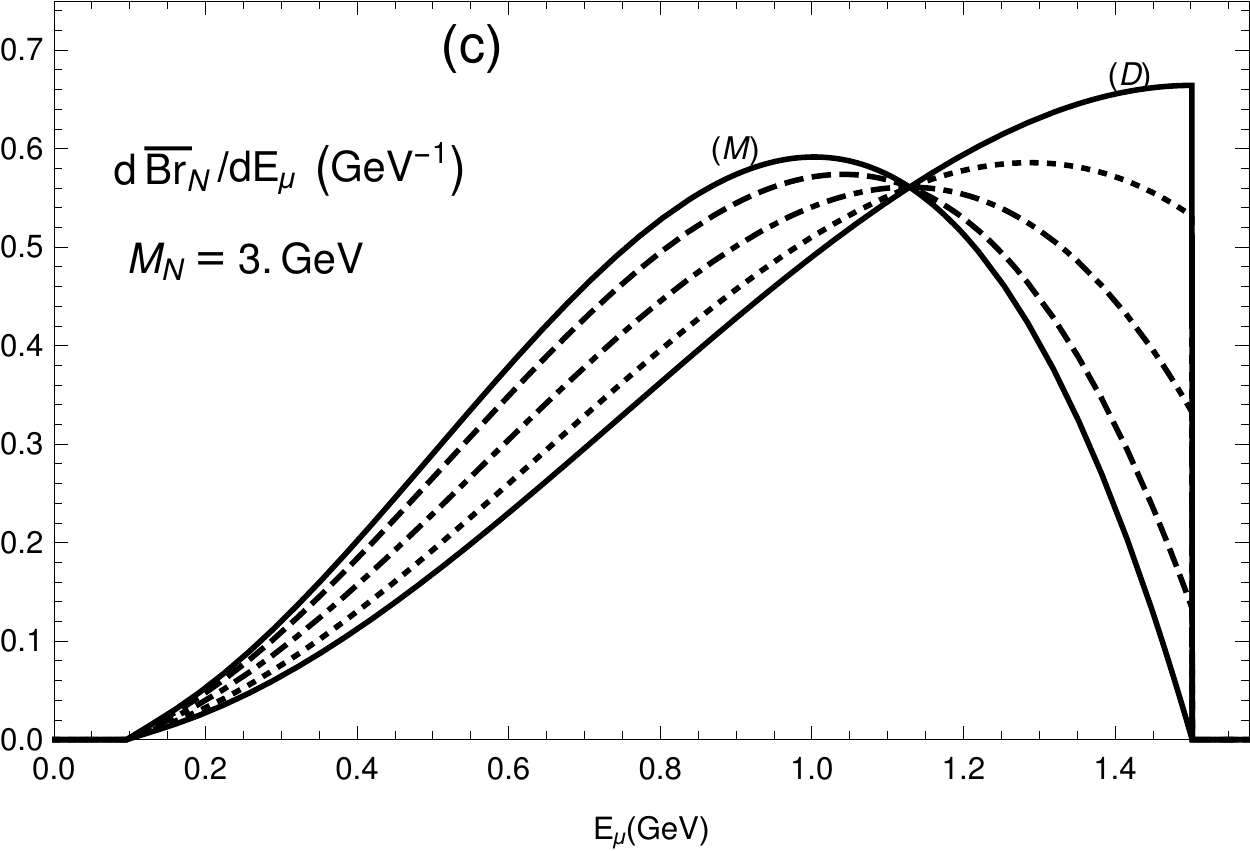}
\end{minipage}
\begin{minipage}[b]{.49\linewidth}
\centering\includegraphics[width=\linewidth]{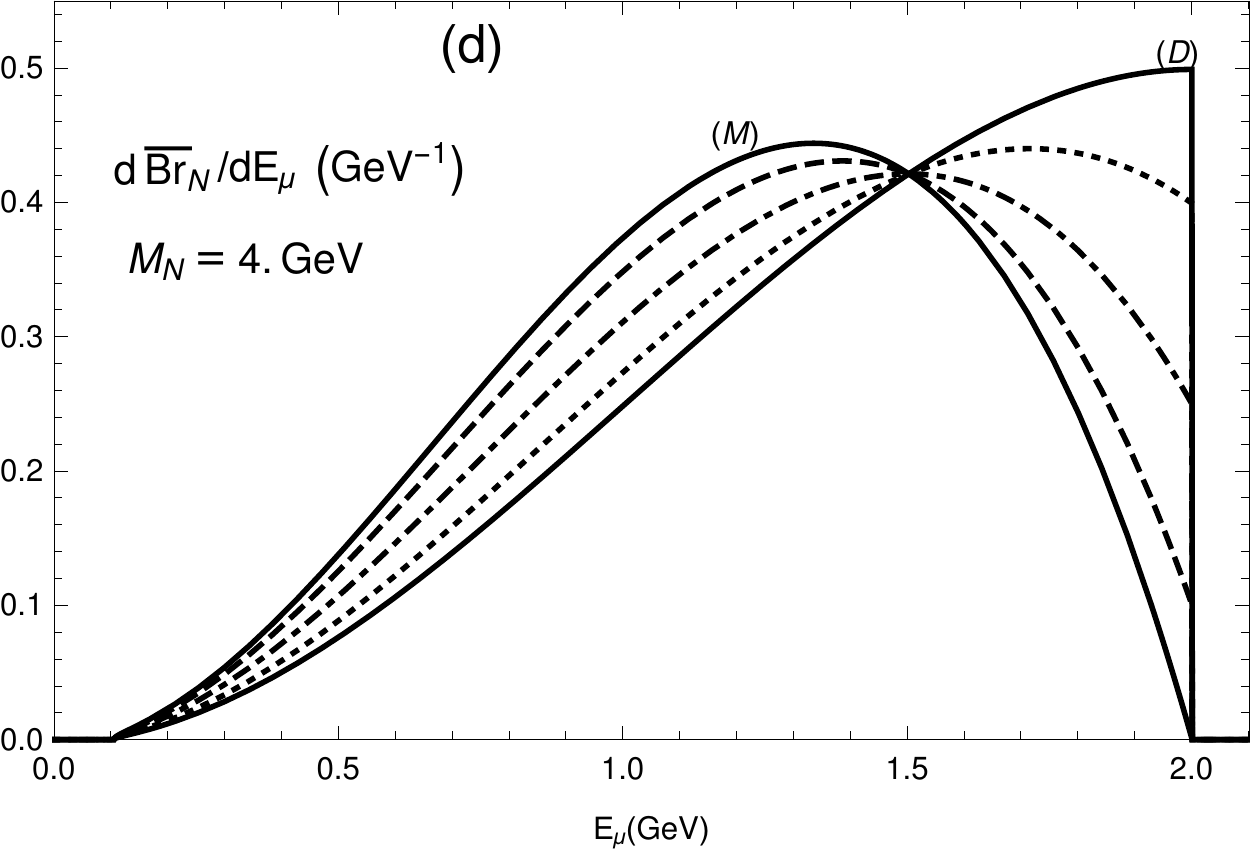}
\end{minipage}
\caption{The same as in Fig.~\ref{FigdBrNdE2}, but now for the decays with $e \leftrightarrow \mu$, i.e., the
canonical differential branching ratio
  $d {\overline {\rm Br}}_{N}(\alpha)/d E_{\mu}$
  as a function of the muon energy in the neutrino $N$ rest frame,
  relevant for the decays
  $B_{(c)} \to   (D^{(*)}) e^{\pm} N \to (D^{(*)}) e^{\pm} e^{\pm} \mu^{\mp} \nu$.}
\label{FigdBrNdE2inv}
\end{figure}
 In Figs.~\ref{FigdBrNdE2}(a)-(d) we present the results for the differential branching ratios (\ref{dBrbar}), i.e., the processes depicted in Figs.~\ref{FigNLNCV} with $\ell_2=e$ and $\ell_3=\mu$, for four different values of $M_N$ ($= 1, 2, 3, 4$ GeV, respectively), for various values of $\alpha = 1.0, 0.8, 0.5, 0.2$ and $\alpha=0$, where $\alpha=0$ is the case of Dirac. We can see clearly differences in the form of the differential branching ratios when $N$ is Dirac and when it is Majorana. If we consider the differential decay rates of $d {\overline {\rm Br}}_N/d E_{\mu}$ for the decays $B_{(c)} \to ( D^{(*)} ) e^{\pm} e^{\pm} \mu^{\mp} \nu$ (i.e., the processes of Figs.~\ref{FigNLNCV} with $\ell_2=\mu$ and $\ell_3=e$), the curves turn out to be very close to those presented in Figs.~\ref{FigdBrNdE2}(a)-(d).
In Figs.~\ref{FigdBrNdE2inv}(a)-(d) we present the analogous differential branching ratios, but now for the process with $e$ and $\mu$ interchanged: $d {\overline {\rm Br}}_{N}(\alpha)/d E_{\mu}$ relevant for the decays $B_{(c)} \to \to  (D^{(*)}) e^{\pm} N \to (D^{(*)}) e^{\pm} e^{\pm} \mu^{\mp} \nu$.

\section{Effective branching ratios due to long lifetime of $N$}
\label{sec:effBr}

For the considered decays to be measured in the experiment, the produced on-shell neutrino $N$ must decay within the detector. However, if the sterile neutrino $N$ is long-lived, only a small fraction of the produced neutrinos $N$ will decay within the detector. Therefore, their theoretical branching ratios should be multiplied by the probability $P_N$ of the produced neutrinos $N$ to decay (nonsurvival) within the detector. This effect has been discussed in the context of various processes in Refs.~\cite{CDK,scatt3,CKZ,CKZ2,CERN-SPS,Gronau,commKim,symm}. If the length of the detector is $L$, and the velocity of the on-shell $N$ in the lab frame is $\beta_N$ (often $\beta_N \approx 1$), this nonsurvival probability is
\bes
\label{PN}
\bea
P_N &=& 1 - \exp \left[ - \frac{L}{\tau_N \gamma_N \beta_N} \right]
\label{PNa}
\\
&\approx & \frac{L}{(\tau_N \gamma_N \beta_N)} \equiv \frac{L}{(\gamma_N \beta_N)} \Gamma_N  = \K \frac{L}{1 \ {\rm m}} {\overline P}_N
 \ ,
\label{PNb}
\eea
\ees
where
\be
{\overline P}_N = \frac{1 \ {\rm m}}{(\gamma_N \beta_N)} \bG_N =
\frac{1 \ {\rm m}}{(\gamma_N \beta_N)}  \frac{G_F^2 M_{N}^5}{96 \pi^3}
\label{bPN}
\ee
is the canonical nonsurvival probability, i.e., $P_N$ with $\K \mapsto 1$ and $L = 1 \ {\rm m}$. Here, $\gamma_N=(1 - \beta_N^2)^{-1/2}$ is the Lorentz time dilation factor (in the lab frame), and in Eq.~(\ref{PNb}) we assumed that $P_N$ is significantly smaller than 1, say $P_N < 0.4$. We refer to Sec.~\ref{subs:GN} for details on the total decay width $\Gamma_N$.

\begin{figure}[htb]
\centering\includegraphics[width=90mm]{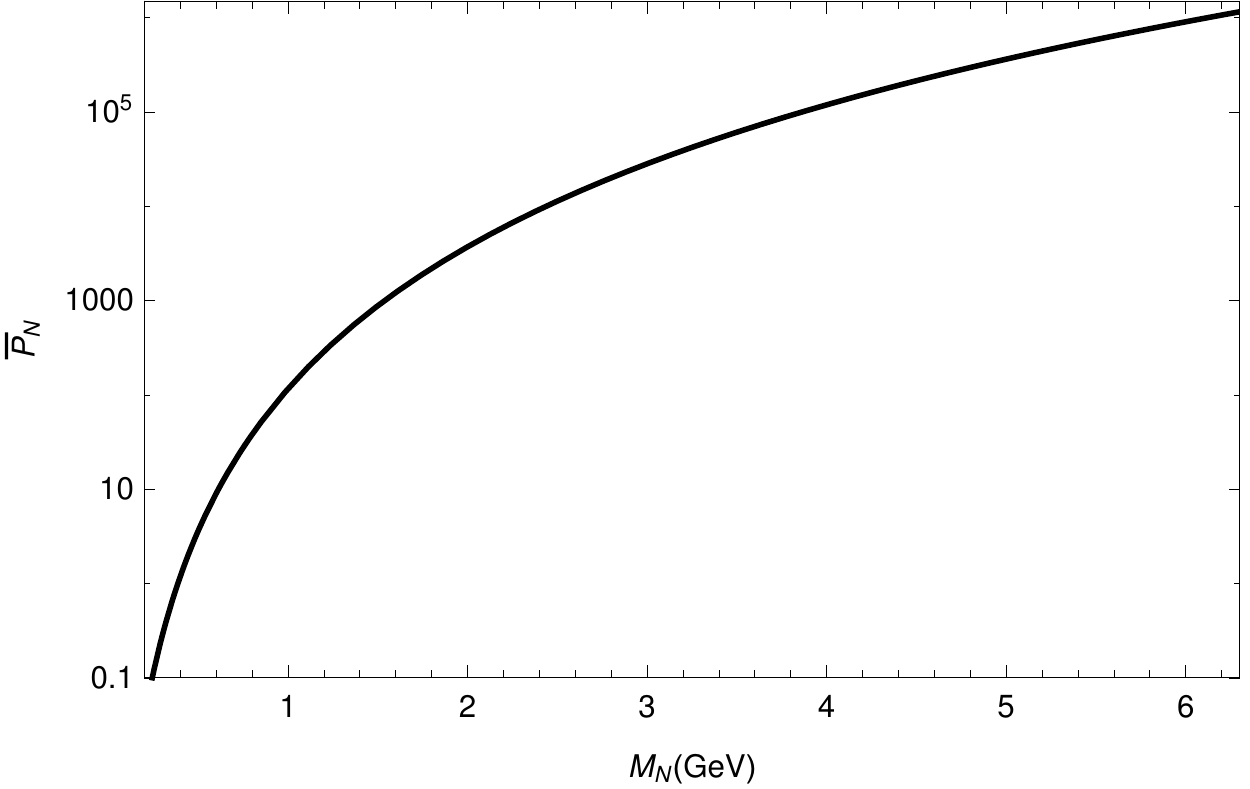}
\caption{The canonical nonsurvival probability ${\overline P}_N$, Eq.~(\ref{bPN}), as a function of mass $M_N$, for the kinematic parameter $\gamma_N \beta_N=2$.}
\label{FigbPN}
\end{figure}

In some cases it is realistic to assume that $P_N \ll 1$ (e.g., $P_N < 0.4$), because the total decay width of the sterile $N$ neutrino, $\Gamma_N$, is proportional to $\K$ which is a linear combination of the (small) heavy-light mixing coefficients $|U_{\ell N}|^2$ ($\ell = e, \mu,\tau$), cf.~Eqs.~(\ref{GNwidth})-(\ref{calKest}) in Sec.~\ref{subs:GN}. Nonetheless, we should check in each considered case of mass $M_N$ whether or not this condition is fulfilled. If it is not, the relevant estimates of the measured branching ratios are the (original) branching ratios presented in Sec.~\ref{sec:Br}, cf.~Figs.~\ref{FigbBrB}-\ref{FigbBrBDPi}. In order to facilitate the checking of this condition, for a given mass $M_N$, we present in Fig.~\ref{FigbPN} the canonical nonsurvival probability ${\overline P}_N$, Eq.~(\ref{bPN}), as a function of mass $M_N$, for the kinematic parameter $\gamma_N \beta_N=2$.

The branching ratio (\ref{Br}) is then multiplied by the nonsurvival probability $P_N$, Eq.~(\ref{PNb}), resulting in the
experimentally measured (effective) branching ratio,\footnote{Please note that the true branching ratio can be
derived from the effective branching ratio by dividing it  by the nonsurvival  factor $P_N$.}
where the total decay width $\Gamma_N$ of the on-shell $N$ neutrino cancels if $P_N \ll 1$
\bes
\label{Breff}
\bea
 {\rm Br}_{\rm eff} \left( B_{(c)} \to (D^{(*)}) \ell_1^{\pm} N \to  (D^{(*)}) \ell_1^{\pm} X Y \right) & \equiv &
 P_N {\rm Br} \left( B_{(c)} \to  (D^{(*)}) \ell_1^{\pm} X Y \right)
 \label{Breffa}
 \\
 & \approx &  \frac{L}{(\gamma_N \beta_N)} \frac{1}{\Gamma_{B_{(c)}}} \Gamma \left( B_{(c)} \to (D^{(*)}) \ell_1^{\pm} N \right) \Gamma(N \to XY)  \ .
\label{Breffb}
\eea
\ees
Eq.~(\ref{Breffb}) is a good approximation to the true value if $P_N < 0.4$ ($1 - e^{-0.4} \approx 0.33$).
We define the canonical effective branching ratios ${\overline {\rm Br}}_{\rm eff}$, containing no heavy-light neutrino mixing factors, as the same expression as Eq.~(\ref{Breffb}), except that now instead of the decay widths $\Gamma$ the canonical decay widths ${\overline \Gamma}$ appear
\be
   {\overline {\rm Br}}_{\rm eff} \left( B_{(c)} \to (D^{(*)}) \ell_1^{\pm} N \to  (D^{(*)}) \ell_1^{\pm} X Y \right) \equiv
 \frac{L}{(\gamma_N \beta_N)} \frac{1}{\Gamma_{B_{(c)}}} {\overline \Gamma} \left( B_{(c)} \to (D^{(*)}) \ell_1^{\pm} N \right) {\overline \Gamma}(N \to XY)  \ .
\label{barBreff}
\ee
We stress that our definition of the canonical branching ratio ${\overline {\rm Br}}_{\rm eff}$ uses the form (\ref{PNb}) as the basis, i.e., it is simply related with the effective branching ratio ${\rm Br}_{\rm eff}$ Eq.~(\ref{Breffa}) only when $P_N \ll 1$ (say, $P_N < 1$).
The widths ${\overline \Gamma}$ were calculated in the previous Sections, cf.~Eqs.~(\ref{bGBlN}), (\ref{GBDNlb}), (\ref{bGBDstNl}) and Figs.~\ref{FigbGBmuN} and \ref{FigbGBDlN} for the first part $B_{(c)} \to (D^{(*)}) \ell_1^{\pm} N$, and Eqs.~(\ref{bGNllnu}) and (\ref{bGNlPi}) and Figs.~\ref{FigbGNXY} for the second part $N \to \ell_2^{\mp} \ell_3^{\pm} \nu$ or $N \to \ell_2^{\mp} \pi^{\pm}$.

The effective branching ratios, ${\rm Br}_{\rm eff}$, in terms of the canonical branching ratios ${\overline {\rm Br}}_{\rm eff}$, are in the case of leptonic decay of $N$ [cf.~Eqs.~(\ref{noDnoPi}) and (\ref{cBr})]\footnote{
We usually have $\ell_j = e, \mu$ ($j=1,2,3$). As explained in Sec.~\ref{subs:bBr}, we take $\ell_1=\ell_3$, because $\ell_1=\ell_2$ ($\equiv \ell$) would imply pairs $\ell^+ \ell^-$ which have strong QED background.}
\bes
\label{cBreff}
\bea
 {\rm Br}_{\rm eff}^{\rm (Dir.)} \left( B_{(c)} \to (D^{(*)}) \ell_1^{\pm} N \to (D^{(*)}) \ell_1^{\pm} \ell_1^{\pm} \ell_2^{\mp} \nu_{\ell_1} \right) & \approx &
 |U_{\ell_1 N}|^2 |U_{\ell_2 N}|^2  {\overline {\rm Br}}_{\rm eff} \left( B_{(c)} \to (D^{(*)}) \ell_1^{\pm} \ell_1^{\pm} \ell_2^{\mp} \nu_{\ell_1} \right) \ ,
 \label{cBreffDir}
\\
 {\rm Br}_{\rm eff}^{\rm (Maj.)} \left( B_{(c)} \to (D^{(*)}) \ell_1^{\pm} N \to (D^{(*)}) \ell_1^{\pm} \ell_1^{\pm} \ell_2^{\mp} \nu \right) & \approx &
 |U_{\ell_1 N}|^2 (|U_{\ell_1 N}|^2+|U_{\ell_2 N}|^2) {\overline {\rm Br}}_{\rm eff} \left( B_{(c)} \to (D^{(*)}) \ell_1^{\pm} \ell_1^{\pm} \ell_2^{\mp} \nu \right).
\nonumber\\
 \label{cBreffMaj}
 \eea
 \ees
 The cases of Majorana and Dirac neutrino $N$ differ in Eqs.~(\ref{cBreff}), as they do in Eqs.~(\ref{cBr}). The explanation for this was given just after Eqs.~(\ref{cBr}).
Here, $\ell_1$ and $\ell_2$ are usually $\mu$ and/or $e$, but could be in principle also $\tau$. We stress that the relations (\ref{cBreff}) are approximate and are applicable only if $P_N \ll 1$ (say, $P_N < 0.4$). This is so because the definition of the effective branching ratio ${\rm Br}_{\rm eff}$, Eq.~(\ref{Breffa}), has the true value of the nonsurvival probability $P_N$ as a factor, Eq.~(\ref{PNa}), while the canonical quantity ${\overline {\rm Br}}_{\rm eff}$, Eq.~(\ref{barBreff}), uses the expression (\ref{PNb}) which reduces to the true $P_N$ only when $P_N \ll 1$ (say, $P_N < 0.4$). We note that the right-hand sides of Eqs.~(\ref{cBreff}) involve factors $\sim |U_{\ell N}|^4$, in contrast to the factors $\sim |U_{\ell N}|^2$ on the right-hand side of the analogous relations (\ref{cBr}). This is so because when the factor $P_N$ is small, it is proportional to $\Gamma_N \propto \K \sim |U_{\ell N}|^2$.

If $N$ decays semileptonically ($N \to \ell_2 \pi$), the relations between
  ${\rm Br}_{\rm eff}$ and ${\overline {\rm Br}}_{\rm eff}$ are somewhat simpler
 \bea
\label{cBreffPi}
  {\rm Br}_{\rm eff} \left( B_{(c)} \to (D^{(*)}) \ell_1 N \to (D^{(*)}) \ell_1 \ell_2 \pi \right) & = &
 |U_{\ell_1 N}|^2 |U_{\ell_2 N}|^2  {\overline {\rm Br}}_{\rm eff}  \left(B_{(c)} \to (D^{(*)}) \ell_1 \ell_2 \pi \right) \ .
 \eea
Here we consider that $\ell_1$ and $\ell_2$ have specific flavors and specific electric charges.

\begin{figure}[htb] 
\begin{minipage}[b]{.49\linewidth}
\centering\includegraphics[width=85mm]{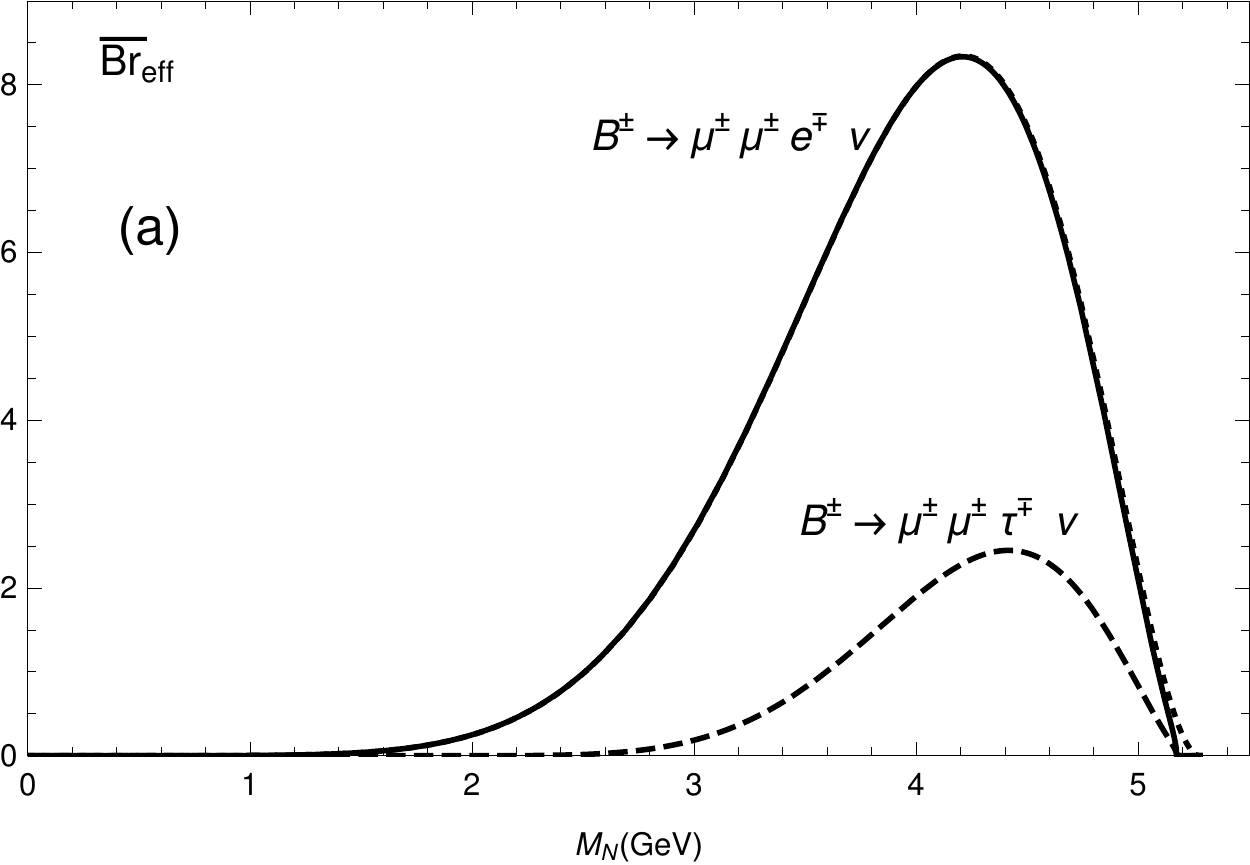}
\end{minipage}
\begin{minipage}[b]{.49\linewidth}
\centering\includegraphics[width=85mm]{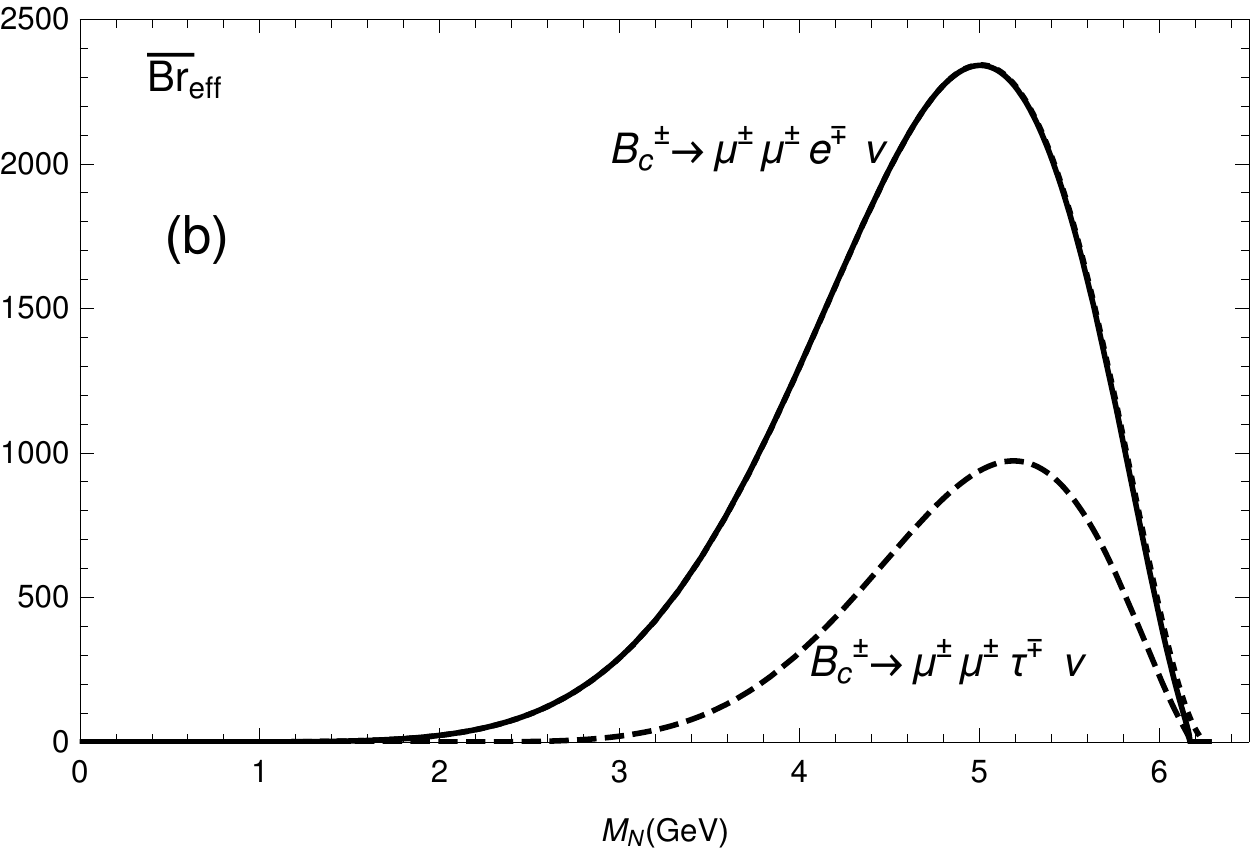}
\end{minipage}
\vspace{-0.2cm}
\caption{The effective canonical branching ratio  ${\overline {\rm Br}}_{\rm eff}$, Eq.~(\ref{barBreff}), as a function of mass of the on-shell neutrino $N$, for the leptonic decays (a) $B^{\pm} \to \mu^{\pm} \mu^{\pm} \ell^{\mp} \nu$, (b) $B_c^{\pm} \to \mu^{\pm} \mu^{\pm} \ell^{\mp} \nu$, where $\ell = e$ (solid) and $\ell=\tau$ (dashed). The length of the detector was taken $L=1$ m, and the Lorentz factor of $N$ neutrino in the lab frame, $\beta_N \gamma_N \equiv \beta_N /\sqrt{1 - \beta_N^2}$, is taken to be $\beta_N \gamma_N=2$, and $P_N \ll 1$ was assumed. Included is also the curve for $B_{(c)}^{\pm} \to e^{\pm} e^{\pm} \mu^{\mp} \nu$ (dotted), which is almost indistinguishable from the solid curve.}
\label{FigbBrefB}
 \end{figure}

\begin{figure}[htb] 
\begin{minipage}[b]{.49\linewidth}
\centering\includegraphics[width=85mm]{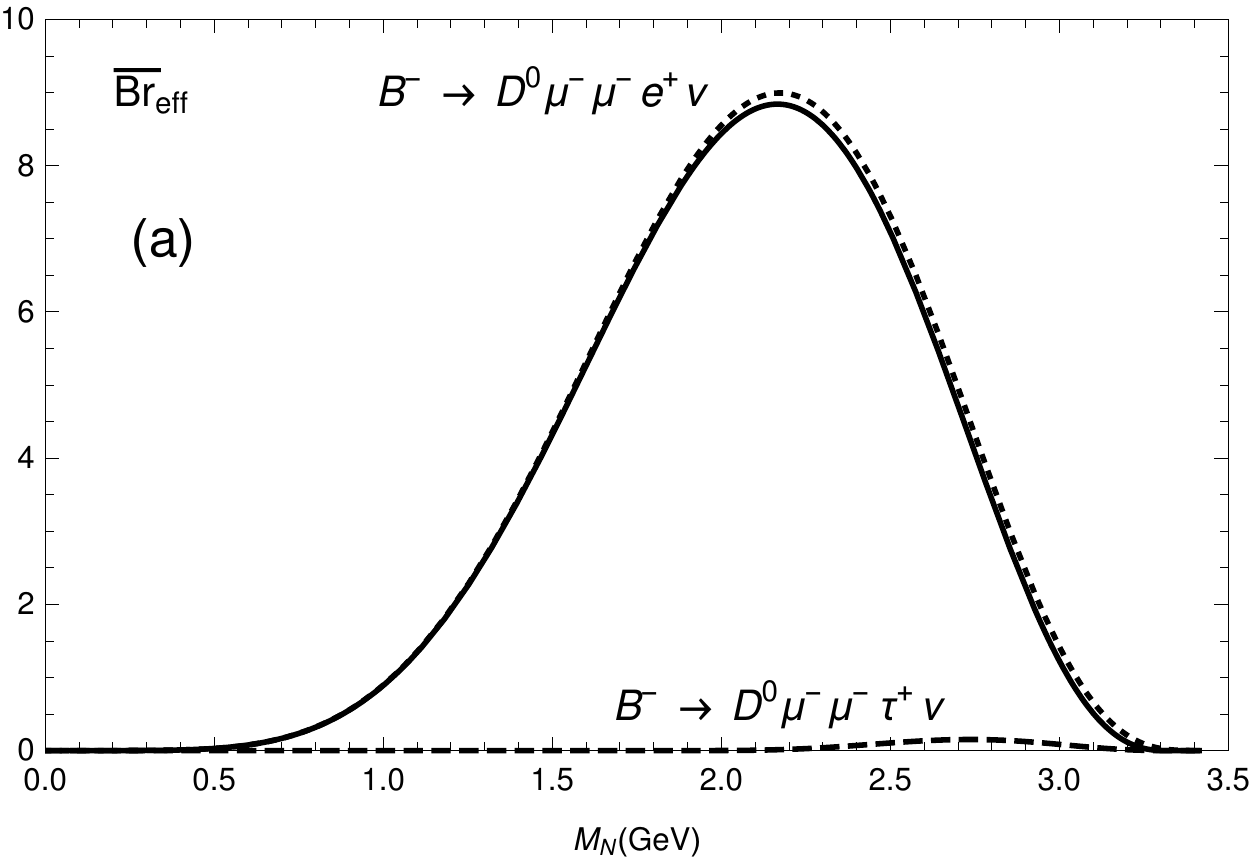}
\end{minipage}
\begin{minipage}[b]{.49\linewidth}
\centering\includegraphics[width=85mm]{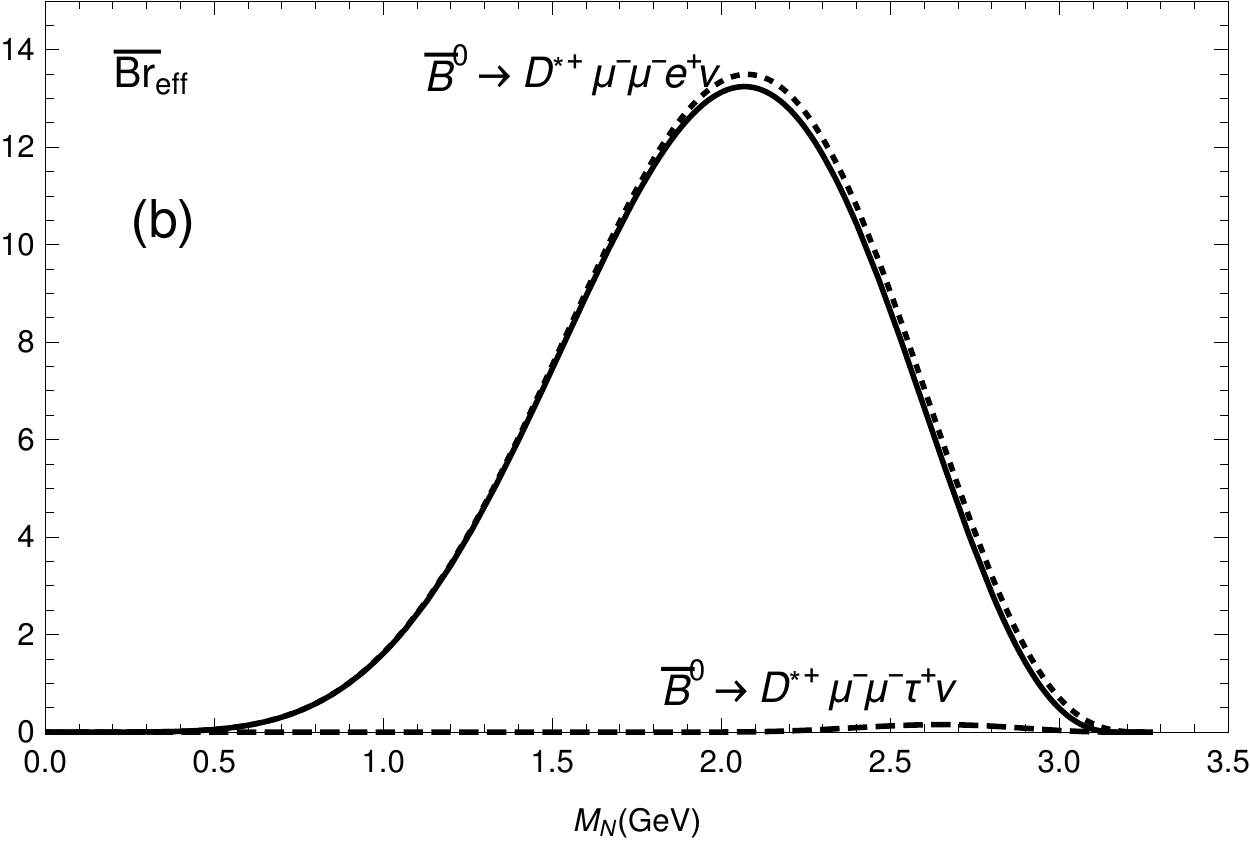}
\end{minipage}
\vspace{-0.2cm}
\caption{The effective canonical branching ratio, as a function of mass of the on-shell neutrino $N$, for the decays (a) $B^{-} \to D^0 \mu^{-} \mu^{-} \ell^+ \nu$, (b) ${\bar B}^0 \to D^{*+} \mu^{-} \mu^{-} \ell^{+} \nu$, where $\ell = e$ (solid) and $\ell=\tau$ (dashed). As in Fig.~\ref{FigbBrefB}, we took $L=1$ m and $\beta_N \gamma_N=2$, and $P_N \ll 1$. The case of $\ell=\tau$ is kinematically strongly suppressed, due to the analogous suppression in Fig.~\ref{FigbGNXY}(a). Included is also the curve for $B \to D^{*+} e^{-} e^{-} \mu^{+} \nu$ (dotted), which is close to the solid curve.}
\label{FigbBrefBD}
 \end{figure}

\begin{figure}[htb] 
\begin{minipage}[b]{.49\linewidth}
\centering\includegraphics[width=85mm]{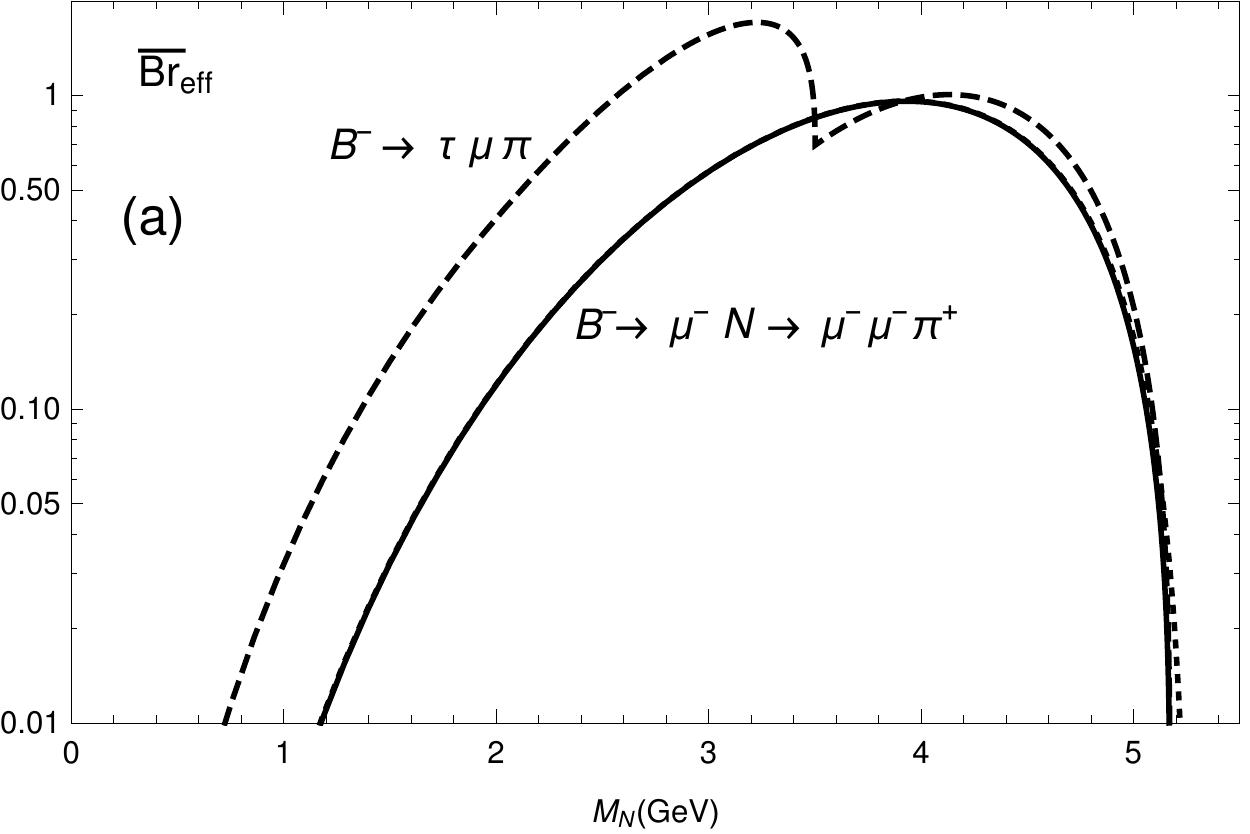}
\end{minipage}
\begin{minipage}[b]{.49\linewidth}
\centering\includegraphics[width=85mm]{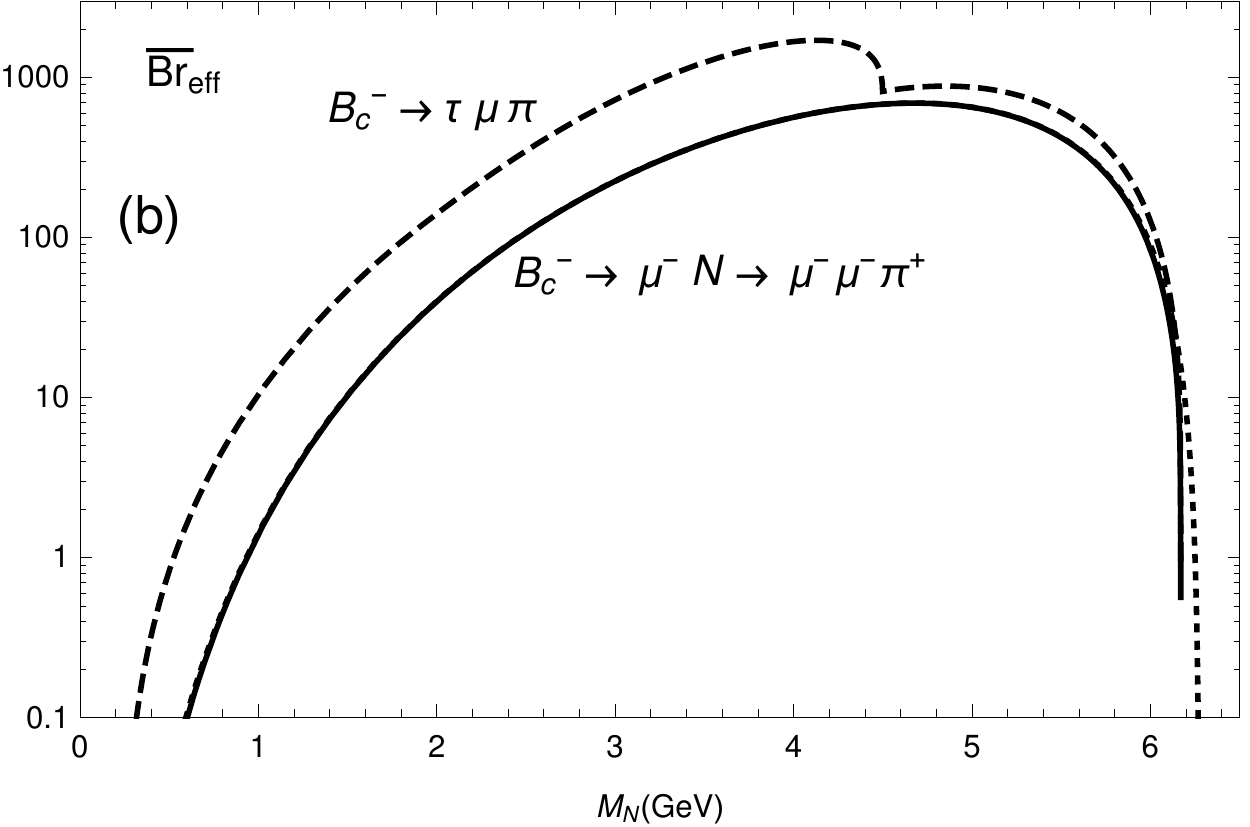}
\end{minipage}
\vspace{-0.2cm}
\caption{The effective canonical branching ratio, as a function of mass of the on-shell neutrino $N$, for the (LNV) decays (a) $B^- \to \mu^- N \to \mu^- \mu^- \pi^+$ (solid) and  $B^- \to e^- N \to e^- e^- \pi^+$ (dotted); (b) $B_c^- \to \mu^- N \to \mu^- \mu^- \pi^+$ (solid) and  $B_c^- \to e^- N \to e^- e^- \pi^+$ (dotted). The dotted curve is practically indistinguishable from the solid one. Included in (a), as a dashed line, is the canonical effective branching ratio for the decays $B^- \to \tau^- N \to \tau^- \mu^{\mp} \pi^{\pm}$ and $B^- \to \mu^- N \to \mu^- \tau^{\mp} \pi^{\pm}$ (sum of all four decays), and in (b) the analogous quantity for $B_c^-$.
  As in Fig.~\ref{FigbBrefB}, we took $L=1$ m and $\beta_N \gamma_N=2$, and $P_N \ll 1$. Logarithmic scale is used for better visibility.}
\label{FigbBrefBPi}
 \end{figure}

\begin{figure}[htb] 
\begin{minipage}[b]{.49\linewidth}
\centering\includegraphics[width=85mm]{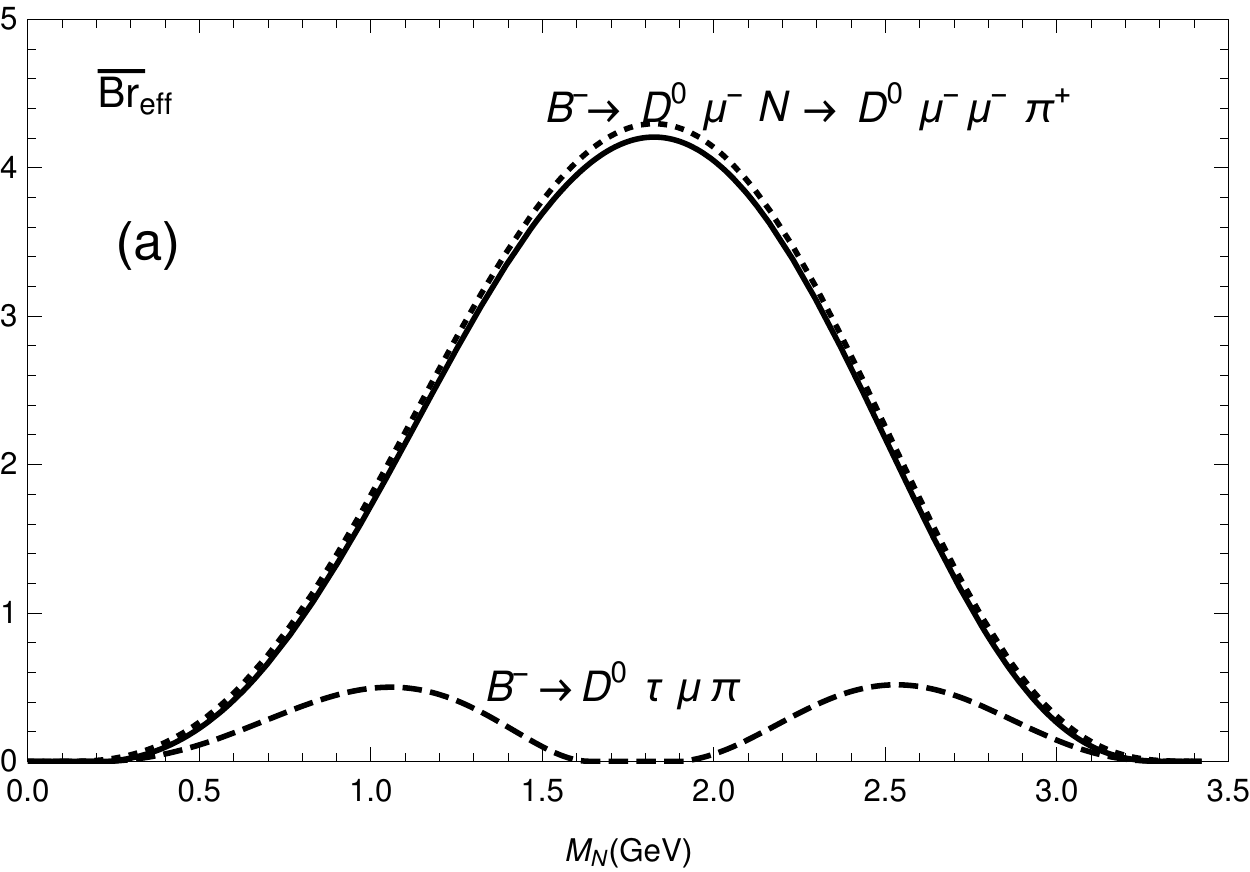}
\end{minipage}
\begin{minipage}[b]{.49\linewidth}
\centering\includegraphics[width=85mm]{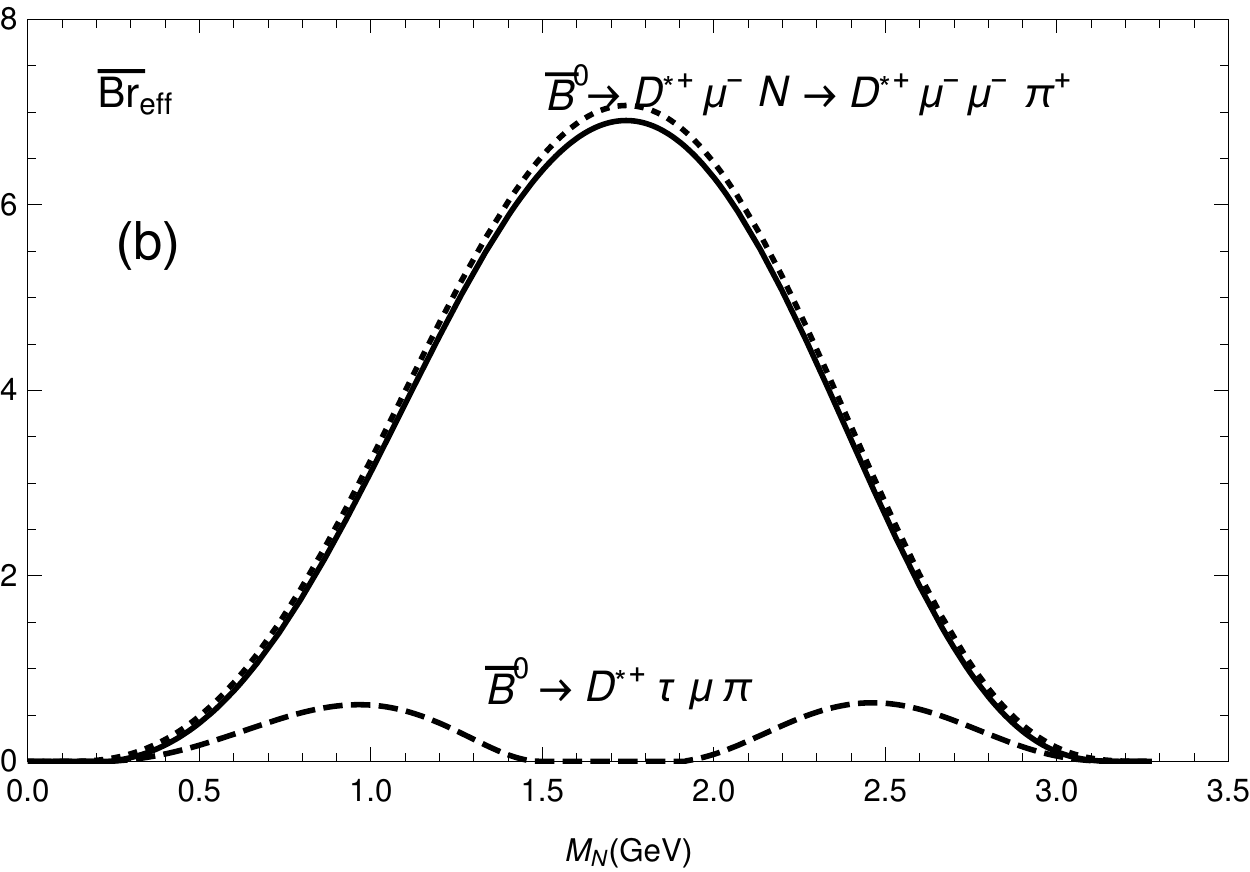}
\end{minipage}
\vspace{-0.2cm}
\caption{The effective canonical branching ratio, as a function of mass of the on-shell neutrino $N$, for the (LNV) decays (a) $B^-  \to D^0 \mu^- N \to D^0 \mu ^-\mu^- \pi^+$ (solid) and $B^-  \to D^0 e^- N \to D^0 e^- e^- \pi^+$ (dotted); (b) ${\bar B}^0 \to D^{*+} \mu^- N  \to D^{*+} \mu^- \mu^- \pi^+$ (solid) and  ${\bar B}^0 \to D^{*+} e^- N  \to D^{*+} e^- e^- \pi^+$ (dotted). Dotted curves are very close to the solid ones. Included in (a), as a dashed line, is the canonical effective branching ratio for the decays $B^- \to D^0 \tau^- N \to D^0 \tau^- \mu^{\mp} \pi^{\pm}$ and $B^- \to D^0 \mu^- N \to D^0 \mu^- \tau^{\mp} \pi^{\pm}$ (sum of all four decays), and in (b) the analogous quantity with $B^0$ and $D^{*+}$. As in Fig.~\ref{FigbBrefB}, we took $L=1$ m and $\beta_N \gamma_N=2$, and $P_N \ll 1$.}
\label{FigbBrefBDPi}
 \end{figure}

The canonical effective branching ratios (i.e., those without the heavy-light mixing coefficients), as a function of the mass of the on-shell neutrino $N$, for some representative considered $B_{(c)}$ meson decays are presented in Figs.~\ref{FigbBrefB}, \ref{FigbBrefBD}, \ref{FigbBrefBPi}, \ref{FigbBrefBDPi}. They are given for the values of the detector width $L=1$ m ($=5.068 \times 10^{15} \ {\rm GeV}^{-1}$) and the kinematic $N$ factor $\beta_N \gamma_N$ ($\equiv \beta_N /\sqrt{1 - \beta_N^2}$) $= 2$.
The results for representative decays  as a function of $M_N$, when the on-shell $N$ neutrino decays leptonically ($N \to \ell_2 \ell_3 \nu$), are presented in Figs.~\ref{FigbBrefB} and \ref{FigbBrefBD}. The results for the analogous decays, when $N$ decays semileptonically $N \to \mu \pi$, are presented in Figs.~\ref{FigbBrefBPi} and \ref{FigbBrefBDPi}.

\section{Discussion of the results and prospects of detection}
\label{sec:disc}

The results of the previous Secs.~\ref{sec:Br} and \ref{sec:effBr} allow us to estimate, for given mass $M_N$ and given values of the heavy-light mixing coefficients $|U_{\ell N}|^2$, the branching ratios of the considered rare decays of $B_{(c)}$ mesons. For example, if we consider $B$ mesons which are to be produced in Belle II experiment in numbers of $\sim 10^{10}$ per year, the rare $B$-decays may be detected there if their predicted measured branching ratios are $\agt 10^{-10}$. And if they are not detected, this will imply a decrease of the present upper bounds for the corresponding $|U_{\ell N}|^2$ coefficients in the considered mass range of $N$.

The present upper bounds for the corresponding mixing coefficients, in the mass range $0.1 \ {\rm GeV} < M_N < 6 \ {\rm GeV}$, were determined by various experiments, cf.~Refs.~\cite{0NuBB,PS191,beamdump,PiENu,DELPHI,L3,CHARMe,CHARMint,CHARMtau,KMuNu,NOMAD,NuTeV} (for a review, see, e.g., \cite{Atre}). Table \ref{t1} gives the present approximate upper bounds for $|U_{\ell N}|^2$ ($\ell=e, \mu, \tau$) for several masses of $N$ in the interval $0.1 \ {\rm GeV} \leq M_N \leq 6 \ {\rm GeV}$.
 \begin{table}
\small
\centering
\caption{The known upper bounds for the squares $|U_{\ell N}|^2$ of the
heavy-light mixing matrix elements, for various specific values of $M_N$. For $|U_{e N}|^2$, as an alternative, the upper bounds with the exclusion of the neutrinoless double beta decay data are included.}
\label{t1}
\begin{tabular}{| c | c | c | c |c|}
\hline
\bf{$M_N [GeV]$} & $|U_{eN}|^2$ & $|U_{eN}|^2$ (excl.$0\nu\beta\beta$) & $|U_{\mu N}|^2$  & $|U_{\tau N}|^2$ \\
\hline
0.1 & $1.5\times10^{-8} $\cite{0NuBB}& $5 \times 10^{-7}$ \cite{PiENu} & $6.0\times10^{-6}$ \cite{KMuNu} & $8.0\times10^{-4}$ \cite{CHARMtau} \\
\hline
0.3 & $2.5\times10^{-9}$ \cite{PS191} & $2 \times 10^{-9}$  \cite{PS191} & $3.0\times10^{-9}$ \cite{PS191} & $1.5\times10^{-1}$ \cite{DELPHI}\\
\hline
0.5 & $2.0\times10^{-8}$\cite{0NuBB}& $1 \times 10^{-6}$ \cite{CHARMe} & $6.5\times10^{-7}$ \cite{NuTeV} & $2.5\times10^{-2}$ \cite{DELPHI}\\
\hline
0.7 & $3.5\times10^{-8}$\cite{0NuBB}& $5 \times 10^{-7}$ \cite{CHARMe} & $2.5\times10^{-7}$  \cite{NuTeV}& $9.0\times10^{-3}$ \cite{DELPHI}\\
\hline
1.0 & $4.5\times10^{-8}$\cite{0NuBB}& $3 \times 10^{-7}$ \cite{CHARMe} & $1.5\times10^{-7}$  \cite{NuTeV}& $3.0\times10^{-3}$ \cite{DELPHI}\\
\hline
2.0 & $1.0\times10^{-7}$\cite{0NuBB}& $1 \times 10^{-7}$ \cite{CHARMe} & $2.5\times10^{-5}$ \cite{DELPHI} & $3.0\times10^{-4}$ \cite{DELPHI}\\
\hline
3.0 & $1.5\times10^{-7}$\cite{0NuBB}& $3 \times 10^{-5}$ \cite{DELPHI} & $2.5\times10^{-5}$ \cite{DELPHI} & $4.5\times10^{-5}$ \cite{DELPHI}\\
\hline
4.0 & $2.5\times10^{-7}$\cite{0NuBB}& $2 \times 10^{-5}$ \cite{DELPHI} & $1.5\times10^{-5}$ \cite{DELPHI} & $1.5\times10^{-5}$ \cite{DELPHI}\\
\hline
5.0 & $3.0\times10^{-7}$\cite{0NuBB}& $2 \times 10^{-5}$ \cite{DELPHI} & $1.5\times10^{-5}$ \cite{DELPHI} & $1.5\times10^{-5}$ \cite{DELPHI}\\
\hline
6.0 & $3.5\times10^{-7}$\cite{0NuBB}& $2 \times 10^{-5}$ \cite{DELPHI} & $1.5\times10^{-5}$ \cite{DELPHI} & $1.5\times10^{-5}$ \cite{DELPHI}\\
\hline
\end{tabular}
\end{table}
For $|U_{e N}|^2$, the most restrictive upper bounds come from the neutrinoless double beta decay ($0\nu\beta\beta$) \cite{0NuBB}. However, due to possible significant uncertainties of the values of the nuclear matrix element in $0\nu\beta\beta$, we present in Table \ref{t1} also alternative upper bounds for $|U_{e N}|^2$ which exclude the $0\nu\beta\beta$ data.

Here we will discuss the results of the previous two Sections, and will illustrate in a few cases how they can be used for predictions, for specific mass ranges of $N$. Once we consider a specific range of $M_N$, and specific possible values of $|U_{\ell N}|^2$, we must first check whether the probability $P_N$ of decay of such $N$ neutrino within the detector is:
\begin{itemize}
\item [(a)] $P_N \approx 1$, i.e., $N$ decays instantly at the same vertex of the production;
\item [(b)] $P_N \ll 1$ (say, $P_N < 0.4$), i.e., $N$ decays within the detector with a displaced secondary vertex;
\item [(c)] $P_N \approx 0$ (practically zero), i.e., $N$ always leaves the detector, resulting in massive missing momenta.
\end{itemize}
We  note that with $P_N \alt 0.4$ the experimentally observed effective branching is getting small, but the decay
with two vertices, if detected, will represent a dramatic detector signature.

This then determines whether the predicted measured branching ratios are:
\begin{itemize}
\item [(a)] $P_N {\rm Br} \approx {\rm Br}$ of Sec.~\ref{sec:Br}, whose canonical values are presented in Figs.~\ref{FigbBrB}-\ref{FigbBrBDPi};
\item [(b)] $\K (L/1{\rm m}) {\bar P}_N {\rm Br}$ of Sec.~\ref{sec:effBr} whose canonical values are presented in Figs.~\ref{FigbBrefB}-\ref{FigbBrefBDPi};
\item[(c)] ${\rm Br}={\rm Br}_{\rm prod.}$ of Sec. II whose canonical decay widths are presented in Figs. 1 and 4.
\end{itemize}
Table \ref{t1} suggests that the processes with muons are at present more probable than those with electrons, although this conclusion is not valid if we exclude the $0\nu\beta\beta$ data for the upper bounds for  $|U_{e N}|^2$.
Nonetheless, in the rare decays with muons and no pion in the final state, we should have at least one electron in the final state.\footnote{The decays with at least one $\tau$ are in general kinematically suppressed, but the mixing coefficients $|U_{\tau N}|^2$ have at the moment less restrictive upper bounds, $|U_{\tau N}|^2 \alt 10^{-4}$ for $2 \ {\rm GeV} < M_N < 3 \ {\rm GeV}$, cf.~Table \ref{t1}. The $\tau$ lepton is difficult to identify in experiments, though.}
This is so because, with three muons in the final state, a pair $\mu^+ \mu^-$ would appear there, and such decays would have QED background from virtual photon decays $\gamma^* \to \mu^+ \mu^-$, as mentioned in the previous Sections. Therefore, among the above rare processes (with no pion), those with possible higher (effective) branching ratio are $B_{(c)} \to (D^{(*)}) \mu^{\pm} \mu^{\pm} e \nu$, which can be LFV or LNC. When measuring ${\rm Br}_{\rm (eff)}$ of these processes, we cannot distinguish between the Majorana and Dirac nature of $N$.

\subsection{The decays with no produced $D^{(*)}$ mesons}
\label{estnoD}

First we will discuss the rare decays where no $D^{(*)}$ mesons are produced (i.e., the cases of Sec. II A and Sec. III).

Comparing Figs.~\ref{FigbBrB}(a) and (b) and \ref{FigbBrefB}(a) and (b), we can see that the purely leptonic rare $B$ decays are suppressed in comparison with the corresponding decays of $B_c$, primarily due to the strong CKM suppression ($|V_{ub}| \approx 10^{-1} |V_{cb}|$). Since only $B$ mesons can be produced at Belle, such rare purely leptonic decays, Figs.~\ref{FigbBrB}(a) and \ref{FigbBrefB}(a), will be difficult to measure at Belle II experiment. For example, if $M_N \approx 4$ GeV, according to Table \ref{t1} we have $|U_{\ell N}|^2 \alt 10^{-5}$ (for all $\ell$). If we assume that $|U_{\ell N}|^2 \sim 10^{-5}$, then it turns out that the decay probability is $P_N \approx 1$. This is so because, according to Eqs.~(\ref{PN})-(\ref{bPN}) we have in general
\be
P_N = 1 - \exp \left( - \K \left( \frac{L}{1 \ {\rm m}} \right) {\overline P}_N  \right) \ .
\label{PN2}
\ee
We have $\K \sim 10^1 |U_{\ell N}|^2$ according to Eq.~(\ref{calKest}), ${\overline P}_N \sim 10^5$ for $M_N \approx 4$ GeV according to Fig.~\ref{FigbPN}, so that for the detector length $L=1$ m we have the expression in the exponential on the right-hand side of Eq.~(\ref{PN2}) $\sim 10^1 |U_{\ell N}|^2 10^5 \sim 10^1$ (where we used $|U_{\ell N}|^2 \sim 10^{-5}$). Since $P_N \approx 1 - \exp(-10^1) \approx 1$ in this chosen case, the relevant canonical branching ratio is ${\overline {\rm Br}}$ from Fig.~\ref{FigbBrB}(a), namely ${\overline {\rm Br}} \sim 10^{-4}$ [${\overline {\rm Br}}_{\rm eff} \sim 10^1$ of Fig.~\ref{FigbBrefB}(a) is now not relevant]. Eqs.~(\ref{cBr}) (with $\ell_1=\ell_3=\mu$, $\ell_2=e$) then imply that the measured branching ratio is
\bea
{\rm Br}(B \to \mu^{\pm} \mu^{\pm} e^{\mp} \nu) &\sim& \frac{1}{\K} |U_{\ell N}|^4
{\overline {\rm Br}}(B \to \mu^{\pm} \mu^{\pm} e^{\mp} \nu)
\sim \frac{1}{10^1 |U_{\ell N}|^2} |U_{\ell N}|^4 {\overline {\rm Br}}(B \to \mu^{\pm} \mu^{\pm} e^{\mp} \nu)
\nonumber\\
& \sim & 10^{-1} 10^{-5} 10^{-4} \sim 10^{-10} \ .
\label{BrLeptEst}
\eea
This implies that at Belle II the number of such decays detected per year will be ${\cal N} \sim 10^{10} 10^{-10} \sim 1$, which is difficult to be observed. If we decrease the mass $M_N$ to, say $M_N \approx 3$ GeV, the results do not change significantly, because $P_N \approx 1$ is still valid, and the canonical branching ratios do not change significantly. At even lower values of $M_N$ we have $P_N \ll 1$, which implies an additional suppression of the measured branching ratios.

If the intermediate on-shell neutrino (with $M_N \approx 4$ GeV) decays to a pion, the relevant figure is Fig.~\ref{FigbBrBPi}(a) for $B^- \to \mu^- \mu^- \pi^+$ [and not  Fig.~\ref{FigbBrefBPi}(a), since $P_N \approx 1$], and the resulting rate (at $M_N = 4$ GeV) is by about one order of magnitude too small for the detection. 

On the other hand, LHC-b can produce $B_c$ mesons copiously, and the rare leptonic decays of $B_c$ may be detected there due to significantly higher branching ratios, cf.~Figs.~\ref{FigbBrB}(b) and \ref{FigbBrefB}(b). Similar conclusion can be made for the corresponding decays of $B_c$ when the intermediate on-shell neutrino decays to a pion, Figs.~\ref{FigbBrBPi}(b) and \ref{FigbBrefBPi}(b). For the latter processes, LHC-b can be sensitive down to the branching ratios ${\rm Br}(B_c^- \to \mu^- \mu^- \pi^+) \sim 10^{-7}$ in LHC run 2 (collected luminosity $5 \ {\rm fb}^{-1}$) and $\sim 10^{-8}$ in the future LHC run 3 (collected luminosity $40 \ {\rm fb}^{-1}$), cf.~Ref.~\cite{Quint}. If we assume that $|U_{\mu N}|^2$  is numerically either the dominant or a representative heavy-light mixing coefficient, then an estimate similar to that of Eq.~(\ref{BrLeptEst}), now based on the results of Fig.~\ref{FigbBrBPi}(b) for $B_c^- \to \mu^- \mu^- \pi^+$, implies that LHC-b can provide upper bounds on $|U_{\mu N}|^2$ of order $\sim 10^{-4}$ (run 2) and $\sim 10^{-5}$ (run 3), in the mass range $M_N$ somewhere between $3$ and $5$ GeV (in such cases $P_N \approx 1$).\footnote{On the other hand, for $M_N < 2$ GeV we have $P_N \ll 1$ if $|U_{\mu N}|^2 \alt 10^{-5}$; and for $M_N < 1$ GeV we have $P_N \ll 1$ if $|U_{\mu N}|^2 \alt 10^{-4}$. Namely, according to Fig.~\ref{FigbPN} and Eqs.~(\ref{PNb}) and (\ref{calKest}) we have: $P_N \approx \K {\overline P}_N \sim 10 |U_{\mu N}|^2 {\overline P}_N$, and ${\overline P}_N \sim 10^3$ ($10^2$) for $M_N \approx 2$ GeV ($1$ GeV). Therefore, for $M_N < 2$ GeV, it is the {\it effective} branching ratios to which LHC-b becomes sensitive, i.e., ${\rm Br}_{\rm eff}(B_c \to \mu^- \mu^- \pi^+) \sim 10^{-7}$ (run 2) and $\sim 10^{-8}$ (run 3). In such cases, the relevant quantities are the effective canonical  branching ratio of Fig.~\ref{FigbBrefBPi}(b) and Eq.~(\ref{cBreffPi}) (with $\ell_1=\ell_2=\mu$). For example, if $M_N=1$ GeV, the effective branching ratio is ${\rm Br}_{\rm eff}=|U_{\mu N}|^4 {\overline {\rm Br}}_{\rm eff} \sim |U_{\mu N}|^4 10^0$, and LHC run 3 will be sensitive down to ${\rm Br}_{\rm eff} \sim 10^{-8}$, implying that it can probe the mixings down to $|U_{\mu N}|^2 \sim 10^{-4}$ (and not $10^{-5}$). In that case $P_N \approx \K {\overline P}_N \sim 10 |U_{\mu N}|^2  {\overline P}_N \sim 10 \times 10^{-4} \times 10^2 \sim 10^{-1} \ll 1$ which is consistent with the assumption $P_N \ll 1$.} These conclusions agree with those of Refs.~\cite{Quint,Mand}. The authors of these references also considered other semileptonic decays of $B_c$ via an on-shell Majorana neutrino, with similar conclusions. Similar conclusions can be obtained by using the leptonic channel $B_c^- \to \mu^- \mu^- e^+ \nu$, Fig.~\ref{FigbBrB}(b), if the signal efficiency is similar to that of the semileptonic decays.

There is an interesting aspect of the rare decays with one produced pion (and no $D^{(*)}$). If in such rare decays also one heavy $\tau$ lepton is produced, then the results of Figs.~\ref{FigbBrBPi}(a) and \ref{FigbBrefBPi}(a) indicate that such processes could in principle be detected at Belle II. Namely, the present upper bounds on $|U_{\tau N}|^2$ are less restrictive, $|U_{\tau N}|^2 \alt 10^{-4}$ for the mass interval $0.3 \ {\rm GeV} < M_N < 3 \ {\rm GeV}$ (cf.~\cite{DELPHI,Atre} and Table \ref{t1}). If $|U_{\tau N}|^2 \sim 10^{-4}$ and $2 \ {\rm GeV} < M_N < 3$ GeV, then it can be checked that $P_N \approx 1$ [cf.~Eqs.~(\ref{calKest}) and Fig.~\ref{FigbPN}]. Therefore, for $2 \ {\rm GeV} < M_N < 3 \ {\rm GeV}$ and $|U_{\tau N}|^2 \sim 10^{-4}$ we can use the branching ratio of Sec.~\ref{sec:Br}. The canonical branching ratio for the decays $B \to \tau \mu \pi$ is
${\overline {\rm Br}} \approx 10^{-4}$ by Fig.~\ref{FigbBrBPi}(a) for $2 \ {\rm GeV} < M_N < 3 \ {\rm GeV}$ (dashed line). Therefore, if $|U_{\tau N}|^2 \sim 10^{-4}$ and
$2 \ {\rm GeV} < M_N < 3 \ {\rm GeV}$, the branching ratio for such decays
would be [using Eqs.~(\ref{cBrPi}) and (\ref{calKest})]
\bea
{\rm Br}(B^- \to \tau \mu \pi) & = & \frac{1}{\K} |U_{\tau N}|^2 |U_{\mu N}|^2 {\overline {\rm Br}}(B^- \to \tau \mu \pi) \approx  \frac{1}{3 |U_{\tau N}|^2} |U_{\tau N}|^2 |U_{\mu N}|^2 {\overline {\rm Br}}(B^- \to \tau \mu \pi)
\nonumber\\
& = & \frac{1}{3} |U_{\mu N}|^2 {\overline {\rm Br}}(B^- \to \tau \mu \pi)
\alt  \frac{1}{3} (2.5 \times 10^{-5}) 10^{-4} \sim 10^{-9} \ .
\label{BrTauPiEst}
\eea
In the last steps, we used the upper bounds for $|U_{\mu N}|^2$ in the considered mass interval, cf.~Table \ref{t1}. The estimate (\ref{BrTauPiEst}) suggests that Belle II could detect up to $\sim 10^1$ rare decays of the type
$B \to \tau \mu \pi$.

If $N$ is Dirac, the dashed lines in Figs.~\ref{FigbBrBPi} and \ref{FigbBrefBPi}
[and \ref{FigbBrBDPi}  and \ref{FigbBrefBDPi}]
get reduced by factor $2$, because only two out of four decays contribute, namely the LNC decays: $B \to ( D^{(*)} ) \tau^- {\bar N} \to ( D^{(*)} ) \tau^- \mu^+ \pi^-$ and $B \to ( D^{(*)} ) \mu^- {\bar N} \to ( D^{(*)} ) \mu^- \tau^+ \pi^-$ (where $B = B^-, {\bar B}^0$).
  If such decays can be detected, the nature of the neutrino can be discerned. For example, if the decays $B^- \to \mu^{-} \tau^{-} \pi^{+}$ are detected, such processes violate the lepton number and the neutrinos have to be Majorana.
The situation with such decays is better by several orders of magnitude if the decaying meson is $B_c$ (i.e., in LHC-b), cf.~Figs.~\ref{FigbBrBPi}(b) and ~\ref{FigbBrefBPi}(b). The results will certainly depend on how efficiently the produced $\tau$ leptons can be identified in such decays, and such identification may be difficult.

If $P_N \ll 1$, then in most of the considered rare $B$-decays the produced $N$
travels through the detector and its production is manifested as a massive
missing momentum (we referred to this as the ``$P_N \approx 0$'' case).
The decay rates for such events are higher than those with $N$ decaying
within the detector, but with the negative aspect of no experimental signature
of $N$-decay. When $P_N \ll 1$, we have
$P_N \sim 10 |U_{\ell N}|^2 {\overline P}_N$ by Eqs.~(\ref{PNb}) and
(\ref{calKest}). Therefore, the case $P_N \ll 1$ is in general to be expected
for lighter masses, cf.~Fig.~\ref{FigbPN} where ${\overline P}_N \alt 10^3$
for $M_N \leq 2$ GeV, and for smaller mixing parameters $|U_{\ell N}|^2$.
On the other hand, the decay widths
$\Gamma(B \to \ell N) \equiv |U_{\ell N}|^2 \bG(B \to \ell N)$ are suppressed
by smaller $|U_{\ell N}|^2$. However, there is a window of such ranges of
(low) $M_N$ and (high) $|U_{\ell N}|^2$ where simultaneously $P_N \ll 1$ and
the decay widths $\Gamma(B \to \ell N)$ are appreciable. Namely,
if $M_N \approx 2$ GeV, we can have at present the values of $|U_{\mu N}|^2$
as high as $\sim 10^{-5}$, cf.~Table \ref{t1} and \cite{Atre}
(we assume that $|U_{\tau N}|^2$ is not larger than $\sim 10^{-5}$ either). Then
$P_N \sim 10 |U_{\mu N}|^2 {\overline P}_N  \sim 10 \times 10^{-5} \times 10^3 \sim 10^{-1}$($\ll 1$). According to Fig.~\ref{FigbGBmuN}(a) we have
$\bG(B \to \mu N) \approx 5 \times 10^{-17}$ GeV at $M_N \approx 2$ GeV,
and therefore the following branching ratio is possible at such $M_N$:
\be
{\rm Br}(B \to \mu N) = \frac{|U_{\mu N}|^2 \bG(B \to \mu N)}{\Gamma_B} \sim
\frac{10^{-5} \times (5 \times 10^{-17})}{4 \times 10^{-13}} \sim 10^{-9} \ ,
\label{BrBtomuN}
\ee
where we took into account that $\Gamma_B \approx 4 \times 10^{-13}$ GeV.
The estimate (\ref{BrBtomuN}) implies that Belle II could see, for the
mentioned approximate values of the parameters $M_N$ and $|U_{\ell N}|^2$,
about $\sim 10^1$ decays per year of $B \to \mu$+ missing momentum $p_N$,
with the invariant mass of the missing momentum $\sqrt{p_N^2} \approx 2$ GeV.
This is by one order of magnitude better than the estimate (\ref{BrLeptEst})
which involves the decay of $N$ well within the detector.
For the decays $B \to \tau$+ missing momentum, the favorable ranges of the
parameters $M_N$ and $|U_{\tau N}|^2$ are even wider than for the decays
$B \to \mu +$ missing momentum, primarily because $\bG(B \to \tau N)$ is
appreciable even at lower masses of $M_N$, cf.~Fig.~\ref{FigbGBmuN}(a);
The identification of $\tau$ leptons is, however, difficult \cite{Cvetic:2015roa}.

\subsection{The decays with produced $D^{(*)}$ mesons}
\label{estD}

We now comment on the rare $B$-decays which produce $D^{(*)}$ mesons (i/e. the cases of of Sec. II B and Sec. III).

Figs.~\ref{FigbBrBD} and \ref{FigbBrBDPi}, and the corresponding figures for the effective branching ratio, Figs. \ref{FigbBrefBD} and \ref{FigbBrefBDPi}, are for rare decays of $B$ mesons where at the first vertex a $D^{(*)}$-meson is produced, evading thus the CKM suppression encountered in the processes involving the leptonic decays of $B \to \ell N$. The rare decays of $B$ with produced $D^{(*)}$ mesons are of interest for the Belle II experiment.
In general, for lighter masses $M_N < 2.5$ GeV, we have $P_N <0.5$ if we assume that there $|U_{\ell N}|^2 \alt 10^{-5}$ for all $\ell$. If this is so, we may use the effective branching ratios of Sec.~\ref{sec:effBr}.
The results in Figs.~\ref{FigbBrefBD} and \ref{FigbBrefBDPi} show that
for $1.5 \ {\rm GeV} < M_N < 2.5 \ {\rm GeV}$ we have ${\overline {\rm Br}}_{\rm eff} \sim 10^1$ if no $\tau$ leptons are involved.
The present upper bounds on $|U_{\mu N}|^2$, in the mass range $2 \ {\rm GeV} < M_N < 3 \ {\rm GeV}$, are  $|U_{\mu N}|^2 \alt 10^{-5}$, cf.~Table \ref{t1}.
 The relations (\ref{cBreffMaj})-(\ref{cBreffPi}) then imply that, if $N$ is Majorana and $2 \ {\rm GeV} < M_N < 2.5 \ {\rm GeV}$, Belle II could produce per year a number ${\cal N}$ of rare LNV decays $B \to D^{(*)} \mu^{\pm} \mu^{\pm} e^{\mp} \nu$ and $B \to D^{(*)} \mu^{\pm} \mu^{\pm} \pi^{\mp}$ of the order
\bea
{\cal N}(B \to D^{(*)} \mu^{\pm} \mu^{\pm} e^{\mp} \nu) & \sim &
{\cal N}(B \to D^{(*)} \mu^{\pm} \mu^{\pm} \pi^{\mp})
\sim 10^{10} \times |U_{\mu N}|^4 {\overline {\rm Br}}_{\rm eff}
\nonumber\\
&\sim & 10^{10} \times (10^{-5})^2 10^1 \sim 10^1 \ .
\label{BDmumuXEst}
\eea
If no such rare decays are detected, then Belle II can decrease the upper bound for $|U_{\mu N}|^2$ in that mass interval.
The second of these processes, $B \to D^{(*)} \mu^{\pm} \mu^{\pm} \pi^{\mp}$, is LFV, and is possible only if $N$ is Majorana. The first of these processes,  $B \to D^{(*)} \mu^{\pm} \mu^{\pm} e^{\mp} \nu$, can be either LNC or LNV. If enough of such decays are measured, then the differential branching ratio $d {\rm Br}/d E_e$ can be measured (where $E_e$ is the energy of $e$ in $N$ rest frame), and this quantity is proportional to $d {\overline {\rm Br}}_N/d E_e$ studied in Sec.~\ref{subs:diff}. There it was argued that by measuring this quantity, the Dirac or Majorana nature of $N$ can be discerned.

Another attractive aspect of the rare $B$-meson decays involving $D^{(*)}$
mesons is the possibility of measuring the decays $B \to D^{(*)} \mu N$ with
the $N$ neutrino not decaying within the detector, i.e., what we referred to as the ``$P_N \approx 0$'' case. The neutrino would manifest itself only as
a massive missing momentum. According to Figs.~\ref{FigbGBDlN}, the
corresponding decays widths, for lower masses $M_N \alt 2$ GeV, are significantly
larger than the corresponding decay widths  without $D^{(*)}$ mesons
Fig.~\ref{FigbGBmuN}(a), principally because the CKM-mixing suppression
($|V_{ub}| \approx 0.004$) is made weaker with the presence of $D^{(*)}$
($|V_{cb}| \approx 0.04$). For the
masses $M_N \alt 1.8$ GeV, it turns out that the mixing
coefficients $|U_{\mu N}|^2$ are at present strongly restricted,
$|U_{\mu N}|^2 \alt 10^{-7}$, cf.~\cite{Atre}
(cf.~also Table \ref{t1} here).\footnote{
The upper bounds become much less restrictive for $M_N > 1.8$ GeV:
$|U_{\mu N}|^2 \alt 10^{-5}$, cf.~\cite{Atre,DELPHI}.}
If we, conservatively, assume in addition that the other mixing coefficients
$|U_{\ell N}|^2$ ($\ell=e, \tau$) also fulfill the strong restrictions
$|U_{\ell N}|^2 \alt 10^{-7}$, the condition $P_N \approx 0$ is
strongly fulfilled: we have $P_N =
\K |U_{\ell N}|^2 {\overline P}_N \sim 10 |U_{\ell N}|^2 {\overline P}_N <
10 \times 10^{-7} \times 10^3 \sim 10^{-3}$
[using Eqs.~(\ref{PNb}), (\ref{calKest}) and Fig.~\ref{FigbPN}], i.e.,
$P_N \ll 1$ ($P_N \approx 0$). At $M_N \alt 1.8$ GeV ,
despite the very restricted values $|U_{\ell N}|^2 \alt 10^{-7}$,
the branching ratio for $B \to D^{(*)} \mu N$ can achieve
the following values:
\be
{\rm Br}(B \to D^{(*)} \mu N) =
\frac{|U_{\mu N}|^2 \bG(B \to D^{(*)} \mu N)}{\Gamma_B}
\sim \frac{10^{-7} \times 10^{-14}}{4 \times 10^{-13}} \approx 2 \times 10^{-9}
\sim 10^{-9} \ ,
\label{BtoDmuN}
\ee
where we took into account that $\Gamma_B \approx 4 \times 10^{-13}$ GeV, and
that $\bG(B \to D^{(*)} \mu N) \sim 10^{-14}$ GeV for $M_N \leq 2$ GeV
according to the results of Figs.~\ref{FigbGBDlN}.
The estimate (\ref{BtoDmuN}) suggests that Belle II could possibly
detect rare decays $B \to D^{(*)} +\mu +$ missing energy of invariant mass
$\sqrt{p_N^2} \alt 2$ GeV, at rates of $\sim 10^1$ per year. However,
as argued earlier, if the mass $M_N$ is somewhat higher,
$1.8 \ {\rm GeV} \alt M_N \alt 2.5 \ {\rm GeV}$,
we can still have $P_N \ll 1$ ($P_N <  0.5$)
and at the same time the mixing coefficients
can become larger by two orders of magnitude,
$|U_{\ell N}|^2 \sim 10^{-5}$ (cf.~Table \ref{t1}).
In such a case the branching ratio ${\rm Br}(B \to D^{(*)} \mu N)$
can go up to $\sim 10^{-8}$ because $\bG(B \to D^{(*)} \mu N)$ decreases there
only by at most one order of magnitude,
$\bG(B \to D^{(*)} \mu N) \sim 10^{-15}$ GeV, cf.~Figs.~\ref{FigbGBDlN}.
The resulting estimate ${\rm Br}(B \to D^{(*)} \mu N) \sim 10^{-8}$
is by one order of magnitude better than the corresponding
estimate (\ref{BDmumuXEst}) for the case of $N$ decaying within the detector.
Further, at significantly lower masses $M_N \approx 0.5$ GeV, the
present upper bounds are $|U_{\mu N}|^2 \approx 6.5 \times 10^{-7}$, and
the canonical decay width $\bG(B \to D^{(*)} \mu N)$ is high, e.g.,
$\bG(B \to D^{*} \mu N) \approx 2 \times 10^{-14}$ GeV,
Fig.~\ref{FigbGBDlN}(b). The estimate of the type (\ref{BtoDmuN})
then increases the branching ratio to up to
${\rm Br}(B \to D^{*} \mu N) \sim 10^{-8}$ for
$M_N \approx 0.5$ GeV, leading to up to
$\sim 10^2$ such events per year at Belle II.

\section{Summary}
\label{summ}

In this work we considered rare decays of $B$ and $B_c$ mesons mediated by heavy on-shell neutrinos $N$ with masses
$M_N \sim 1$ GeV. The work was performed especially in view of the upgrade plan for the dedicated Belle experiment (Belle II) in which $\sim 10^{10}$ $B$ mesons are to be produced per year, in addition to the presently ongoing LHC-b. Direct decays of $B$ meson of the type $B \to \ell_1 N \to \ell_1 X Y$ (where $XY = \ell_2 \ell_3 \nu$ or $\ell_2 \pi$, and $\ell_j$ are charged leptons) are strongly suppressed due to the small CKM element $V_{ub} \approx 0.004$, and such decays turn out to be difficult to detect at Belle II. Nonetheless, $B_c$ mesons have significantly weaker CKM suppression ($|V_{cb}| \approx 0.04$), they are copiously produced at LHC-b, and the mentioned decays with $B_c$ could be detected there. However, $B_c$ mesons are not produced at Belle. Therefore, in order to evade the mentioned strong CKM suppression, we also investigated decays of $B$ mesons which produce a $D^{(*)}$ meson at the first vertex, namely $B \to D^{(*)} \ell_1 N$, where the on-shell heavy neutrino $N$ may subsequently decay (within the detector) leptonically $N \to \ell_2 \ell_3 \nu$ or semileptonically $N \to \ell_2 \pi$. In these decays, we took into account the possible effects of the heavy neutrino lifetime. Our calculations and subsequent estimates raise the possibility of detection of such rare decays at Belle II. If such rare decays are detected, in some of such cases there is a possibility to determine the nature of $N$ (Majorana or Dirac), via the identification of the lepton numbers of the final particles (LNV or LNC processes), or even via the measurement of differential decay widths if enough such events are detected. If such rare decays are not detected at Belle II, then the upper bounds on some of the heavy-light mixing coefficients $|U_{\ell N}|^2$ can be decreased, for the relevant mass ranges of the heavy neutrino $N$ ($M_N \sim 1$ GeV). Another attractive possibility is that Belle II detects the decays $B \to D^{(*)} \mu N$ where $N$ does not decay within the detector but manifests itself as a massive missing momentum. We point out that such events could be produced at Belle II in significant numbers for various ranges of the values of $N$ mass and of the heavy-light mixing coefficients $|U_{\ell N}|^2$.

\section*{Acknowledgments}
\noindent
This work of G.C. was supported in part by FONDECYT (Chile) Grant No.~1130599.
The work of C.S.K. was supported in part by the NRF grant funded by the Korean government of the MEST (No. 2011-0017430) and (No. 2011-0020333).

\appendix

\section{General expression for ${\overline \Gamma}(N \to \ell_2 \ell_3 \nu)$}
\label{app1}

The decay width for the process $N \to \ell_2 \ell_3 \nu$ is given in
Eqs.~(\ref{GNllnu}) for the LNC and LNV version, where the canonical
width  (without the heavy-light mixing factor)
${\overline \Gamma}(N \to \ell_2 \ell_3 \nu)$,
Eq.~(\ref{bGNllnu}), contains the factor ${\cal F}(x_2,x_3)$ [with
$x_j$ the dimensionless rescaled masses Eq.~(\ref{xj})]. This factor
has the following expression, cf.~Ref.~\cite{CKZ}:
\bea
\lefteqn{
{\cal F}(x_2,x_3) =
{\Bigg \{}
\lambda^{1/2} (1, x_2, x_3) {\big [} (1 + x_2) (1 -8 x_2 + x_2^2)  -
x_3 (7 - 12 x_2 + 7 x_2^2)
}
\nonumber\\
&&
- 7 x_3^2 (1 + x_2)  + x_3^3 {\big ]}
- 24 (1 - x_3^2) x_2^2 \ln 2
\nonumber\\
&&
+  12 {\bigg [} - x_2^2 (1 - x_3^2) \ln x_2
+ (2 x_2^2 -x_3^2 (1 + x_2^2)) \ln (1 + x_2
+ \lambda^{1/2} (1, x_2, x_3)  - x_3)
\nonumber\\
&&
+ x_3^2 (1 - x_2^2)
\ln \left( \frac{(1 - x_2)^2 + (1-x_2) \lambda^{1/2} (1, x_2, x_3) - x_3 (1+x_2)}{x_3}
\right) {\bigg ]}
{\Bigg \}} \ ,
\label{calF}
\eea
and the function $\lambda^{1/2}$ is given by
\be
\lambda^{1/2}(x,y,z) = \left[ x^2 + y^2 + z^2 - 2 x y - 2 y z - 2 z x \right]^{1/2} \ .
\label{sqlam}
\ee
It can be checked that ${\cal F}$ is symmetric under the exchange of the two arguments: ${\cal F}(x_2,x_3)={\cal F}(x_3,x_2)$.
In the limit when one of the charged leptons is massless,
the above expression reduces to the well-known expression
\be
{\cal F}(x,0) =  {\cal F}(0,x)= f(x) = 1 - 8 x + 8 x^3 - x^4 - 12 x^2 \ln x \ .
\label{fx}
\ee
This expression is a good approximation when, e.g., one of the charged leptons is an electron and the other is a muon ($x_1=x= M_{\mu}^2/M_N^2$).

\section{Decay width of $B \to D \ell_1 N$}
\label{appD}

\begin{figure}[htb]
\centering\includegraphics[width=90mm]{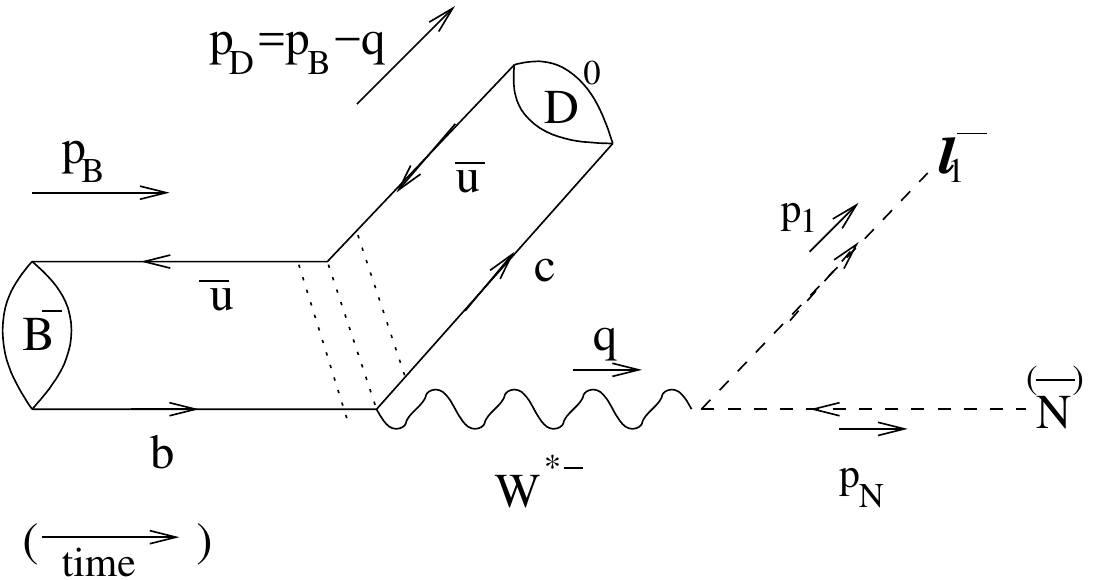}
\caption{Schematical representation of the decay $B^- \to D^0 \ell_1^- {\bar N}$.}
\label{FigBDW}
\end{figure}

In this Appendix, we outline the derivation of the decay
widths $\Gamma(B \to D \ell_1 N)$, with massive neutrino $N$
and (massive) charged lepton $\ell_1$, which may be relevant especially
for the search of sterile neutrinos at Belle(II). The process is schematically
presented in Fig.~\ref{FigBDW}, for the case of $B^{-} \to D^0 \ell_1^- N$.
The decay width is
\be
\Gamma(B^- \to D^0 \ell_1 N) = \frac{1}{2 M_B} \frac{1}{(2 \pi)^5}
\int d_3 |{\cal T}|^2 \ ,
\label{BDNl}
\ee
where $d_3$ is the usual integration differential of the final three-particle
phase space
\bea
d_3 &\equiv& \frac{d^3 {\vec p}_D}{2 E_{D}({\vec p}_D)}
 \frac{d^3 {\vec p}_1}{2 E_{\ell_1}({\vec p}_1)}
 \frac{d^3 {\vec p}_N}{2 E_N({\vec p}_N)}
 \delta^{(4)} \left( p_B - p_D - p_1 - p_N \right)
 \nonumber\\
 & = & d_2(B^- \to D^0(p_D) W^*(q)) d q^2 d_2(W^*(q) \to \ell_1(p_1) {\overline N}(p_N)) \ ,
\label{d3}
\eea
and ${\cal T}$ is the reduced decay amplitude
\be
{\cal T}  =  U_{\ell_1 N} V_{c b} \frac{G_F}{\sqrt{2}} \left[{\overline u}_{(\ell_1)}(p_1) \gamma_{\mu} (1 - \gamma_5) v_{(N)}(p_N) \right]
\left\{ \left[ (2 p_D + q)^{\mu} - \frac{(M_B^2-M_D^2)}{q^2} q^{\mu} \right] F_1(q^2) + \frac{(M_B^2-M_D^2)}{q^2} q^{\mu} F_0(q^2) \right\} \ .
\label{TBDNl}
\ee
Here, $F_1(q^2)$ and $F_0(q^2)$ are the form factors of the
$B$-$D$ transition
\be
\langle D(p_D) | {\overline c}(0) \gamma^{\mu} b(0) | B^{-}(p_B) \rangle
= \left[ (2 p_D + q)^{\mu} - \frac{(M_B^2-M_D^2)}{q^2} q^{\mu} \right] F_1(q^2) + \frac{(M_B^2-M_D^2)}{q^2} q^{\mu} F_0(q^2) \ ,
\label{FFBD}
\ee
where $q=p_B-p_D$ is the momentum of the virtual $W^-$ ($q=p_1+p_N$,
cf.~Fig.~\ref{FigBDW}), e.g., cf.~Refs.~\cite{NeuPRps,CaNeu}.

Squaring the absolute value of ${\cal T}$, summing over the final helicities,
and integrating over the two-particle phase spaces
$d_2(W^*(q) \to \ell_1(p_1) {\overline N}(p_N))$ and $d_2(B^- \to D^0 W^*(q))$
[cf.~Eq.~(\ref{d3})] then results in the following differential decay width
($M_1$ is the mass of $\ell_1$):
\be
\frac{d \Gamma (B \to D \ell_1 N)}{d q^2} =  |U_{\ell_1 N}|^2
\frac{d {\overline \Gamma} (B \to D \ell_1 N)}{d q^2} \ ,
\label{dGBDNl}
\ee
where the canonical (i.e., with no heavy-light mixing coefficient) decay
width is
\bea
\lefteqn{
\frac{d {\overline \Gamma} (B \to D \ell_1 N)}{d q^2} =
\frac{1}{384 \pi^3} G_F^2 |V_{c b}|^2 \frac{1}{(q^2)^2 M_B}
\lambda^{1/2} \left(1, \frac{q^2}{M_B^2}, \frac{M_D^2}{M_B^2} \right)
\lambda^{1/2} \left(1, \frac{M_1^2}{q^2}, \frac{M_N^2}{q^2} \right)
}
\nonumber\\
&& \times
    {\bigg \{} F_1(Q^2)^2 \left[ 2 (q^2)^2 - q^2 M_N^2  + M_1^2 (2 M_N^2- q^2) - M_N^4 - M_1^4 \right] \left[ (q^2 - M_D^2)^2 - 2 M_B^2 (q^2 + M_D^2) + M_B^4 \right]
    \nonumber\\
    && + F_0(q^2)^2 3 (M_B^2 - M_D^2)^2 \left[ q^2 M_N^2 + M_1^2 (2 M_N^2 + q^2) - M_N^4 - M_1^4 \right]
        {\bigg \}} \ .
\label{dbGBDNl}
\eea
We assumed that the form factors are real. In such a case, the (differential) decay widths have the same expression (\ref{dGBDNl})-(\ref{dbGBDNl}) for all the processes $B \to D \ell_1 N$ irrespective of the electric charges involved:
$B^{-} \to D^{0} \ell_1^- {\overline N}$;
$B^{+} \to {\overline D^{0}} \ell_1^+ N$;
${\overline B^0} \to D^+ \ell_1^- {\overline N}$;
$B^0 \to D^- \ell_1^+ N$. Furthermore, the expressions
are the same irrespective of the nature of $N$ (Dirac or Majorana).
The form factors $F_1(q^2)$ and $F_0(q^2)$ are practically the same
in all these cases.
The total decay width is obtained by integrating the differential decay
width in the kinematically allowed interval:
$(M_N+M_1)^2 \leq q^2 \leq (M_B-M_D)^2$,
where we have $(M_B-M_D) \approx 3.414$ GeV for charged $B$, and
$(M_B-M_D) \approx 3.410$ GeV for neutral $B$ decays.

When $M_N=M_1=0$ (the case investigated in the literature), then the form factor $F_0(q^2)$ does not contribute to $d \Gamma/d q^2$, and our formula reduces to the known expression for $B \to D e \nu$, \cite{NeuPRps} and references therein.

\section{Decay width of $B \to D^{*} \ell_1 N$}
\label{appDst}

Since $D^*$ is a vector meson, the expressions are more complicated than in the case of the (pseudoscalar) $D$ meson. Here we will follow the approach of Ref.~\cite{GiSi}, where this type of decay width was calculated in the case of massless $N$ and $\ell_1$. We obtain here the result for the general case of massive $N$ and $\ell_1$.

\begin{figure}[htb]
\centering\includegraphics[width=80mm]{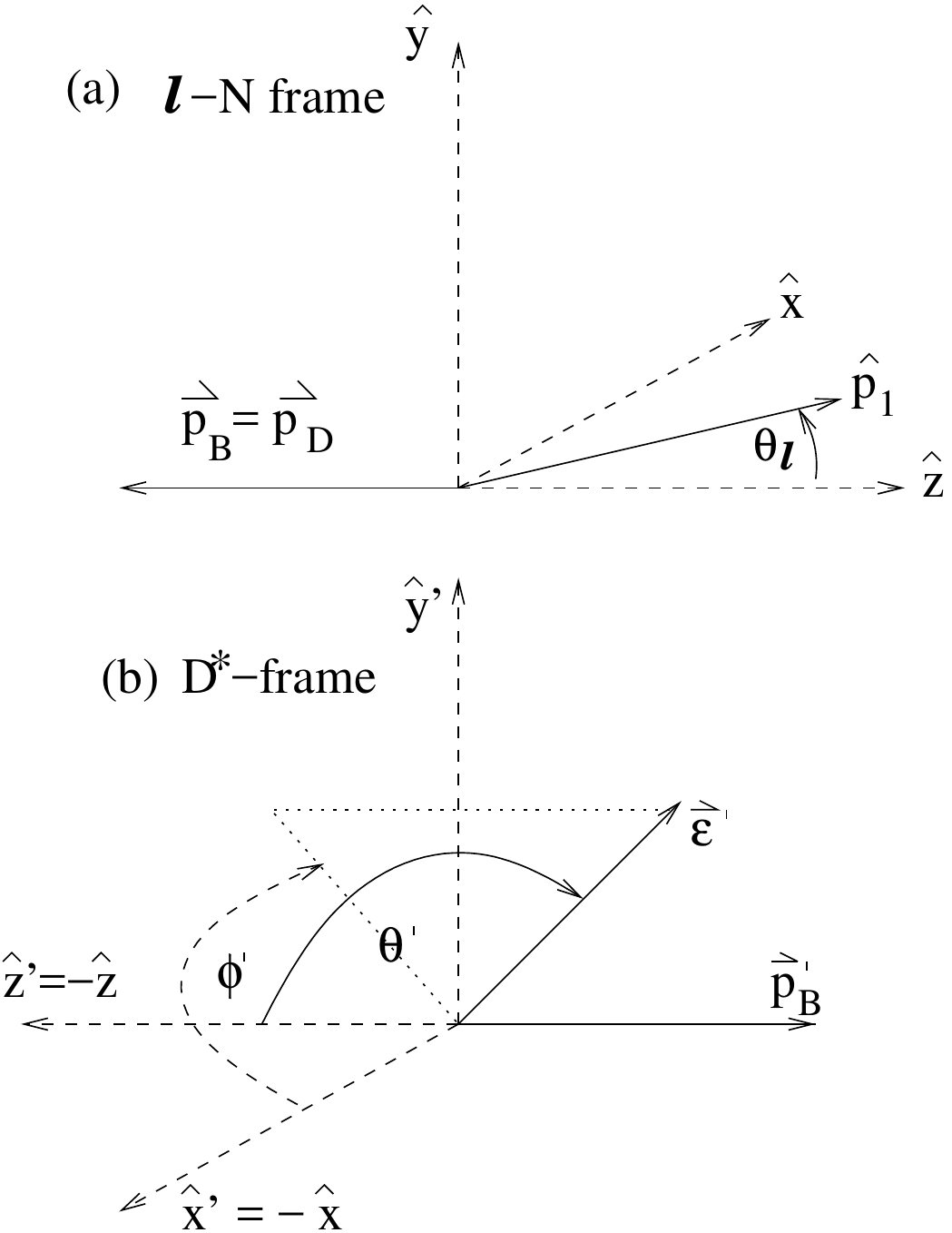}
\caption{(a) Coordinate system in $\Sigma$ frame ($\ell_1$-$N$ rest system), and the direction ${\hat p}_1$ of the charged lepton $\ell_1$ in it; (b) Coordinate system in
  $\Sigma'$ frame ($D^*$ rest system), and the $D^*$ polarization vector
${\vec {\epsilon}}'$ in it.}\
\label{FigNellSys}
\end{figure}

The schematical Figure \ref{FigBDW} applies also this time.
The main difference from the decay $B \to D \ell_1 N$ discussed in
Appendix \ref{appD} is that now the $B$-$D^*$ matrix element is more
complicated than Eq.~(\ref{FFBD}), e.g., cf.~Ref.~\cite{NeuPRps}\footnote{
  We use the convention $\varepsilon^{0123}=+1$ \cite{IZ}, while Refs.~\cite{GiSi,NeuPRps} use $\varepsilon^{0123}=-1$. Further, we use the definition of $B^0$ and $D^0$ of Ref.~\cite{PDG2014}.}
\bes
\label{Hmu}
\bea
H^{\mu}_{(\eta=-1)} &\equiv& \langle D^{*-}(p_D) | {\overline c}(1 - \gamma_5) \gamma^{\mu} b | B^{0}(p_B) \rangle =
\langle {\overline D^{*0}}(p_D) | {\overline c}(1 - \gamma_5) \gamma^{\mu} b | B^{+}(p_B) \rangle
\label{Hetam}
\\
H^{\mu}_{(\eta=+1)} &\equiv& \langle D^{*+}(p_D) | {\overline b}(1 - \gamma_5) \gamma^{\mu} c | {\overline B^{0}}(p_B) \rangle =
\langle D^{*0}(p_D) | {\overline b}(1 - \gamma_5) \gamma^{\mu} c | B^{-}(p_B) \rangle \ ,
\label{Hetap}
\eea
\ees
and these matrix elements are written in terms of the form factors as
\bea
H^{\mu}
& = &
i 2 \eta  \frac{\varepsilon^{\mu \nu \alpha \beta}}{(M_B+ M_{\Dst})} \epsilon^*_{\nu} (p_D)_{\alpha} (p_B)_{\beta} V(q^2) - \left[ (M_B+M_{\Dst}) \epsilon^{* \mu} A_1(q^2)
  - \frac{\epsilon^* \cdot q}{(M_B+M_{\Dst})} (p_B+p_D)^{\mu} A_2(q^2) \right]
\nonumber\\
&& + 2 M_{\Dst} \frac{\epsilon^* \cdot q}{q^2} q^{\mu} \left( A_3(q^2) - A_0(q^2) \right) \ ,
\label{FFBDst}
\eea
where $A_3(q^2)$ is not independent
\be
A_3(q^2) = \frac{(M_B+M_{\Dst})}{2 M_{\Dst}} A_1(q^2) -
\frac{(M_B-M_{\Dst})}{2 M_{\Dst}} A_2(q^2) \ .
\label{A3}
\ee
We note that the first term in Eq.~(\ref{FFBDst}) has a factor $\eta = \pm 1$.
In the considered processes, we have $\eta=-1$ when $\ell_1^+$ is produced, and $\eta=+1$ when $\ell_1^-$ is produced.
The reduced decay amplitude for the processes $B \to D^{*-} \ell_1 N$ is
\bes
\label{TBDstNl}
\bea
    {\cal T}_{(\eta=-1)}  &=&  U^*_{\ell_1 N} V^*_{c b} \frac{G_F}{\sqrt{2}} \left[{\overline u}_{(N)}(p_N) \gamma_{\mu} (1 - \gamma_5) v_{(\ell_1)}(p_1) \right]
    H^{\mu}_{(\eta=-1)} \ ,
    \label{TBDstNletm}
    \\
 {\cal T}_{(\eta=+1)}  &=&  U_{\ell_1 N} V_{c b} \frac{G_F}{\sqrt{2}} \left[{\overline u}_{(\ell_1)}(p_1) \gamma_{\mu} (1 - \gamma_5) v_{(N)}(p_N) \right]
    H^{\mu}_{(\eta=+1)} \ ,
    \label{TBDstNletp}
    \eea
\ees
Square of the absolute value, summed over the final leptonic helicities,
gives
\be
|{\cal T}|^2 = |U_{\ell_1 N}|^2 |V_{c b}|^2 \frac{G_F^2}{2} L^{\mu \nu} H_{\mu} H_{\nu}^* \ ,
\label{Tsq1}
\ee
where $L^{\mu \nu}$ is the lepton tensor
\bea
L^{\mu \nu} & = & 2 {\rm tr} \left[ \fsl{p}_N \gamma^{\mu} \fsl{p}_1 \gamma^{\nu} (1 - \gamma_5) \right]
\nonumber\\
& = & 8 \left[ p_{N}^{\mu} p_1^{\nu} +  p_{N}^{\nu} p_1^{\mu} - g^{\mu \nu} p_N \cdot p_1 + i \eta \; \varepsilon^{\mu \nu \delta \eta} (p_N)_{\delta} (p_1)_{\eta} \right] \ .
\label{Lmunu}
\eea
The evaluation will be performed, as in Ref.~\cite{GiSi}, in
$W^*$-frame (i.e., in $\ell_1$-$N$ frame $\Sigma$), in which the momenta
will be denoted generically (without primes).
We have
\bea
L^{i j} & = & 4 q^2 \left[ \delta^{i j} \left( 1 - \frac{(M_N^2+M_1^2)}{q^2} \right) - {\hat e}^i {\hat e}^j \lambda\left(\frac{M_N^2}{q^2}, \frac{M_1^2}{q^2}, 1 \right) - i \eta \; \varepsilon^{i j k} {\hat e}^k \lambda^{1/2}\left(\frac{M_N^2}{q^2}, \frac{M_1^2}{q^2}, 1 \right) \right] \ ,
\label{Lij}
\eea
where $i,j=1,2,3$, $\varepsilon^{ijk}=\varepsilon^{0ijk}$ ($\varepsilon^{123}=+1$) is the antisymmetric 3-tensor, $\delta^{ij}$ is Kronecker delta, and $\lambda$ is the square of the function $\lambda^{1/2}$ of Eq.~(\ref{sqlam}). Further, ${\hat e}^j$ are the components of the
unitary spatial vector along the charged lepton direction in $\ell_1$-$N$ frame: ${\hat e} = {\hat p_1} = {\vec p}_1/| {\vec p}_1|$.
In $\ell_1$-$N$ frame, the ${\hat z}$-axis is defined to be the direction of
$W^*$: ${\hat z} = {\hat q}$ (where $q$ is in $B$ frame),
and ${\hat x}$ axis is in the same half-plane with ${\hat e} = {\hat p_1}$
and ${\hat z}$.
In this system of coordinates in $\ell_1$-$N$ frame, we have  [cf.~Fig.~\ref{FigNellSys}(a)]
\be
   {\hat e} \equiv {\hat p_1} = ({\rm sin}\theta_{\ell}, 0, {\rm cos} \theta_{\ell}) \ , \qquad (0 \leq \theta_{\ell} \leq \pi) \ .
   \label{hate}
   \ee

  Since now $N$ (and $\ell_1$) are massive particles, other components of the lepton tensor will contribute as well
  \bes
  \label{Ls}
  \bea
  L^{0 j} = L^{j 0} & = & 4  {\hat e}^j (M_N^2 - M_1^2) \lambda^{1/2}\left(\frac{M_N^2}{q^2}, \frac{M_1^2}{q^2}, 1 \right)  \ ,
  \label{L0j}
  \\
  L^{00} & = & 4 \left[ (M_N^2+M_1^2) - \frac{(M_N^2-M_1^2)^2}{q^2} \right] \ .
  \label{L00}
  \eea
  \ees
  On the other hand, in the mesonic expressions $H^{\mu}$, the $D^*$
  polarization 4-vector $\epsilon$ appears, whose general form is simple
  in $D^*$ frame ($\Sigma'$, primed). The coordinate system
  in $D^*$ frame is defined, in analogy with Ref.~\cite{GiSi}, in such a way
  that the ${\hat z}'$ axis is in the direction of $D^*$ meson in $B$ frame,
  i.e.,  ${\hat z}' = - {\hat z} = - {\hat q}$. Further, it is convenient
  to define ${\hat y}' = {\hat y}$, i.e., the $y$-axis in $D^*$ frame coincides with the $y$-axis in $\ell_1$-$N$ frame; as a consequence, ${\hat x}' = -{\hat x}$ [cf.~Fig.~\ref{FigNellSys}(b)].  The polarization vector in $D^*$ rest frame in this coordinate system is thus
  \be
  \epsilon'^{T} = (0, \sin \theta' \cos \phi', \sin \theta' \sin \phi', \cos \theta') \ .
  \label{epsp}
  \ee
  This polarization vector, when boosted to $\ell_1$-$N$ frame (and written in the coordinate system of $\ell_1$-$N$) is then
  \bes
  \label{eps}
  \bea
  \epsilon^0 &=& \frac{\sqrt{q^2}}{2 M_{\Dst}} \lambda^{1/2} \left( 1,
  \frac{M_{\Dst}^2}{q^2}, \frac{M_B^2}{q^2} \right) \cos \theta' \ ,
  \label{eps0}
  \\
  \epsilon^1 &=& - \sin \theta' \cos \phi' \ , \quad
  \epsilon^2 = \sin \theta' \sin \phi' \ ,
  \label{eps12}
  \\
  \epsilon^3 & = & - \frac{(M_B^2 - q^2 - M_{\Dst}^2)}{2 M_{\Dst} \sqrt{q^2}}
  \cos \theta' \ .
  \label{eps3}
  \eea
  \ees
  In this frame, it is useful to expand the hadronic components of
  $H^{\mu}$, Eq.~(\ref{FFBDst}), in the helicity basis of the $\Sigma$-frame (of the virtual $W$)
  \be
  H \equiv (H^{\mu})_{(\mu=0,\ldots,3)} = H_{-\eta} {\hat e}_{+} +
  H_{+\eta} {\hat e}_{-} + H^3 {\hat e}_3 + H^0 {\hat x}_0 \ ,
  \label{Hexp}
  \ee
  where
  \bea
      {\hat e}_{\pm} & = & \frac{1}{\sqrt{2}} ( \pm {\hat x} - i {\hat y}), \quad
      {\hat e}_3 = {\hat z}, \quad {\hat x}_0^T=(1,{\vec 0}) \ ,
      \label{helbas}
      \eea
 and we recall that $\eta=\pm 1$ if $\ell_1^{\mp}$ is produced, respectively.
 In terms of the form factors (\ref{FFBDst}), we have
  \bea
 H_{\pm 1} & = & \mp \frac{1}{\sqrt{2}} \sin \theta' e^{\pm i \phi'} {\bar H_{\pm}} \ , \quad H^3 = \cos \theta' {\bar H^3} \ ,
 \quad H^0 = - \cos \theta' {\bar H^0} \ ,
 \label{Hs}
 \eea
where
\bes
 \label{bHs}
 \bea
 {\bar H_{\pm 1}} &=& (M_B+M_{\Dst}) A_1(q^2) \mp V(q^2) \frac{|{\vec q}| 2 M_B}{(M_B+M_{\Dst})} \ ,
 \label{bHpm}
 \\
 {\bar H^3} & = & \frac{M_B^2}{2 M_{\Dst} \sqrt{q^2}} \left[
       (M_B+M_{\Dst}) A_1(q^2) \left(1 - \frac{(q^2+M_{\Dst}^2)}{M_B^2} \right)
         - 4 A_2(q^2) \frac{|{\vec q}|^2}{(M_B+M_{\Dst})} \right] \ ,
 \label{bH3}
 \\
 {\bar H^0} & = & \frac{M_B |{\vec q}|}{M_{\Dst} \sqrt{q^2}} \left[
        (M_B+M_{\Dst}) A_1(q^2)  - (M_B- M_{\Dst}) A_2(q^2) + 2 M_{\Dst} \left( A_0(q^2) - A_3(q^2) \right) \right] \ .
 \label{bH0}
 \eea
\ees
Here
\be
|{\vec q}| = \frac{1}{2} M_B \lambda^{1/2} \left( 1, \frac{q^2}{M_B^2}, \frac{ M_{\Dst}^2}{M_B^2} \right)
\label{magq}
\ee
is the magnitude of the 3-vector ${\vec q}$ of the virtual $W$ in the $B$-frame (note: in $\ell_1$-$N$ it is zero).
The absolute square $|{\cal T}|^2$ of the reduced amplitude (\ref{Tsq1}), summed over the final fermionic helicities (but not yet over the polarizations of $D^*$) is then obtained.   This then gives the following differential cross section with respect to the direction of $\ell_1$ in $\ell_1$-$N$ frame (${\hat p}_1$), and with respect to $q^2$ and the direction of the virtual $W$ in $B$-frame (${\hat q}$)
  \bea
  \frac{d \Gamma(\epsilon(\theta',\phi'))}{dq^2 d \Omega_{\hat q} d \Omega_{{\hat p}_1}} & = &
  \frac{1}{8^4 \pi^5} \frac{|U_{\ell_1 N}|^2 G_F^2 |V_{cb}|^2}{M_B^2}
  \blam^{1/2} 2 |{\vec q}| q^2 {\bigg \{}
  \left[
    2 \left(1 - \frac{(M_N^2+M_1^2)}{q^2} \right) -\blam \sin^2 \theta_{\ell})
    \right]
\frac{1}{2} \sin^2 \theta' \left( ({\bar H_{+1}})^2 + ({\bar H_{-1}})^2  \right)
\nonumber\\
&& +
\blam^{1/2} \cos \theta_{\ell} \sin^2 \theta' \left( ({\bar H_{+1}})^2 - ({\bar H_{-1}})^2  \right)
\nonumber\\
&& +
2 \left[ \left(1 - \frac{(M_N^2+M_1^2)}{q^2} \right) - \blam \cos^2 \theta_{\ell} \right] \cos^2 \theta' ({\bar H^3})^2
- \blam \sin^2 \theta_{\ell} \sin^2 \theta' \cos(2 \phi') {\bar H_{+1}} {\bar H_{-1}}
\nonumber\\
&& - \sin \theta_{\ell} \sin (2 \theta') \cos \phi' \left[ \blam^{1/2} {\bar H^3} ({\bar H_{+1}} - {\bar H_{-1}})
  + \blam \cos \theta_{\ell} {\bar H^3} ( {\bar H_{+1}} +{\bar H_{-1}}) \right]
\nonumber\\
&& +
\left( \frac{M_N^2-M_1^2}{q^2} \right) \blam^{1/2} \left[ \sin \theta_{\ell} \sin(2 \theta') \cos \phi' {\bar H^0} ({\bar H_{+1}} + {\bar H_{-1}}) +
  4 \cos \theta_{\ell} \cos^2 \theta' {\bar H^0}{\bar H^3} \right]
\nonumber\\
&&
2 \left[ - \left(\frac{M_N^2-M_1^2}{q^2} \right)^2 + \frac{(M_N^2+M_1^2)}{q^2}
    \right] \cos^2 \theta' ({\bar H^0})^2 {\bigg \}} \ ,
  \label{dGdq2domdom1}
  \eea
  where we denoted
  \be
  \blam \equiv \lambda \left( 1, \frac{M_N^2}{q^2}, \frac{M_1^2}{q^2} \right) \ .
  \label{blam}
  \ee
  We notice that the expression (\ref{dGdq2domdom1}) is independent of $\eta= \pm1$, i.e., the result is the same when $\ell_1^+$ or $\ell_1^-$ is produced in the $B \to D^{*} \ell_1 N$ processes.\footnote{
    This is a consequence of the fact that not only the leptonic tensors $L^{\mu \nu}$ depend on $\eta=\pm 1$, but also the hadronic matrix elements
    $H^{\mu}$, cf.~Eqs.~(\ref{FFBDst}) and (\ref{Lmunu}).}
  Summing over the three polarizations of $D^*$ then gives\footnote{
    This means, summing the cases of $\epsilon(\theta',\phi')$ for: (1) $\theta'=\pi/2$, $\phi'=0$; (2) $\theta'=\pi/2$, $\phi'=\pi/2$; (3) $\theta'=0$ (and $\phi'$ arbitrary).}
 \bea
  \frac{d \Gamma}{dq^2 d \Omega_{\hat q} d \Omega_{{\hat p}_1}} & = &
  \frac{1}{8^4 \pi^5} \frac{|U_{\ell_1 N}|^2 G_F^2 |V_{cb}|^2}{M_B^2}
  \blam^{1/2} 2 |{\vec q}|  q^2 {\bigg \{}
\left[2 \left(1 - \frac{(M_N^2+M_1^2)}{q^2} \right) -
  \blam \sin^2 \theta_{\ell}  \right]
\left( ({\bar H_{+1}})^2 + ({\bar H_{-1}})^2 \right)
\nonumber\\
&& + 2 \blam^{1/2} \cos \theta_{\ell}
\left( ({\bar H_{+1}})^2 - ({\bar H_{-1}})^2 \right)
+ 2 \left[ \left(1 - \frac{(M_N^2+M_1^2)}{q^2} \right) - \blam \cos^2 \theta_{\ell} \right] ({\bar H^3})^2
\nonumber\\
&& +
4 \left( \frac{M_N^2-M_1^2}{q^2} \right) \blam^{1/2} \cos \theta_{\ell} {\bar H^0}{\bar H^3}
+ 2 \left[ - \left(\frac{M_N^2-M_1^2}{q^2} \right)^2 + \frac{(M_N^2+M_1^2)}{q^2}
  \right] ({\bar H^0})^2 {\bigg \}} \ .
    \label{dGdq2domdom2}
\eea
The integration over $d \Omega_{{\hat p}_1} = 2 \pi d \cos \theta_{\ell}$ is then straightforward, and the subsequent integration over $d \Omega_{\hat q}$ gives factor $4 \pi$. This then leads to the following final result for the differential cross section with respect to the square $q^2$ of the virtual $W^2$ momentum $q$, summed over all final state helicities and polarizations:
\bes
\label{dGdq2}
\bea
\lefteqn{\frac{d \Gamma(B \to D^* \ell_1 N)}{d q^2} =
\frac{1}{64 \pi^3} \frac{|U_{\ell_1 N}|^2 G_F^2 |V_{cb}|^2}{M_B^2}
\blam^{1/2} |{\vec q}| q^2
{\Bigg \{}
\left( 1 - \frac{(M_N^2+M_1^2)}{q^2} - \frac{1}{3} \blam \right)
\left( ({\bar H_{+1}})^2 + ({\bar H_{-1}})^2 +({\bar H^3})^2  \right)
}
\nonumber\\
&&
+ \left[ - \left(\frac{M_N^2-M_1^2}{q^2} \right)^2 + \frac{(M_N^2+M_1^2)}{q^2} \right] ({\bar H^0})^2
{\Bigg \}}
\label{dGdq2a}
\\
&=&
\frac{1}{64 \pi^3} \frac{|U_{\ell_1 N}|^2 G_F^2 |V_{cb}|^2}{M_B^2}
 \blam^{1/2} |{\vec q}| q^2 {\Bigg \{}
 \left( 1 - \frac{(M_N^2+M_1^2)}{q^2} - \frac{1}{3} \blam \right)
 {\bigg [} 2 (M_B+M_{\rm D})^2 A_1(q^2)^2
 \nonumber\\
 &&
 + \frac{8 M_B^2 |{\vec q}|^2}{(M_B+M_{\Dst})^2} V(q^2)^2 +  \frac{M_B^4}{4 M_{\Dst}^2 q^2} \left( (M_B+M_{\Dst})
   \left( 1 - \frac{(q^2+M_{\Dst}^2)}{M_B^2} \right) A_1(q^2) -  \frac{4 |{\vec q}|^2}{(M_B+M_{\Dst})} A_2(q^2) \right)^2 {\bigg ]}
 \nonumber\\
 &&
 + \left[ - \left(\frac{M_N^2-M_1^2}{q^2} \right)^2 + \frac{(M_N^2+M_1^2)}{q^2} \right] \frac{M_B^2 |{\vec q}|^2}{M_{\Dst}^2 q^2}
 \left(2 M_{\Dst} A_0(q^2) \right)^2
 {\Bigg \}} \ .
\nonumber\\
 \label{dGdq2b}
 \eea
 \ees
 The last expression was obtained from the expression (\ref{dGdq2a}) by
 using the relations (\ref{bHs}) and (\ref{A3}).
 We recall that $\blam$ and $|{\vec q}|$ are given in Eqs.~(\ref{blam}) and (\ref{magq}), respectively.

 When the masses of the final state fermions $\ell_1$ and $N$ are both zero, the terms containing hadronic components ${\bar H^0}$ reduce to zero everywhere, because the components $L^{0j}$ and $L^{00}$ of the lepton tensor disappear, cf.~Eqs.~(\ref{Ls}). In such a case, the final result (\ref{dGdq2}), with $|U_{\ell_1 N}| \mapsto 1$, reduces to the corresponding (zero fermion mass) result of Refs.~\cite{GiSi} and \cite{NeuPRps}.

 \section{The differential decay widths}
 \label{appdiff}

The differential decay widths for the decays $d \Gamma(\pi^{\pm} \to e^{\pm} e^{\pm} \mu^{\mp} \nu)/d E_{\mu}$, with intermediate on-shell $N$ neutrino, were written in Refs.~\cite{CDK,CKZ,symm}. Here we write a somewhat generalized variant of these differential decay widths, namely those corresponding to the processes of Fig.~\ref{FigNLNCV} in Sec.~\ref{subs:diff}. The specific case considered in Sec.~\ref{subs:diff}, in Eqs.~(\ref{diff1}) and (\ref{diff2}) refers to $\ell_2 = e$ and $\ell_3 = \mu$.  We will denote the (total) energy of lepton $\ell_2$ in $N$ rest frame as $E_2 \equiv E_{\ell_2}$. The masses of the two charged leptons $\ell_2$ and $\ell_3$ are denoted as $M_2$ and $M_3$, respectively.

For the LNC decay $N \to \ell_2^- W^{*+} \to \ell_2^- \ell_3^+ \nu_{\ell_3}$, Fig.~\ref{FigNLNCV}(a), we have
\bes
\label{dGLNC}
\bea
\lefteqn{
\frac{d \Gamma^{\rm (LNC)} ( N \to \ell_2^- W^{*+} \to \ell_2^- \ell_3^+ \nu_{\ell_3})}{d E_2} = |U_{\ell_2 N}|^2 \frac{d \bG^{\rm (LNC)} (N \to  \ell_2^- \ell_3^+ \nu_{\ell_3})}{d E_2}
}
\label{dGLNCa}
\\
\frac{d \bG^{\rm (LNC)} (N \to  \ell_2^- \ell_3^+ \nu_{\ell_3})}{d E_2} & = &
\frac{G_F^2}{12 \pi^3} \sqrt{(E_2^2-M_2^2)} \frac{\left[ (M_N - 2 E_2) M_N + M_2^2- M_3^2 \right]^2}{\left[ (M_N - 2 E_2) M_N + M_2^2 \right]^3}
\nonumber\\
&& \times {\big \{} 8 E_2^3 M_N^2 - 2 M_2^2 M_N (M_N^2 + M_2^2 + 2 M_3^2) -
2 M_N E_2^2 \left[ 5 (M_N^2 + M_2^2) + M_3^2 \right]
\nonumber\\
&& + E_2 \left[ 3 M_N^4 + 10 M_2^2 M_N^2 + 3 M_3^2 (M_N^2 + M_2^2) + 3 M_2^4 \right] {\big \}} \ ,
\label{dGLNCb}
\eea
\ees
The energy $E_2$ varies in the interval $M_2 \leq E_2 \leq (M_N^2 + M_2^2 - M_3^2)/(2 M_N)$.

Analogously, for the LNV decay $N \to \ell_3^+ W^{*-} \to \ell_3^+ \ell_2^- \nu_{\ell_2}$, Fig.~\ref{FigNLNCV}(b), we have a somewhat simpler expression
\bes
\label{dGLNV}
\bea
\frac{d \Gamma^{\rm (LNV)} ( N \to \ell_3^+ W^{*-} \to \ell_3^+ \ell_2^- \nu_{\ell_2})}{d E_2} &=& |U_{\ell_3 N}|^2 \frac{d \bG^{\rm (LNV)} (N \to  \ell_2^- \ell_3^+ \nu_{\ell_2})}{d E_2}
\label{dGLNVa}
\\
\frac{d \bG^{\rm (LNV)} (N \to  \ell_2^- \ell_3^+ \nu_{\ell_2})}{d E_2} & = &
\frac{G_F^2}{2 \pi^3} E_2 \sqrt{(E_2^2-M_2^2)} \frac{\left[ (M_N - 2 E_2) M_N + M_2^2- M_3^2 \right]^2}{\left[ (M_N - 2 E_2) M_N + M_2^2 \right]} \ .
\label{dGLNVb}
\eea
\ees
The energy $E_2$ varies in the same interval as in the LNC case: $M_2 \leq E_2 \leq (M_N^2 + M_2^2 - M_3^2)/(2 M_N)$.

\end{document}